\documentclass[aps,pre,showpacs,twocolumn,showkeys,longbibliography]{revtex4-1}
\usepackage{amssymb}
\usepackage{amsmath}
\usepackage{graphicx}
\graphicspath{ {Machintosh_HD/Utenti/admin/Scrivania} }

\newcommand{\Section}[1]{\section{#1}}
\newcommand{\Pick}{\,^{|}}
\newcommand{\DoublePick}{\,^{|\hspace{-1pt}|}}
\newcommand{\PickNeutered}{\,^{\nmid}}
\newcommand{\Eq}[1]{Eq. (\ref{#1})}

\begin{document}
\title{How to Select Observers}
\author{Robert Garisto}
\affiliation{American Physical Society, 1 Research Road, Ridge, NY 11961}
\begin{abstract}
A number of problems in physics, mathematics, and philosophy involve observers in given situations which lead to debates about whether observer-specific information should affect the probability for some outcome or hypothesis.  Our purpose is \emph{not} to advocate for such observer selection effects, but rather to show that any such effects depend greatly on the assumptions made.
We focus on the debate about the existence of a `Doomsday effect'---whether observer index information should cause one to favor possibilities with fewer observers, which has been argued to have implications for models of cosmology.  
Our central goal is to reconcile the apparent inconsistencies in the literature by introducing a
formalism to 
lay bare assumptions made and 
address a  
key
issue that 
has not been clearly articulated
in such problems: whether the observer is selected by \textit{picking from}  or \textit{being in} a set of worlds. In the former there generally are observer selection effects, and in the latter there generally are not.  This leads us to differentiate what we call \emph{inclusive} from \emph{exclusive} selection, and how they relate to the concept of a multiverse.  Then we relax the assumption that all  observers are equally typical, and consider the problem of Boltzmann brains, showing that typicality can play a role in solving the problem.  We then stress the need for scale-invariant questions, which causes us to analyze J. Richard Gott's approach to the problem.  This all allows us to analyze the Doomsday and Universal Doomsday arguments.  We find that there is no Doomsday effect, absent a set of assumptions we find somewhat unreasonable. Then we use our formalism to resolve a debate in the philosophy community called the `Sleeping Beauty Problem.'  Finally, we conclude with a heuristic summary, free from equations, and point to possible future directions of this line of research.
\end{abstract}
\maketitle


\Section{Introduction} \label{Section_Introduction}

Physicists usually shun observer-specific information, and for good reason.  Our theories are based on invariances, such as those with respect to space and time, and should not depend on who is testing them.  Emmy Noether showed that conservation laws are rooted in symmetries \cite{noether1918}.  Yet we accept boundary conditions and symmetry breaking because of the constraints of the real world.  And sometimes just being an observer can bias our viewpoint.  It took millennia for humans to realize that we were not the center of the Universe and that we are atypical collections of matter in being confined to the surface of a habitable planet.  Some of the apparent coincidences which seem necessary for life to have evolved may be due to generalizing this notion of us being atypical \cite{carter_1974,Weinberg:CC}. But our purpose here is to focus on one particular type of observer effect: that probabilities we assign to the selection of an entity may differ if the entity is an observer because the observer has the capacity to self-select.  We will see that changing assumptions can completely change these effects, so, at a minimum, anyone invoking them, or decrying them, should carefully lay out all assumptions made.  

The quintessential example is the `Doomsday argument' \cite{Carter1983}, about which there is much debate 
\cite{DieksSIA,Bostrom96,Bostrom_book,OlumSIA,BostromSIA,Knobe-Olum-Vilenkin,Gerig-Olum-Vilenkin}.
Suppose you assign some prior probability $p$ for case $S$, that the `world' of which you are a part (and we will define `world' in various ways) will persist only for a short time, with a relatively small number of `people' ever living in that world.  The other possibility, $L$, is that it will persist longer, with more total `people,' to which you assign probability $1-p$.  But you realize that in your guess for $p$, you have neglected to take into account any possible observer selection effects (OSEs).  The Doomsday argument says that you should adjust $p$ upward because the probability is small that you would just happen to live very, very early in the life of a world, and thus you are more likely to live in a short-lived world for which you would be more typical.  Is that right?  It depends on your assumptions.

Throughout most of the paper, we will be talking about probabilistic situations where there is a set $P$ of `people' (entities capable of being observers, though not always the primary observer in the situation) from which one is selected, and we want to know the probability that the `person' belongs to a subset of $P$ associated with some property, \emph{e.g.} ``born before the year 2100."  
A key question is whether the `person' self-selects directly from set $P$ (which is generally embedded in enclosing sets such as worlds), which we call a `Be-selection' (a Be for short), or whether they are selected in some other way, which we call a `Pick-selection' (a Pick for short).  In most of our scenarios, the latter entails more than one selection because in order to pick an element of set $P$ one must generally first pick an element of one of the sets that encloses $P$ (\emph{e.g.}, to pick a nut from a set of jars, one must first pick one of the jars). 
The posterior probabilities for Be and Pick selection differ: OSEs tend to arise in the latter but not the former.

Philosopher Nick Bostrom has written much about the Doomsday argument \cite{Bostrom96,Bostrom_book,BostromSIA}.
He too discusses two possible ways an observer could be selected, often using problems of prisoners, which make good toy models because they entail observers confined to specific enclosing sets (cells in cellblocks in prisons).  We will assume through most of the paper what he calls the Self-Sampling Assumption (SSA), which just means that you assume you are equally likely to be any member of the set of possible observers you define in your problem,
\emph{i.e.,} it is an assumption of \emph{typicality}.
He also considers something called the Self-Indication Assumption (SIA), which says you should weight the probability of your existence by the number of people in the world in which you exist  \cite{DieksSIA,OlumSIA,GarrigaVilenkin2008}.  
This is essentially a kludge factor, and why it has rightly been found to be problematic \cite{BostromSIA,GarrigaVilenkin2008,Gerig-Olum-Vilenkin,Carroll:2017BB}.  
In fact, the SIA gives the wrong answer whenever there is a selection from an enclosing set, such as in the Warden problem we discuss in Section \ref{Section_Warden}, or when we take theories to be mutually exclusive, as in Section \ref{Section_Theories}.
Nevertheless, we will see that the weighting factor associated with the SIA appears naturally with the SSA if we assume observers are Be-selected rather than Pick-selected.

So there are conflicting and problematic results and apparent misunderstandings in the literature, and much of this is due to there being no universal notation. Our goal in writing this paper is to resolve these issues. Central to doing so is our novel nested-set notation, which we hope will allow authors to make clear their assumptions on how observers are selected, so readers can judge for themselves whether the assumptions made, and the results they lead to, are reasonable.

The paper is structured as follows.  In the next two sections, we consider the selection of observers within `worlds' (prisoners in cellblocks), first via a Be-selection, and then via a Pick-selection, showing how OSEs arise in the latter. In the following two sections, we discuss what happens if we embed the worlds in an enclosing set $E$, and there is just one Be-selection on $P$ (an \emph{inclusive selection}), or an additional Pick on set $E$ (an \emph{exclusive selection}), again with OSEs in the latter.  
If we take set $E$ to comprise `everything' then we term the inclusive case \emph{the inclusiverse} and the exclusive case \emph{an exclusiverse}. The key difference between them is that in the former we assume that all hypothesized things exist, and in the latter we do not. This leads to a general principle: it is effects of the latter which lead to OSEs.
Later we discuss whether it is possible to distinguish these two cases, and relate them to the term `multiverse', but our purpose is to lay out how to calculate probabilities given certain assumptions, not to posit the nature of reality.
Next we discuss spaces of theories, typicality, and the issue of `freak' observers in cosmology called Boltzmann brains and how our analysis can frame that problem.  
Then we consider an analysis by J. Richard Gott \cite{Gott1993}, which lets us phrase the Doomsday argument in a scale-invariant way.  We are then ready to fully address the Doomsday argument, and what has been called `Universal Doomsday.'  We show that while many sets of assumptions lead to no Doomsday effect, it is possible to come up with a set of assumptions, however implausible, which leads to one. Then we address a related problem in philosophy called the `Sleeping Beauty Problem.'
Finally, we summarize our results and point to future directions.  

In an effort to make the paper readable to the wider world, the summary is comprehensive of our results without equations. We have also put details of 
our nested-set notation and a table that summarizes our results into Appendix A.  And in the body of the paper, we spell out many intermediate steps in our equations since some interested in the results here may include those less familiar with working out such steps.

\Section{To Be: Prisoner Problem} \label{Section_Prisoner}
Imagine you are a prisoner and have the following information: The prison you are in has two types of cellblocks, small ($S$) and large ($L$), which contain $\bar{n}_S$ and $\bar{n}_L$ cells per cellblock respectively. You want to estimate the probability that you are in an $S$ cellblock.

\begin{figure}[htbp]
\begin{center}
\includegraphics[width = 85mm]{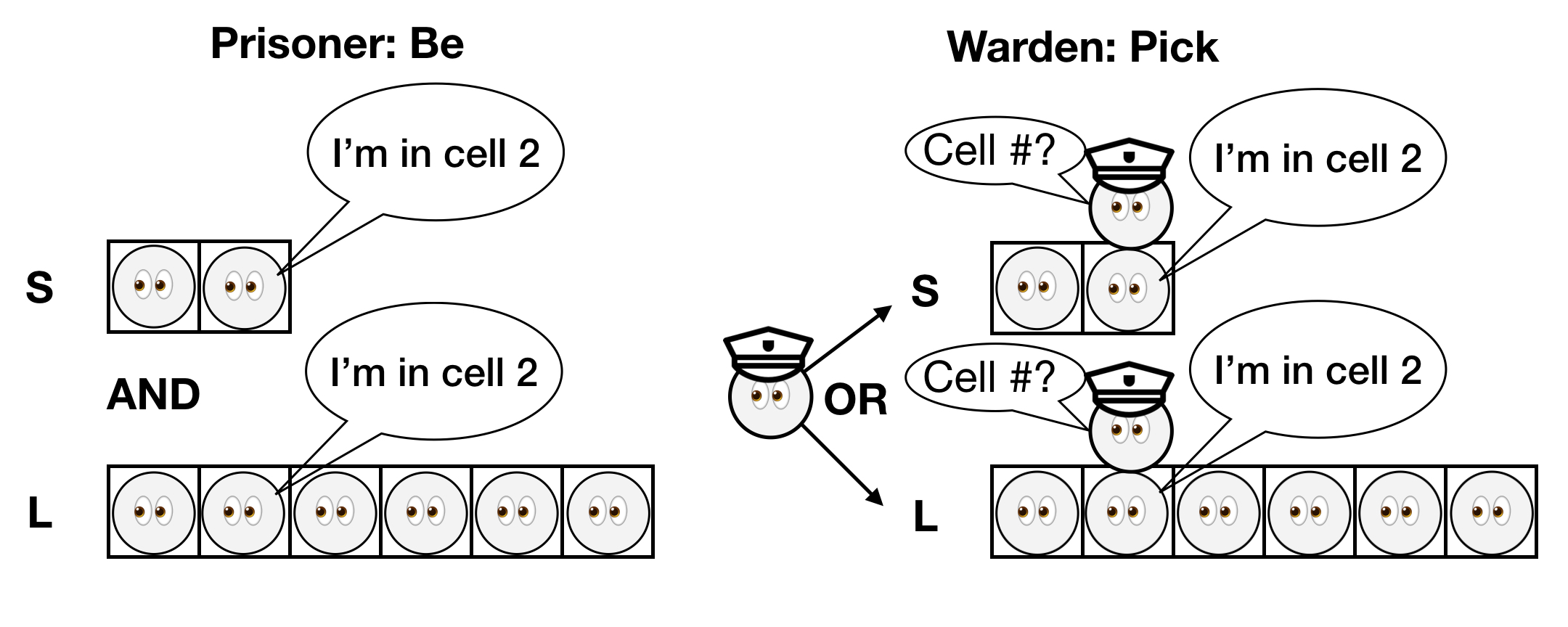}
\caption{Why the Warden Problem (with a Pick selection)  leads to an OSE and the Prisoner Problem (with a Be selection) does not: There are two cellblocks, $S$ and $L$.  Prisoners all simply ask themselves, ``Which cellblock am I in?" and then observe their cell number to answer.  There are more prisoners in the $L$ cellblock to ask the question, which cancels the rank factor that a smaller faction of prisoners are in the first two cells in $L$ than in $S$, so those in cell 2 are equally likely to be in either cellblock. The warden first must Pick a cellblock at random, then select a cell at random within that cellblock.  If the selected prisoner is in cell 2, it is more likely that the warden picked the $S$ than the $L$ cellblock because the number of prisoners per cellblock did \emph{not} affect the odds that she picked that cellblock, and so the rank factor is not canceled as it was in the Be case.}
\label{Fig_Prisoner_Warden}
\end{center}
\end{figure}

Before we dive into a lot of notation, let us consider a simple numerical example, where there is one cellblock of each type, with $\bar{n}_S=2$ and $\bar{n}_L=6$ (see the left side of Fig. \ref{Fig_Prisoner_Warden}). You do not know your cell number at the outset, so you could be in either the $S$ or $L$ cellblock. Now, you look at your door and learn your cell number. If it is greater than 2, you know you are in the $L$ cellblock. Let's assume that it is cell number 2, so you could be in either cellblock. What is the probability that you are in the $S$ cellblock? Well, there are exactly two cells with cell number 2, one in each cellblock. And you have no reason to favor one over the other, so you should assign a probability of 1/2 for being in the $S$ cellblock. Note that this is equal to the probability of picking the $S$ cellblock at random. In other words, the posterior probability for being in cellblock $S$, given the cell-number datum that you could be in either cellblock, is the same as the prior probability of randomly picking cellblock $S$---there is no observer selection effect.

Now, let us formalize the problem for a general number of prisoners and cellblocks. You assign labels $N_S$ and $N_L$ to the number of cellblocks of each type, but all you know is that there is at least one cellblock (since you are in one), \emph{i.e.}, $N \equiv N_S+N_L \geq 1$.  You also know that the prison is full and that each prisoner was assigned a random cell in the prison, with exactly one prisoner per cell. 
Let the ratio of cells in $L$ and $S$ cellblocks be 

\begin{equation}
\rho \equiv \bar{n}_L/ \bar{n}_S,
\end{equation}

\noindent
which is by assumption greater than 1. The bar just indicates we have normalized to the number of cellblocks.  The total number of prisoners in all cellblocks of type $J=S$ or $L$ is $n_J$, which is equal to the number of cells per cellblock of that type times the number of cellblocks of that type:

\begin{equation}
n_J=\bar{n}_J N_J.
\end{equation}

Let us call the set of prisoners $P$ (for `person,' the set that will usually hold our observers), and the set of cellblocks $W$ (for `world,' since this problem is an analogue to one of observers in worlds).  
$W_S$ and $W_L$ are the subsets of $W$ containing all $S$ and $L$ cellblocks, respectively.  Since there are only two types of cellblocks, the set $W$ is the union of them: $W=W_S \cup W_L$.  You assign some prior probability for what the fraction of small cellblocks $P(W_S)=N_S/N $ might be (we assume that the probability of picking any given cellblock is simply $1/N$, and these $P(W_S)$ and $P(W_L)$ are \emph{fixed} inputs---we will explore varying ratios of them in Section \ref{Section_Inclusive}).  Note that $P$ is nested within $W$, \emph{i.e.}, every element of $P$ (a prisoner) is associated with a particular element of $W$ (a cellblock).  
The compound set $PW_S$ contains the set of $S$ cellblocks, and the set of prisoners in $P$ who are in $S$ cellblocks (see Appendix A for details on notation).  

We will assume the Self-Sampling Assumption (SSA) \cite{Bostrom_book},

\begin{quote}
SSA: One should reason as if one is a random sample from the set of all observers in one’s reference class.
\end{quote}

\noindent
This is simply assuming \emph{typicality},
that the probability of you being in a subset of a larger set is simply equal to the fraction of observers of the reference class (which we call set $P$) who are in that subset.  For example, the probability to Be in subset $P_x$ of set $P$ is just $P(P_x | P)=n_x/n$.

You learn one datum, your cell number.  Divide the datum into two categories: $d$ if your cell number is $\leq \bar{n}_S$, and ${\neg d}$ if it is $>\bar{n}_S$.  The corresponding subsets of $P$ are $P_d$ and $P_{\neg d}$ ($P=P_d \cup P_{\neg d}$).
If your datum is ${\neg d}$, you know for sure that you are in an $L$ cellblock (because your cell number is greater than $\bar{n}_S$).  The case of interest is when the datum is $d$, where you could still be in either type of cellblock.  The question we want to answer in the Prisoner Problem is,

\begin{quote}
What is the posterior probability that a prisoner is in an $S$ cellblock, given that they match datum $d$?
\end{quote}

For convenience we define the number of people matching datum $d$ to be $m \equiv n_d$, and the number of people matching datum $d$ within a cellblock type $J$ to be $m_J \equiv n_{d,J}$, where $J= L$ or $S$.
All observers with cell numbers $\leq \bar{n}_S$ match datum $d$, so the number of people per cellblock matching datum $d$ is $\bar{m}=\bar{n}_S$, and this holds for both $S$ and $L$ cellblocks, so,

\begin{equation}
\label{equal_m}
\bar{m}=\bar{m}_S=\bar{m}_L=\bar{n}_S.
\end{equation}

We want to calculate the probability of you being in a cellblock type $S$ (\emph{i.e.}, in subset $P W_S$ of $PW$) given the datum, $d$, that you are in a low cell number (\emph{i.e.}, in subset $P_d W$ of $PW$), which we write at the conditional probability $P(P W_S | P_d W)$.  We will calculate this using Bayes' Law, so we need the likelihood of matching the datum given that we are in a cellblock type $S$,

\begin{equation}
\label{likelihood_P(DgivenS)}
P(P_d W | P W_S)=\frac{m_S}{n_S}=\frac{\bar{m}_S}{\bar{n}_S}=1,
\end{equation}

\noindent
and the probability \cite{endnotenotprior} to Be in cellblock type $S$,

\begin{equation}
\label{Beprior}
P(PW_S)=\frac{n_S}{n}=\frac{\bar{n}_S N_S}{\bar{n} N}=\frac{\bar{n}_S}{\bar{n}} P(W_S),
\end{equation}

\noindent
where $P(W_S)$ is the prior probability to Pick a cellblock of type $S$ (which, assuming random typical selection, is equal to our prior value for fraction of worlds, $N_S/N$).  

We need to pause here because \Eq{Beprior}, despite its simplicity, is the key to most of our results.  We have simply taken the SSA at face value.  Since the prisoner has an equal chance of being in any cell, the probability to Be in the subset of prisoners in $S$ cellblocks is simply the fraction of prisoners in such cellblocks, $n_S/n$, which as we show in \Eq{Beprior} is equal to the prior $P(W_S)$ weighted by the average number of prisoners $\bar{n}_S$ per cellblock of this type.
We should at this point note the competing assumption, the Self-Indication Assumption \cite{Bostrom_book}:

\begin{quote}
SIA: Given the fact that you exist, you should (other things equal) favor hypotheses according to which many observers exist over hypotheses on which few observers exist.
\end{quote}

\noindent
This \emph{does} giving the weighting factor seen in \Eq{Beprior}, but it is a kludge factor because it gives that factor regardless of how the observer is selected, which, as we shall see, is inappropriate whenever the first selection is from a set that encloses the observer.  
(Some may take the SIA to mean that this weighting factor should be applied {\it where appropriate}---not in any situation where you are an observer. If so, then a way to think of our formalism is that it shows when that weighting factor is appropriate.)
In contrast, we derived the weighting factor in \Eq{Beprior} simply using typicality (the SSA) and the recognition that we are selecting the observer directly from set $P$. The effect from how the observer is selected 
is made transparent by our nested-set notation.  
There are a number of places in the literature which simply refer to ``$P(S)$'' and let it equal to the prior probability for picking a world type $S$, when to be a prisoner requires $P(PW_S)$ with its weighting factor $\bar{n}_S/\bar{n}$.  Failing to include this factor leads to erroneous support for a Doomsday effect.

Here is another way to understand this weighting factor.  If you use the information that you are an observer in a random cell before also applying datum $d$, you are more likely to be in an $L$ cellblock than your prior for the fraction of $L$ cellblocks would suggest.  For example, if $P(W_S)=P(W_L)=1/2$, there are $\rho$ times as many observers in $L$ cellblocks as in $S$ cellblocks, and so the probability of being in a cellblock type $L$ (before knowing $d$) is $\rho$ times that of being in a cellblock type $S$. This factor of $\bar{n}_S$ in Eq. (\ref{Beprior}) will exactly cancel a factor of $1/\bar{n}_S$ in the likelihood Eq. (\ref{likelihood_P(DgivenS)}).  
(As we shall see in the next section, this factor is absent if there is a Pick on the world set $W$.  We should also note that by our formulation of the problem we are assuming that the prisoner could be in both types of cellblocks.  We will later consider the cases where there are mutually exclusive `universes' (Section \ref{Section_Exclusive}) and hypotheses (Section \ref{Section_Exclusive_Theory}).)

So the posterior probability of you being in a cellblock type $S$ given datum $d$ is given by Bayes' Law,

\begin{eqnarray}
\label{Prisoner_P(SgivenD)}
P&&(PW_S | P_d W) =  \frac{P(P_d W | P W_S) P(PW_S)} {P(P_dW)}  \nonumber \\
=&&\frac{P(P_d W | P W_S) P(PW_S)}{\sum_J P(P_d W | P W_J) P(PW_J)}
\nonumber \\
=&&\frac{\frac{\bar{m}_S}{\bar{n}_S} \frac{\bar{n}_S}{\bar{n}} P(W_S)}
{\sum_J \frac{\bar{m}_J}{\bar{n}_J} \frac{\bar{n}_J}{\bar{n}} P(W_J)}=
\frac{\bar{m}_S}{\bar{m}} P(W_S)= P(W_S),
\end{eqnarray}

\noindent
where $J=S$ or $L$, $\sum_J  \bar{m}_J P(W_J)=\bar{m}$, and $\bar{m}=\bar{m}_S=\bar{m}_L$.  The righthand side is the prior probability for picking a cellblock of type $S$---\emph{i.e.}, the probability before we have any observer information at all.  As we noted before, the prior here to \emph{pick} a world type $S$, $P(W_S)$, is a fixed value $N_S/N$, not updated by the datum.  What is updated is our posterior probability to \emph{be} in such a world.
(Note that we can also write this more compactly using the shorthand notation described in Appendix A, see \Eq{Prisoner_P(SgivenD)_Be_short}.)
We can express the fact that there is no net observer selection effect by comparing the ratio of probabilities after ($R_P$) and before ($R_W$) observer information:

\begin{eqnarray}
\label{Rpw_Be}
R_P &\equiv& \frac{P(PW_L | P_d W)}{P(PW_S | P_d W)} =\frac{P(W_L)}{P(W_S)}
\nonumber \\
R_W &\equiv& \frac{P(W_L)}{P(W_S)}, R_{P/W} \equiv \frac{R_P}{R_W} =1.
\end{eqnarray}

In the Prisoner Problem, using observer information, which includes the effect of you being in a small cellblock, as well as the likelihood of you being in a low-numbered cell, you obtain the prior probability to Pick a cellblock type $S$.  In short, in the Prisoner Problem, when your datum is $d$, there is no net observer selection effect ($R_{P/W}=1$).

\Section{To Pick: Urn and Warden Problems} \label{Section_Warden}

Now let $W$ be a set of urns, and $P$ a set of ping-pong balls in them.  Each urn contains either a large ($\bar{n}_L$) or small ($\bar{n}_S$) number of consecutively numbered balls---defining subsets $W_L$ and $W_S$.  You pick an urn at random, and a ball at random from the urn.  Before picking the ball, in fact before you actually picked an urn, you had a prior probability that the urn you picked is of type $S$, $P(W_S)$.  After seeing the ball, \emph{what is the posterior probability that the urn is type $S$?}, i.e.,

\begin{quote}
What is the posterior probability that you pick an $S$ urn and then a random ball in it, given that the ball you pick matches datum $d$?
\end{quote}

Again, let us first use a numerical example to build intuition. Suppose there are two urns, one $S$ and one $L$, with $\bar{n}_S=2$ and $\bar{n}_L=6$. You pick a random urn, and then pick a random ball from it (we shall see that this is the same as the Warden problem on the right side of Fig. \ref{Fig_Prisoner_Warden}). If the ball number is greater than 2, the urn you picked was the $L$ urn. So let's assume the same datum as before, that it is ball number 2, which corresponds to datum $d$. Now, before you knew the ball number, there was an equal chance that you picked the $S$ or $L$ urn. But once you have datum $d$, your posterior probability of having picked the $S$ urn has greatly increased because all the balls in the $S$ urn match $d$, whereas that is true only of 1/3 of the balls in the $L$ urn.  In fact, while your prior for picking the urns was equal, your posterior probability of picking the $S$ urn is 3 times that of picking the $L$ urn (3/4 vs. 1/4). Though the setup seems the same as in Section \ref{Section_Prisoner}, the fact that there was an initial selection of the urn makes all the difference.

Let us now go into the details. Obviously, if the ball's number is $>\bar{n}_S$, you will know that it is an $L$ urn and that posterior probability is 0.  So let's assume that the datum $d$ you get is that the ball's number is  $\leq \bar{n}_S$.  It is tempting to say that the situation is identical to the Prisoner example, and that we learn nothing about the urn.  After all, both kinds of urns have the same number of balls with number less than $\bar{n}_S$.  But the situation is different because \emph{in order to pick the ball from the urn, we first had to pick the urn}.  To denote that selection, we put a Pick sign ``$\Pick$'' between sets (see Appendix A for more on our set notation).  So to Pick any ball from any urn is $P \Pick W$, and to Pick a ball matching datum $d$ from an $S$ urn is $P_d \Pick W_S$.  Thus what we seek is $P(P \Pick W_S | P_d \Pick W)$, the probability of picking a ball from an $S$ urn given that we picked a ball matching datum $d$.

The probability of matching datum $d$ given the urn is type $S$ is exactly the same as Eq. (\ref{likelihood_P(DgivenS)}) because if it is \emph{given} that you picked an S urn, the Pick has no effect on the likelihood, it is `neutered' (see Appendix A) and we put a slash through the Pick sign to indicate this:

\begin{equation}
\label{likelihood_P(DgivenS)_Pick}
P(P_d \PickNeutered W | P \PickNeutered W_S)=P(P_d  W | P  W_S)=\frac{m_S}{n_S}=\frac{\bar{m}_S}{\bar{n}_S}=1,
\end{equation}

\noindent 
and with $\bar{m}_S=\bar{n}_S$ (grouping all the balls matching datum $d$ together), $P(P_d \Pick W | P \Pick W_S )=1$.  However, the probability of picking a ball from an urn of type $S$ is \emph{not} the same as Eq. (\ref{Beprior}) because there is no weighting for the number of balls.  The probability of picking an $S$ urn and then picking a ball from it is same as the prior probability for picking an $S$ urn,

 \begin{equation}
\label{Pickprior}
P(P \Pick W_S)= P(W_S).
\end{equation}

\noindent
Because of this, there is no factor of $\bar{n}_S$ in the numerator to balance the $1/\bar{n}_S$ rank factor in the likelihood, so Bayes' Law does not just return the prior as it did in the Be case in \Eq{Prisoner_P(SgivenD)}:

\begin{eqnarray}
\label{Prisoner_P(SgivenD)_Pick}
P&&(P \Pick W_S | P_d \Pick W)=\frac{P(P_d \PickNeutered W | P \PickNeutered W_S) P(P \Pick W_S)}{\sum_J P(P_d \PickNeutered W | P \PickNeutered W_J) P(P \Pick W_J)}
\nonumber \\
=&&\frac{\frac{\bar{m}_S}{\bar{n}_S} P(W_S)}
{\sum_J \frac{\bar{m}_J}{\bar{n}_J}  P(W_J)}= 
\frac{P(W_S)}
{\sum_J \frac{\bar{n}_S}{\bar{n}_J}  P(W_J)} \nonumber \\
=&& \frac{P(W_S)}
{P(W_S) + \frac{1}{\rho}  P(W_L)}.
\end{eqnarray}

\noindent
(Note, for shorthand notation, see \Eq{Prisoner_P(SgivenD)_Pick_short}.)  For $P(W_L)/\rho$ small, this goes to 1.  

The posterior probability for $L$ given $d$ is

\begin{equation}
\label{Prisoner_P(LgivenD)_Pick}
P(P \Pick W_L | P_d \Pick W)= \frac{\frac{1}{\rho} P(W_L)}
{P(W_S) + \frac{1}{\rho}  P(W_L)},
\end{equation}

\noindent
which, for equal priors, goes to $1/\rho$ for $P(W_L)/\rho$ small. As in Section \ref{Section_Prisoner}, the prior here is a fixed input $N_L/N$ that is unchanged by the datum.  Our posterior is the probability of the \emph{urn that we picked} to be type $L$.  To see how data can update a multivalued prior with Pick selection, see Sections \ref{Section_Exclusive} and \ref{Section_Universal}.

The ratios for $P$ and $W$ become, 

\begin{eqnarray}
\label{Rpw_Pick}
R_{P \Pick} &\equiv& \frac{P(P \Pick W_L | P_d \Pick W)}{P(P \Pick W_S | P_d \Pick W)} 
=\frac{1}{\rho} \frac{P(W_L)}{P(W_S)}
\nonumber \\
R_W &\equiv& \frac{P(W_L)}{P(W_S)}, R_{P \Pick/W} \equiv \frac{R_{P\Pick}}{R_W} =\frac{1}{\rho}.
\end{eqnarray}

\noindent
There is thus a very strong selection effect when one has to first Pick the urn ($R_{P \Pick/W}=1/\rho$).

Of course balls are not people, so it is tempting to think that it is the nature of the elements of set $P$ that causes the difference with the Prisoner Problem.  To counter that, consider what we call the Warden Problem, where $P$ is again a set of prisoners in cellblocks $W$.  But this time, instead of the prisoner just \emph{being} the observer within a cellblock, a warden selects a prisoner by first \emph{picking} a random cellblock, and then picking a random prisoner within the cellblock, all without noting which type of cellblock she has picked.  So the question in the Warden Problem is,

\begin{quote}
What is the posterior probability that a warden picks an $S$ cellblock and then a random prisoner in it, given that the prisoner they pick matches datum $d$?
\end{quote}

\noindent
Then all follows exactly as in the Urn problem, and the posterior probability we seek is $P(P \Pick W_S | P_d \Pick W)$.  The warden has a prior probability $P(W_S)$ for having picked a cellblock type S, the likelihood that she gets datum $d$ given that she picked a cellblock type $S$ is one (\emph{i.e.}, $P(P_d \Pick W | P \Pick W_S )=1$), and by Bayes' Law, her posterior probability given datum $d$ is given by Eq. (\ref{Prisoner_P(SgivenD)_Pick}), with a large selection effect, $R_{P \Pick/W}=1/\rho$.  

The reason the Warden Problem differs from the Prisoner Problem is that the warden has to first Pick a cellblock, whereas the prisoner is there without needing to be picked by anyone else.  See Fig. \ref{Fig_Prisoner_Warden}. (It may help your intuition to imagine $\bar{n}_L$ huge, say 2000 so $\rho=1000$. The Prisoner problem is unchanged since if you satisfy $d$  you are still in cell 1 or 2 of your cellblock, but in the Warden problem she is certain to pick cell 1 or 2 if she picks the $S$ cellblock but there is only one chance in 1000 that she she will do that in the $L$ cellblock.)

We note that if we try to use the SIA in this problem, we will get the wrong answer. If you are a prisoner and a warden picks your cell at random after having picked your cellblock at random, and you learn you match datum $d$, you should conclude that you are likely in an $S$ cellblock. But the SIA would have you weight your prior probability to be in a given cellblock by the number of cells, as in \Eq{Beprior}, falsely leading you to conclude that there is no OSE, whereas typicality (the SSA) gives you the correct unweighted prior of \Eq{Pickprior}.

Just to highlight further, it is the Pick on the nesting set $W$ that causes a change in the posterior probability. Consider the Warden Cafeteria problem, where all the prisoners are in a cafeteria, and the warden Picks a prisoner at random.  If that prisoner is from a cell number  $\leq \bar{n}_S$, what is the probability that they came from an $S$ cellblock?  Now the selection is directly from set $P$, or equivalently, from inside of the nested set $PW$, so that the posterior probability is $P(P W_S | P_d W)$, just as in the Be case---there is no observer selection effect in the Warden Cafeteria problem.  A Pick directly from the observer set is the same as a Be on that set (see Appendix A).  What causes a change in the posterior probability is a Pick on a set in which $P$ is nested, such as $W$.

\Section{Inclusive Selection}  \label{Section_Inclusive}

However many nested sets we have, there are two possibilities: either there is just a selection on the innermost set (a Be, unless there is a way to directly Pick from it as in the Warden Cafeteria problem), which we call \emph{inclusive} selection, or there is also at least one selection on one of the enclosing sets (a Pick in all of our examples because we do not consider any sets enclosed by (to the left of) $P$), which we call \emph{exclusive} selection. The selection in the Prisoner Problem is inclusive and in the Warden Problem it is exclusive. 

Suppose we have a larger enclosing set, $E$, in which $P$ and $W$ are nested.  For the Prisoner and Warden Problems, this could be the set of all prisons, each of which has their own small-to-large cellblock ratio.  We can even take $E$ to encompass everything that we deem possible---such as a set of universes in all possible configurations.  Then we define two possibilities for the reality:

\begin{quote}
The \emph{inclusiverse}:  All things we deem possible are realized.

An \emph{exclusiverse}: Only some of the things we deem possible are realized.
\end{quote}

\noindent
The key question is whether all things to which we assign a nonzero probability actually occur (inclusive selection), or there are some mutually exclusive possibilities (exclusive selection).   Perhaps a quantum example is useful.  If one assumes that quantum theory is unitary and all pieces of the wave function with nonzero amplitude are realized, so that Schr\"odinger's cat is both alive and dead (as in the Many Worlds case), that is inclusive selection.  If one assumes that the wave-function collapses to a specific eigenvalue, so that Schr\"odinger's cat is alive or dead, not both, that is an exclusive selection. In the rest of this section we study inclusive selection, though not its implications for reality.

Let's consider inclusive selection for the Prisoner Problem, but with a much more modest set, where $E$ is the set of all prisons we consider and the only selection is the self-selection of the prisoner.  If we think that there are exactly two types of prisons, say with all $S$ cellblocks or all $L$ cellblocks, then the key to inclusiveness is that we calculate probabilities under the assumption that both types of prisons exist---there is no Pick on the selection of $E$ needed. We explicitly show the sum over subsets of $E$, $e$, so when we do the same calculation for the exclusive case, the difference will be apparent.  For simplicity we will assume that the number of prisoners for any $J=S$ or $L$ cellblock is the same across all prisons, so
$\bar{n}_{J,e}=\bar{n}_J$, and similarly we assume the number of prisoners per cellblock matching datum $d$ is the same, $\bar{m}_{J,e}=\bar{m}_J$.  The subsets $E_e$ differ only in their fractions of $S$ and $L$ worlds. The likelihood for the inclusive case comes out the same as in the Be case, \Eq{likelihood_P(DgivenS)}:

\begin{eqnarray}
\label{likelihood_P(DgivenS)_inclusive}
P&&(P_d W E | P W_S E) \nonumber \\
=&&
\sum_e P(P_d W E_e | P W_S E_e) P(P W_S E_e | P W_S E) \nonumber \\
=&&\frac{\bar{m}_S}{\bar{n}_S} \sum_e P(P W_S E_e | P W_S E) 
=\frac{\bar{m}_S}{\bar{n}_S}=1.
\end{eqnarray}

\noindent
There is no $e$ dependence in the first term, since we assumed that $\bar{n}_S$ and $\bar{m}_S$ do not depend on $e$. The prior to Be in cellblock type $S$ with inclusive selection of $E$ is 

\begin{eqnarray}
\label{Beprior_inclusive}
&&P(PW_SE)=
\sum_e P(P W_S E_e | P W E_e) P(P W E_e) \nonumber \\
&&=\sum_e \frac{\bar{n}_{S,e}}{\bar{n}_{,e}} P(W_S E_e | W E_e) \frac{\bar{n}_{,e}}{\bar{n}} P(W E_e) 
 \\
&&=\frac{\bar{n}_S}{\bar{n}} \sum_e P(W_S E_e | W E_e) P(W E_e)=\frac{\bar{n}_S}{\bar{n}} P(W_S E), \nonumber
\end{eqnarray}

\noindent
which is the same as \Eq{Beprior}, just the prior probability of picking a world of type $S$ weighted by the number of observers per world type $S$.  Note that a factor of $1/ \bar{n}_{,e}$ and $\bar{n}_{,e}$ cancel here.  Therefore, the posterior probability of you being in a cellblock type $S$ given datum $d$ with an inclusive selection of $E$ is the same as \Eq{Prisoner_P(SgivenD)},

\begin{eqnarray}
\label{Prisoner_P(SgivenD)_inclusive}
P&&(PW_S E | P_d W E)=\frac{P(P_d W E| P W_S E) P(PW_S E)}{\sum_J P(P_d W E| P W_J E) P(PW_J E)}
\nonumber \\
=&&\frac{\frac{\bar{m}_S}{\bar{n}_S} \frac{\bar{n}_S}{\bar{n}} P(W_S E)}
{\sum_J \frac{\bar{m}_J}{\bar{n}_J} \frac{\bar{n}_J}{\bar{n}} P(W_J E)}=
\frac{\bar{m}_S}{\bar{m}} P(W_S E) \nonumber \\
=&& P(W_S E),
\end{eqnarray}

\noindent
just the prior probability of picking a world of type $S$, and we again get $R_{P/W}^E=1$ as in \Eq{Rpw_Be}.  There is no net observer selection effect for the Prisoner Problem in the inclusive case ($R_{P/W}^{E}=1$).  Generalizing, if we are considering a problem where observers are selected only by being, and there is no other selection---all allowed possibilities are realized, as in the inclusiverse---then there is no OSE.

\Section{Exclusive Selection}  \label{Section_Exclusive}

Let us analyze the Prisoner Problem with exclusive selection.  The key difference from the inclusive case is that we must Pick a subset $E_e$: although we posit that there are multiple possibilities $E_e$, only one of them is actually realized.  As we said in the previous section, if $E$ is the set of everything possible, and we take reality to correspond to a smaller subset, then we live in an exclusiverse.  But we will focus on a more mundane set:  for the Prisoner Problem, those subsets of $E$ are prisons.  

The defining characteristic of these subsets $E_e$ is the fraction of worlds of type $S$ they contain, which we define as $y$. So the probability of picking an $S$ world, 

\begin{equation}
\label{define_y}
y \equiv P(W_{S} E_e | W E_e), 
\end{equation}

\noindent
and a world of type $L$, $1-y =P(W_{L} E_e | W E_e)$, is the same for all elements of  a given $E_e$.  That is, $E_e$ is completely specified by its $y$---in fact we will simply label these subsets by $y$.
Again we assume for simplicity that the number of prisoners per type of world is independent of $e$: $\bar{n}_{J,e}=\bar{n}_J$ and $\bar{m}_{J,e}=\bar{m}_J$.
But note that the average number of prisoners per cellblock in a given prison, $\bar{n}_{,e}$ varies from prison to prison:

\begin{eqnarray}
\label{n_e}
\bar{n}_{,e} = \bar{n}_S P(W_{S,e}) + \bar{n}_L P(W_{L,e}) \nonumber \\
\equiv \bar{n}_y= \bar{n}_S (y + \rho (1-y)). 
\end{eqnarray}

The likelihood in the exclusive case is the same as in inclusive case \Eq{likelihood_P(DgivenS)_inclusive} because the Pick of subset $E_e$ on the first term in the sum is neutered:

\begin{eqnarray}
\label{likelihood_P(DgivenS)_exclusive}
P&&(P_d W \Pick E | P W_S \Pick E) \nonumber \\
&&=
\sum_e P(P_d W \PickNeutered E_e | P W_S \PickNeutered E_e) P(P W_S \Pick E_e | P W_S \Pick E) \nonumber \\
&&=\frac{\bar{m}_S}{\bar{n}_S} \sum_e P(P W_S \Pick E_e | P W_S \Pick E) 
=\frac{\bar{m}_S}{\bar{n}_S}=1.
\end{eqnarray}

\noindent
However, the prior is different because now we have to first Pick a subset $E_e$, and there is not a $\bar{n}_{,e}$ to cancel the $1/\bar{n}_{,e}$ as there was in \Eq{Beprior_inclusive},

\begin{eqnarray}
\label{Beprior_exclusive}
&&P(PW_S \Pick E)=
\sum_e P(P W_S \PickNeutered E_e | P W \PickNeutered E_e) P(P W \Pick E_e) \nonumber \\
&&=\sum_e \frac{\bar{n}_{S,e}}{\bar{n}_{,e}} P(W_S \PickNeutered E_e | W \PickNeutered E_e) P(E_e) \nonumber \\
&&= \sum_y \frac{y}{y + \rho (1-y)} P(\Pick y).
\end{eqnarray}

\noindent
For the last line, we have assumed again $\bar{n}_{S,e}=\bar{n}_S$, relabeled the subsets $E_e$ by $y$, and used the definitions for $y$ in \Eq{define_y} and $\bar{n}_{,e}$ in \Eq{n_e}.  The sum covers all values of $y$ from 0 to 1 with nonzero $P(\Pick y)$, which is the probability of picking an ensemble element of type $y$ (it is shorthand for $P(PW \Pick E_y)$---see Eqs. (\ref{shorthandalpha}-\ref{Pickswithy})). 
(Note that as with the Warden problem, the SIA gives the wrong answer here because $P(E_e)$ should \emph{not} be weighted by $\bar{n}_{,e}$ in \Eq{Beprior_exclusive} since we are first Picking subsets of $E$.)
Similarly for $L$,

\begin{eqnarray}
\label{L_exclusive}
&&P(P_d W \Pick E | P W_L \Pick E) = \frac{\bar{m}_L}{\bar{n}_L}, \nonumber \\
&&P(PW_L \Pick E) = \frac{\bar{n}_L}{\bar{n}_S} \sum_y \frac{1-y}{y + \rho (1-y)} P(\Pick y).
\end{eqnarray}

Let us use Bayes' Law again to obtain the posterior probability  of you being in a cellblock type $S$ or 
$L$ given datum $d$ in the exclusive case, which has the same form as the inclusive case \Eq{Prisoner_P(SgivenD)_inclusive} except with Picks on $E$, which we obtain from Eqs. (\ref{likelihood_P(DgivenS)_exclusive}--\ref{L_exclusive}):

\begin{eqnarray}
P&&(PW_S \Pick E | P_d W \Pick E)  \nonumber \\
&&=\frac{P(P_d W \Pick E | P W_S \Pick E) P(PW_S \Pick E)}{\sum_J P(P_d W \Pick E| P W_J \Pick E) P(PW_J \Pick E)}
\nonumber \\
\label{Prisoner_P(SgivenD)_exclusive}
&&=\frac{\sum_y \frac{y}{y + \rho (1-y)} P(\Pick y)}
{\sum_y \frac{y +\frac{\bar{m}_L}{\bar{m}_S} (1-y)}{y + \rho (1-y)} P(\Pick y)} 
= \frac{\sum_y \frac{y}{\rho - (\rho -1)y} P(\Pick y)}
{\sum_y \frac{1}{\rho - (\rho -1)y} P(\Pick y)} , \\
P&&(PW_L \Pick E | P_d W \Pick E) \nonumber \\ 
&&=\frac{P(P_d W \Pick E | P W_L \Pick E) P(PW_L \Pick E)}{\sum_J P(P_d W \Pick E| P W_J \Pick E) P(PW_J \Pick E)}
\nonumber \\
\label{Prisoner_P(LgivenD)_exclusive}
&&=\frac{\frac{\bar{m}_L}{\bar{m}_S} \sum_y \frac{1-y}{y + \rho (1-y)} P(\Pick y)}
{\sum_y \frac{y +\frac{\bar{m}_L}{\bar{m}_S} (1-y)}{y + \rho (1-y)} P(\Pick y)}
= \frac{\sum_y \frac{1-y}{\rho - (\rho -1)y} P(\Pick y)}
{\sum_y \frac{1}{\rho - (\rho -1)y} P(\Pick y)} ,
\end{eqnarray}

\noindent where we use $\bar{m}_S=\bar{m}_L$ of \Eq{equal_m} and we rewrote the denominators to collect the $y$ dependence.  We are again interested in the ratio of $L$ to $S$ posterior probabilities,

\begin{equation}
\label{R_P_exclusive}
R_P^{\Pick E} \equiv \frac{P(PW_L \Pick E | P_d W \Pick E)}{P(PW_S \Pick E | P_d W \Pick E)}
=\frac{\sum_y \frac{1-y}{\rho - (\rho -1)y} P(\Pick y)}{\sum_y \frac{y}{\rho - (\rho -1)y} P(\Pick y)}.
\end{equation}

\noindent
We want to normalize this to,

\begin{eqnarray}
\label{R_W_exclusive}
R&&_W^{\Pick E} \equiv \frac{P(W_L \Pick E)}{P(W_S \Pick E)}=\\
&&\frac{\sum_e P(W_L \PickNeutered E_e | W \PickNeutered E_e) P(W \Pick E_e)}
{\sum_e P(W_S \PickNeutered E_e | W\PickNeutered E_e) P(W \Pick E_e)}
=\frac{\sum_y (1-y) P(\Pick y)}{\sum_y y P(\Pick y)}. \nonumber
\end{eqnarray}

We can see immediately that if there is only one value $Y$ for which $P(\Pick y=Y)$ is nonzero, both $R_P^{\Pick E}$ and $R_W^{\Pick E}$ are equal to $(1-Y)/Y$ and their ratio,  $R_{P/W}^{\Pick E}$ is 1---no observer selection effect.  That's because that is really the inclusive case---while there is a Pick on $E$, it is neutered, and all of the values ({\it i.e.}, the one value) are realized.  So for the exclusive case, there needs to be more than one allowed value of $y$.

So let us explore different assumptions for the function $P(\Pick y)$, which, to remind you, is our prior probability for elements of $E$ with $S$-world fraction $y$.  For simplicity, let us define the probability density,

\begin{equation}
p(\Pick y)\equiv P(\Pick [y,y+dy])/dy
\end{equation}

\noindent
where now $y$ is not a set of discrete values, but all real numbers in $[0,1]$.  
We can then write the sums in Eqs. (\ref{R_P_exclusive}) and (\ref{R_W_exclusive}) as integrals:

\begin{eqnarray}
\label{RPE}
R_P^{\Pick E} = \frac{\int_0^1 dy \frac{1-y}{\rho - (\rho -1)y} p(\Pick y)}{\int_0^1 dy \frac{y}{\rho - (\rho -1)y} p(\Pick y)}, \\
\label{RWE}
R_W^{\Pick E} = \frac{\int_0^1 dy (1-y) p(\Pick y)}{\int_0^1 dy \ y p(\Pick y)}.
\end{eqnarray}

\subsection{Near a Single Point}
Let us first explore the case where we take $y$ to have a nonzero probability near a single point $Y$, in particular that $p(\Pick y)$ is constant over the range $Y-\sigma$ to $Y+\sigma$, where of course $\sigma$ is no larger than $Y$ or $1-Y$ so that the points are on the range 0 to 1:

\begin{equation}
\label{Py_near}
p(\Pick y)_{near} = \frac{1}{2 \sigma} (\Theta(y-(Y-\sigma)) - \Theta(y-(Y+\sigma))).
\end{equation}

\noindent
($\Theta(x)$ is the step function, equal to 0 for $x<0$ and 1 for $x\ge1$.)  Plugging this into \Eq{RWE}, for the prior ratio probabilities or picking $L$ worlds to $S$ worlds, we get

\begin{equation}
\label{RWE_near}
R_W^{\Pick E} = 
\frac{[y -\frac{1}{2} y^2]^{Y+\sigma}_{Y-\sigma}}{[\frac{1}{2} y^2]^{Y+\sigma}_{Y-\sigma}} = 
\frac{1-Y}{Y},
\end{equation}

\noindent
just as we obtained for a single point. (This is true because the integrand in the numerator and denominator of $R_W^{\Pick E}$  are linear in $y$.)  The expression for $R_P^{\Pick E}$  is more complicated because of the denominator of the integrands.
In the limit of $\sigma \to 0$, $R_P^{\Pick E}$ is,

\begin{equation}
\label{RPE_near}
R_P^{\Pick E} \simeq 
\frac{1-Y}{Y} 
\bigglb [1 - \frac{1}{3} \sigma^2 
\frac{\rho -1}{(\rho (1-Y) +Y) Y(1-Y)} \biggrb ],
\end{equation}

\noindent
and thus their ratio is,

\begin{equation}
\label{RPWE_near}
R_{P/W}^{\Pick E}  \simeq 
1 - \frac{1}{3} \sigma^2 
\frac{\rho -1}{(\rho (1-Y) +Y) Y(1-Y)}.
\end{equation}

Thus if $p(\Pick y)$ is nonzero within $\pm \sigma$ of a single point $Y$, there is a small observer selection effect of order $\sigma^2$.  In the limit that $\rho \to \infty$ (actually one must be careful when $Y$ is near 1, so really we take $\rho (1-Y) \to \infty$), 

\begin{equation}
\label{RPWE_near_limit}
R_{P/W}^{\Pick E} \to
1 - \frac{1}{3} \sigma^2 
\frac{1}{Y(1-Y)^2}.
\end{equation}

\noindent
So the closer we restrict our prior to be near a single point $Y$, the less $R_{P/W}^{\Pick E}$ differs from 1, and this behavior is independent of $\rho$.

\subsection{Flat Prior}

The simplest prior assumption is that every value of $y$ is equally likely,

\begin{equation}
\label{Py_flat}
p(\Pick y)_{flat} = 1.
\end{equation}

\noindent
From \Eq{RWE} this gives equal probability of picking $S$ and $L$ worlds,

\begin{equation}
\label{RWE_flat}
R_W^{\Pick E} = 
\frac{[y -\frac{1}{2} y^2]^1_0}{[\frac{1}{2} y^2]^1_0} = 1,
\end{equation}

\noindent
which we also could have obtained from \Eq{RWE_near} for $Y=\sigma=1/2$.  The posterior ratio of being in $L$ and $S$ worlds, $R_P^{\Pick E}$, is thus unchanged when normalized to $R_W^{\Pick E}=1$, and for their ratio we obtain, 

\begin{equation}
\label{RPWE_flat}
R_{P/W}^{\Pick E} =
\frac{1-(\ln \rho + 1)/\rho}{\ln \rho -1 +1/\rho}
\to \frac{1}{\ln \rho -1},
\end{equation}

\noindent
where we take the limit of $\rho \to \infty$ (this approximation is good only for $\rho \gtrsim 100$).  So for a flat prior, we get an observer selection effect which goes roughly as $1/\ln\rho$, in between the original Prisoner Problem, $R_{P/ W}=1=\rho^0$, and Warden Problem, $R_{P \Pick/ W}=\rho^{-1}$.  

If the point of choosing a flat prior is to minimize the effect of assumptions on the outcome, it might make more sense to use inclusive selection instead of a flat-prior exclusive selection---to say that all values of $y$ are realized rather than one of them is realized with equal probability for each.  Assuming the latter leads to a small observer selection effect while the former does not.

\subsection{Two Separated Points}

To get a sense of how much the Prisoner Problem in the exclusive case can approach the Warden Problem, it suffices to consider a prior with nonzero probabilities at two points, $Y \pm \sigma$, where $0 <Y < 1$ and $0< \sigma \leq \min{(1/2,Y,1-Y)}$, so that both points lie in the range $[0,1]$:

\begin{equation}
\label{Py_two}
p(\Pick y)_{two} = \frac{1}{2} (\delta(y-(Y-\sigma)) + \delta(y-(Y+\sigma))).
\end{equation}

\noindent
($\delta(x) =1$ for $x=0$ and is 0 otherwise.)  Since the integrands in $R_W^{\Pick E}$ are linear the $\sigma$ terms cancel, and we again get  $R_W^{\Pick E} = (1-Y)/Y$.  For $R_P^{\Pick E}$, we obtain,

\begin{equation}
\label{RPE_two}
R_P^{\Pick E} =
\frac{1 -Y(2-1/\rho)+ (Y+\sigma)(Y-\sigma)(1 - 1/\rho)}
{Y - (Y+\sigma)(Y-\sigma)(1 - 1/\rho)}.
\end{equation}

\noindent
If we assume $Y=1/2$, and define $k\equiv 2 \sigma$, then $R_W^{\Pick E}= 1$ and \Eq{RPE_two} reduces to,

\begin{equation}
\label{RPE_two_Yhalf}
R_{P/W}^{\Pick E}(Y=1/2) =
\frac
{1 - k^2 + (1+k^2)/\rho}
{1 + k^2 + (1-k^2)/\rho}.
\end{equation}

\noindent
Note that $0< k \leq1$.  For $k$ near 0, $R_P^{\Pick E}$ approaches 1---two points very close together is very much like the inclusive case.  For $Y=1/2$ and $k=1$, \textit{i.e.} when the two points are $y=0$ and $y=1$,

\begin{equation}
\label{RPE_two_0_or_1}
R_{P/W}^{\Pick E}(y=0 \text{ or } 1) = \frac{1}{\rho}.
\end{equation}

\noindent
In other words, the Prisoner Problem in the exclusive case where the prior is that the prison is either all $L$ cellblocks ($y=0$) or all $S$ cellblocks ($y=1$), has the same observer selection effect as the Warden Problem in \Eq{Rpw_Pick}.  By insisting on an either-or-Pick on the enclosing set $E$, we have, in essence, turned a Be for the Prisoner into a Pick on which top-level subset she is in.

\begin{figure}[htbp]
\begin{center}
\includegraphics[width = 85mm]{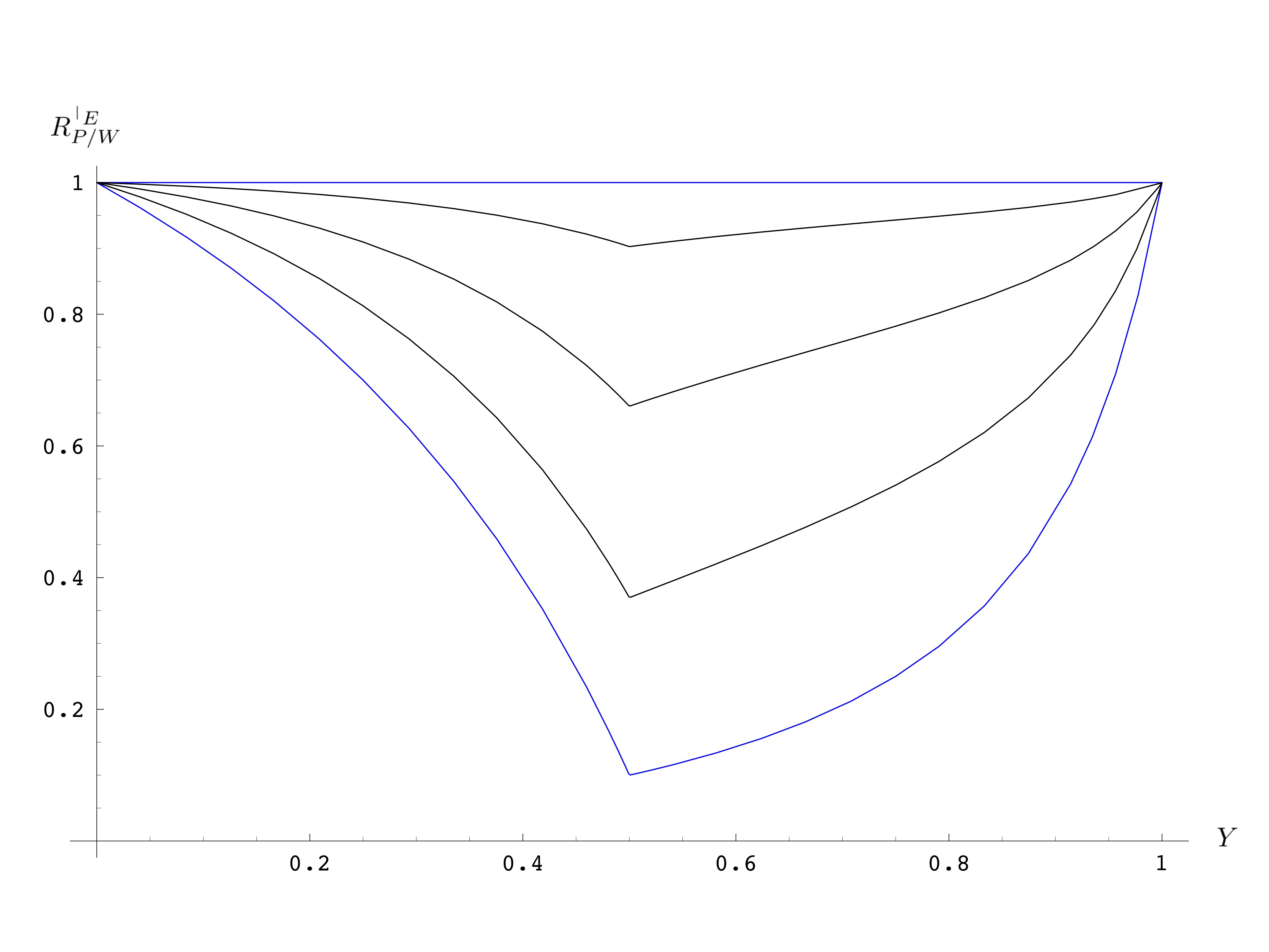}
\caption{
How to interpolate between the Prisoner (no OSE) and Warden ($1/\rho$ OSE) cases: For exclusive selection over an ensemble $\{E_y\}$ ($y$ is the fraction of worlds of type $S$ in that ensemble element) which consists of two separated points $y=Y\pm \sigma$, we plot a measure of the OSE, $R_{P/W}^{\Pick E}$ (the ratio of the ratios of posteriors to priors for $L$ and $S$ worlds for the exclusive Pick over ensemble $E$), versus $Y$ for $\rho=10$ (the ratio of the number of people per world of type $L$ to that of type $S$).  The OSE depends on how far apart the points are, which is characterized by $k \in (0,1]$  defined in \Eq{define_k}. Contours top to bottom are for $k=$ 0, 0.25, 0.5, 0.75 and 1.  There is no OSE for $k \to 0$ (akin to the Prisoner case).  The maximal OSE (minimal value of $R_{P/W}^{\Pick E}$) is for $k=1$ at $Y=1/2$ (akin to the Warden case), with a value $R_{P/W}^{\Pick E}(Y=1/2,k=1)=1/\rho=0.1$.
}
\label{Fig_Plot}
\end{center}
\end{figure}

So we can go anywhere from no OSE, as in the Prisoner case, to a Warden-level $1/\rho$ OSE simply by adjusting our prior assumptions.  In Fig. \ref{Fig_Plot}, we plot $R_{P/W}^{\Pick E}$ as a function of $Y$ for different values of $k$, which we more generally define as 

\begin{equation}
\label{define_k}
k \equiv 
\begin{cases}
\frac{\sigma}{Y} & Y \leq \frac{1}{2}, \\
\frac{\sigma}{1-Y} & Y \geq \frac{1}{2}.
\end{cases}
\end{equation}
  
\noindent
For $Y$ near 0 or 1, or $k$ near 0, $R_{P/W}^{\Pick E} \simeq 1 =\rho^0$, and the exclusive case is like the inclusive one.  The observer selection effect is maximized for $Y= 1/2$ and $k=1$, yielding  $R_{P/W}^{\Pick E}=\rho^{-1}$ of \Eq{RPE_two_0_or_1}.

\Section{Exclusive Theory Selection and the Presumptuous Philosopher} \label{Section_Theories}

\subsection{Exclusive Theory Selection}  \label{Section_Exclusive_Theory}

Instead of taking $E$ to be the top-level set, consider a set of theories, $\Theta$.  This set of theories might include very different hypotheses about reality, or they might simply specify different enclosed subsets, such as,

\begin{eqnarray}
\label{two_theories}
\Theta_L: \text{``All cellblocks are type $L$'"} \nonumber \\
\Theta_S: \text{``All cellblocks are type $S$"}  
\end{eqnarray}

\noindent
These two theories could have been encoded in $E$: they are $E_{y=0}$ and  $E_{y=1}$ respectively.  But we tend to approach theories differently from ensembles, notably that usually one assumes that only one theory is true, that we have to Pick a theory before proceeding further.  This is exclusive theory selection, and the probabilities are the same as in Section \ref{Section_Exclusive}.  For example, if our prior for the two theories in \Eq{two_theories} are equal, 
$R_{\Theta} =P(\Theta_L)/P(\Theta_S)= 1$, then 

\begin{equation}
\label{RPW_Theta}
R_{P \Pick/\Theta}= \frac{P(PW \Pick \Theta_L | P_d W\Pick \Theta)}{P(PW \Pick \Theta_S | P_d W\Pick \Theta)} = \frac{1}{\rho},
\end{equation}

\noindent
just as in \Eq{RPE_two_0_or_1}.
(This is assuming typicality (the SSA). Again the SIA gives the wrong answer because it does not take into account selections on enclosing sets, here the Pick selection on mutually exclusive theories.)

It is possible to have inclusive selection of a theory, where one assumes multiple theories are realized.  For example, one could posit that prisons vary from country to country, so both theories in \Eq{two_theories} would be realized somewhere.  There is then no Pick on $\Theta$, and one recovers the probabilities in the inclusive section, where there were no observer selection effects ($R_{P/\Theta}=1$).  One can even have a seemingly fundamental theory be part of an inclusive selection.  For example, the landscape in string theory allows different regions of the larger universe to manifest different low energy theories with their own fundamental constants.  If one posits that one can be an observer in any region of the landscape that has observers, that is inclusive theory selection.

As we said, the main point of this paper is to show that the conclusions one draws depend on the assumptions made.  If we assume exclusive selection, such as the theories in \Eq{two_theories} being mutually exclusive, we will conclude that there are observer selection effects, but if we assume an inclusive case, such as half the prisons have only $S$ cellblocks and half have only $L$ cellblocks, we will conclude that there are no such effects.

\subsection{Probing a Multiverse?}  \label{Section_Probing_Multiverse}

Suppose we consider both possibilities about the selection from set $P$ through set $E$: that it is inclusive as discussed in Section \ref{Section_Inclusive}, or exclusive, as discussed in Section \ref{Section_Exclusive}, and treat these as competing hypotheses, $\Theta_{in}$ or $\Theta_{ex}$.  If we treat these hypotheses as mutually exclusive, with a Pick on set $\Theta$, the overall selection is exclusive.  But let us focus on the rest of the selection, from $P$ to $E$, which is inclusive or exclusive.  We can then in principle use our data to alter our posterior probabilities for each hypothesis.
Suppose we define $E$ to be everything, so that the inclusive (exclusive) case corresponds to the inclusiverse (an exclusiverse). How do these terms relate to the term `multiverse'?  If taken literally, multiverse simply means that there are more realities than the one we perceive, either via something like parallel universes or just the universe being so large that realities similar to ours occur in some other part of it.  That does not actually imply that all possible universes are realized.  A set of a few parallel universes, which we will call \emph{a partial multiverse}, is an exclusiverse, since not everything possible is realized.  If \emph{all} possibilities are realized, to avoid ambiguity we will call it \emph{the complete multiverse}. So,

\begin{quote}
\emph{The inclusiverse} is the same as \emph{the complete multiverse}:  All things we deem possible are realized.

\emph{An exclusiverse} is the same as a universe or \emph{a partial multiverse}: Some things we deem possible are not realized.
\end{quote}

The question of this subsection is

\begin{quote}
Can we determine whether we live in the inclusiverse or an exclusiverse simply by using a datum such as the date? 
\end{quote}

To get a handle on this, let us consider the Prisoner Problem again, where our selection in sets $PW$ is a Be.  Let $PW$ again be embedded in a larger set $E$, which itself is considered in the context of one of two hypotheses,

\begin{eqnarray}
\label{two_selection_theories}
\Theta_{in}: \text{``Inclusive selection on $E$"} \nonumber \\
\Theta_{ex}: \text{``Exclusive selection on $E$"}  
\end{eqnarray}

\noindent
We need new notation to combine these hypotheses in a single probability, with a ``controlled-Pick" on $E$, so that there is a Pick on $E$ for hypothesis $ex$, but not for hypothesis $in$.  For this we put a left arrow pointing from $\Theta$ to the Pick on $E$:

\begin{eqnarray}
P(PW \Pick  \hspace{-2pt} \overleftarrow{\, E \Theta_{in} \hspace{-9pt}}\hspace{9pt}) = P(PWE\Theta_{in}), \nonumber\\
P(PW \Pick  \hspace{-2pt} \overleftarrow{\, E \Theta_{ex} \hspace{-9pt}}\hspace{9pt}) = P(PW\Pick E\Theta_{ex}).
\end{eqnarray}

\noindent
Using this notation, what we want to calculate is the posterior probability for hypotheses $h=in$ or $ex$ given datum $d$: 

\begin{eqnarray}
P_{h|d} &\equiv& P(PW \Pick \hspace{-2pt} \overleftarrow{\, E \Pick \Theta_{h} \hspace{-8pt}}\hspace{8pt} | P_d W \Pick \hspace{-2pt} \overleftarrow{\, E \Pick \Theta \hspace{-4pt}}\hspace{4pt}) 
 = \frac{P_{d|h} P_{h}}{P_d}  \\
 &=& \frac{P(P_dW \Pick \hspace{-2pt} \overleftarrow{\, E \PickNeutered \Theta_{h} \hspace{-8pt}}\hspace{8pt} | P W \Pick \hspace{-2pt} \overleftarrow{\, E \PickNeutered \Theta_h \hspace{-8pt}}\hspace{8pt}) P(PW \Pick \hspace{-2pt} \overleftarrow{\, E \Pick \Theta_{h} \hspace{-8pt}}\hspace{8pt} )}
 {P(P_dW \Pick \hspace{-2pt} \overleftarrow{\, E \Pick \Theta \hspace{-4pt}}\hspace{4pt})}. \nonumber
 \end{eqnarray}

\noindent
If we define our prior probabilities for $h=in$ and $ex$ to be $\alpha$ and $\beta$, respectively, \emph{i.e.},

\begin{eqnarray}
P_{in} &\equiv& P(PW \Pick \hspace{-2pt} \overleftarrow{\, E \Pick \Theta_{in} \hspace{-9pt}}\hspace{9pt}) \equiv \alpha, \nonumber \\
P_{ex} &\equiv& P(PW \Pick \hspace{-2pt} \overleftarrow{\, E \Pick \Theta_{ex} \hspace{-9pt}}\hspace{9pt}) \equiv \beta,
\end{eqnarray}

\noindent
then our posteriors are simply,

\begin{eqnarray}
\label{P_h_d_alpha_beta}
P_{in|d} &=& \frac{\alpha P_{d|in}} {\alpha P_{d|in} + \beta P_{d|ex}}, \nonumber \\
P_{ex|d} &=& \frac{\beta P_{d|ex}} {\alpha P_{d|in} + \beta P_{d|ex}}.
\end{eqnarray}

Note that we also need priors for the probabilities of the elements of $E$.
For simplicity, let us assume that the only ensembles with nonzero probability are $y=0$ (all $L$-type cellblocks) or $y=1$ (all $S$-type cellblocks), which we saw in \Eq{RPE_two_0_or_1} gives maximal OSE for the $ex$ case.  There is of course no OSE in the $in$ case.  For the inclusive case, let's assume equal probabilities for $y=0$ and $1$:

\begin{equation}
\label{prior_in_case}
P(E_0 \PickNeutered \Theta_{in}| E \PickNeutered \Theta_{in}) 
= P(E_1 \PickNeutered \Theta_{in}| E \PickNeutered \Theta_{in}) = \frac{1}{2},
\end{equation}

\noindent
but for the exclusive case let's allow them to vary,

\begin{equation}
\label{prior_ex_case}
P(E_0 \PickNeutered \Theta_{ex}| E \PickNeutered \Theta_{ex})  = q,\  
P(E_1 \PickNeutered \Theta_{ex}| E \PickNeutered \Theta_{ex})= p,
\end{equation}

\noindent
where $p+q=1$.  Our likelihoods are then

\begin{eqnarray}
\label{P_d_h}
P_{d|in} &=& \frac{\bar{m}} {\bar{n}}  = \frac{2} {\rho+1}, \nonumber \\
P_{d|ex} &=&q \frac{\bar{m}} {\bar{n_0}}+ p \frac{\bar{m}} {\bar{n_1}} =p+ \frac{q} {\rho}.
\end{eqnarray}

\noindent
We can then plug these likelihoods into \Eq{P_h_d_alpha_beta} to obtain the posterior probabilities.  It is clear that they depend on $p$ (with $q=1-p$).  

For $p=1/2$, so that the $y=0$ and $y=1$ weights in the $ex$ case match those of the $in$ case in \Eq{prior_in_case}, we obtain posterior probabilities,

\begin{eqnarray}
\label{P_h_d_p_onehalf}
P_{in|d} &\equiv& \alpha^\prime \vert_{p=1/2} = \frac{\alpha} {\alpha +  \beta \frac{(\rho+1)^2}{4\rho}}, \nonumber \\
P_{ex|d} &\equiv& \beta^\prime  \vert_{p=1/2} =\frac{\beta} {\beta + \alpha \frac{4\rho}{(\rho+1)^2}}.
\end{eqnarray}

\noindent
Since $\alpha$ and $\beta$ are $\geq 0$ and $\rho>1$ (so that $(\rho +1)^2>4 \rho$), the denominator for $\alpha^\prime$ ($\beta^\prime$) is larger (smaller) than $1$, and datum $d$ seems to decrease (increase) our credence in inclusive (exclusive) selection on $E$, except in the trivial case where 
$\alpha$ or $\beta$ is zero. This would seem to argue that if $E$ is a set of universes (not just prisons), we could use observer data to alter our probability that we live in the inclusiverse!  

But there is a second prior in this problem, that of $p$ (with $q=1-p$). 
We chose $p=1/2$ to make the probabilities for $y=0$ and $y=1$ the same as those in the inclusive case.  An equally reasonable hypothesis would be to set $p$ equal to the value that gives the same value for datum $d$ for each hypothesis, so that $P_{d|in}=P_{d|ex}=2/(\rho+1)$.  With a little algebra, we see that this holds for

\begin{equation}
\label{p_solution}
p=\frac{1} {\rho +1}.
\end{equation}

\noindent
For this value of $p$, the denominators in \Eq{P_h_d_alpha_beta} are $1$ (since $\alpha + \beta =1$), and

\begin{eqnarray}
\label{P_h_d_p_solution}
P_{in|d} &\equiv& \alpha^\prime \vert_{p=1/(\rho+1)}=\alpha, \nonumber \\
P_{ex|d} &\equiv& \beta^\prime \vert_{p=1/(\rho+1)}=\beta,
\end{eqnarray}

\noindent
so for this value of $p$ we gain no information about hypotheses $in$ and $ex$ from datum $d$.  

What happened?  When we thought, due to \Eq{P_h_d_p_onehalf}, that we had obtained information about hypotheses $in$ and $ex$ from datum $d$, what we really learned about was the probability of getting datum $d$ based on two factors, whether the selection from $E$ was inclusive or exclusive, \emph{and} the priors we had for the elements of $E$ in each case.  To the extent that $d$ tells us anything about these cases, it is about a combination of these factors.  We cannot disentangle these factors here.  In general, one cannot claim that data tell us about whether we are in the inclusiverse (aka the complete multiverse) or not unless one can show that all other factors which separate the inclusiverse from exclusiverse hypotheses are fixed.

\subsection{Presumptuous Philosopher}  \label{Section_Presumptuous}

In the Introduction, we noted that some authors argued against the Doomsday argument by assuming the Self-Indication Assumption (SIA): that we should weight the probability of some situation by the number of observers in it.  
As we have discussed,
this is essentially a kludge, adding the factor that we found in Be choices without the clearcut mathematical rationale we presented (based on applying the SSA---typicality---properly).  This is perhaps why it has been referred to as ``controversial'' \cite{GarrigaVilenkin2008,Gerig-Olum-Vilenkin}.

Nick Bostrom argues against the SIA with the following problem \cite{Bostrom_book,BostromSIA}.  A philosopher is told that theories $\Theta_L$ and $\Theta_S$ have equal probabilities prior to taking into account any observer information. This is like the problem of exclusive theory selection we considered in Section \ref{Section_Exclusive_Theory}, except that there is no datum $d$ favoring $S$ over $L$. The philosopher states that there is no need to test which is right (and since this is exclusive selection, only one is right) because, by the SIA, $\Theta_L$ is $\rho$ times more likely than $\Theta_S$ because there are $\rho$ times as many observers in that case.

Bostrom is right that the philosopher is being presumptuous here, and this is a good argument against the SIA---that if one is to Pick between $\Theta_S$ and $\Theta_L$, there should be no effect from there being more observers in the latter case, because we are {\it picking} a theory.  
This is simply an example of what we have found regarding the SIA---that it gives the wrong answer when there is a selection from an enclosing set, here $\Theta$.
But there is no reason to have invoked the SIA in the first place.

In short, the Presumptuous Philosopher has no bearing on our results because it argues against the SIA, which we did not use.

We note, however, that if the philosopher correctly uses the SSA and is asked about an inclusive problem, whether she is more likely to be in a domain of the inclusiverse governed by theory  $\Theta_L$ or $\Theta_S$, she would be correct to answer that she is more likely to be in the former due the SSA weighting by number of observers. In that case she is not presumptuous at all \cite{Scott_private1}.

\Section{Typicality}  \label{Section_Typicality}

All of the probabilities we have discussed thus far assume that the selection, Be or Pick, is \emph{typical}, that, for example, if the fraction of observers in some subset $P_a$ of $P$ is $n_a/n$, then the probability of selecting a person in that subset is also $n_a/n$.  Suppose we relax that assumption and allow \emph{atypical selection}, where the probability of selecting a person from subset $P_a$ differs from $n_a/n$---some values of $a$ are intrinsically more likely to be selected than others \cite{HartleSrednicki_Typical}.  For example, observers at CERN are not typical of Earth's population---they are more likely to be scientists than the population overall.  Srednicki and Hartle\cite{SrednickiHartle_LargeUniverse} describe an atypical selection in their Eq. (6.1):

\begin{equation}
\label{SrednickiHartleEq1}
``P(q_1 | T,\xi,D_0) = \sum_A \xi_A P(q1 @ A | T,D_0 @A)"
\end{equation}

\noindent
where $q_1$ is a posterior result, $T$ is a given theory, $D_0$ is data, $\xi$ is a `xerographic distribution,' which is a set of copies $A$ of $q_1$ at different locations meeting data $D_0$, and $\xi_A$ is the probability weight of xerographic occurrence $A$ which is not necessarily what we would obtain from a typical selection.  We need to translate this all into our notation.

\subsection{Atypical notation}

Let us define  $\xi^0$ to be a \emph{Typical Be}, a typical selection on the set $P$ (embedded in set $W$).  We are interested in subsets $P_a$ of $P$ for some property $a$ of the people in $P$:

\begin{equation}
\label{define_xi0}
\xi^0_a \equiv P(P_a W) = \frac{\bar n_a}{\bar n},\ \xi^0_{a|d} \equiv P(P_a W |P_d W)= \frac{\bar m_{a}}{\bar m}.
\end{equation}

\noindent 
Now let us define an \emph{Atypical Be} using $\xi$ to mark the atypical selection point,

\begin{equation}
\label{define_xi}
\xi_a \equiv P(^\xi P_aW),\ \xi_{a|d} = P(^\xi P_a W |^\xi P_d W),
\end{equation}

\noindent
which may not simply be a ratio of numbers of elements of set $P$.  However, for a given atypical selection $\xi$ on $P$, we will show that we can always find a new set $\tilde P$, with number of people per world $\tilde{\bar{n}}$, upon which a typical selection $\tilde \xi^0$,

\begin{equation}
\label{define_tilde_xi0}
\tilde{\xi}^0_a \equiv P(\tilde{P}_a W) = \frac{\tilde{\bar n}_a}{\tilde{\bar n}},\ \tilde{\xi}^0_{a|d} = P(\tilde{P}_a W |\tilde{P}_d W)= \frac{\tilde{\bar m}_{a}}{\tilde{\bar m}},
\end{equation}

\noindent
gives the same answer.  Here the tilde quantities are related to their counterparts by some scaling factors $\kappa_a$ and $\kappa_{a|d}$:

\begin{eqnarray}
\label{tilde-n_a-intof-kappa_a}
\tilde n_a &\equiv& \kappa_a n_a, \tilde n_{aK} \equiv \kappa_a n_{aK},  \nonumber \\
\tilde m_a &\equiv& \kappa_{a|d} m_a, \tilde m_{aK} \equiv \kappa_{a|d} m_{aK}.
\end{eqnarray}

\noindent
We claim that the Atypical Be on $P$, $\xi$, is equal to the Typical Be on $\tilde{P}$, $\tilde{\xi}^0$,

\begin{equation}
\label{xi_eq_tilde_xi0}
\xi_a=\tilde{\xi}_a^0,\ \xi_{a|d}=\tilde{\xi}_{a|d}^0,
\end{equation}

\noindent
if we define $\kappa_a$ as the ratio of atypical to typical selection,

\begin{equation}
\label{def_kappas}
\kappa_a \equiv c \frac{\xi_a}{\xi_a^0},\ \kappa_{a|d} \equiv c_d \frac{\xi_{a|d}}{\xi_{a|d}^0},
\end{equation}

\noindent
where constants $c$ and $c_d$ are independent of $a$.   We have the freedom to vary $c$ and $c_d$ because the overall numbers of people in $\tilde{P}$ do not matter, just the ratios we are interested in.  However, they do affect the values for $\tilde{\bar{n}}$ and $\tilde{\bar{m}}$:

\begin{eqnarray}
\label{n_and_tilde_n}
\tilde{\bar{n}}&=& \sum_a \tilde{\bar{n}}_a = \sum_a \kappa_a \bar{n}_a = c \bar{n} \sum_a \xi_a = c \bar{n},
\\
\tilde{\bar{m}}&=& \sum_a \tilde{\bar{m}}_a = \sum_a \kappa_{a|d} \bar{m}_a = c_d \bar{m} \sum_a \xi_{a|d} = c_d \bar{m}, \nonumber
\end{eqnarray}

\noindent
using the fact that probabilities for even atypically selected people sum to 1.  Note that we can choose to set $c$ and $c_d$ equal 1 and have $\tilde{\bar{n}}=\bar{n}$ and $\tilde{\bar{m}}=\bar{m}$, but we need not do this.  Now we can show Eq. (\ref{xi_eq_tilde_xi0}) does in fact hold,

\begin{eqnarray}
\label{proof_xi}
\xi_a &=& \frac{1}{c} \kappa_a \xi_a^0 = \frac{1}{c} \kappa_a \frac{\bar{n}_a}{\bar{n}}
=\frac{\tilde{\bar{n}}_a}{c \bar{n}}=\frac{\tilde{\bar{n}}_a}{\tilde{\bar{n}}} = \tilde{\xi}_a^0,\\
\xi_{a|d} &=& \frac{1}{c_d} \kappa_{a|d} \xi_{a|d}^0 = \frac{1}{c_d} \kappa_{a|d} \frac{\bar{m}_a}{\bar{m}}
=\frac{\tilde{\bar{m}}_a}{c_d \bar{m}}=\frac{\tilde{\bar{m}}_a}{\tilde{\bar{m}}} = \tilde{\xi}_{a|d}^0,
\nonumber
\end{eqnarray}

\noindent 
and we can write our atypical selection on $P$ as a typical selection on $\tilde{P}$ with number of elements defined by Eq. (\ref{tilde-n_a-intof-kappa_a}) with $\kappa$ defined in Eq. (\ref{def_kappas}).

\subsection{Posterior probability}

We can now write Srednicki and Hartle's Eq. (\ref{SrednickiHartleEq1}) in our notation.  We want the posterior probability $P(PW_K | P_d W)$, but with an Atypical Be, \emph{i.e.}, $P(^\xi PW_K | ^\xi P_d W)$:

\begin{eqnarray}
\label{atypical_posterior}
P(^\xi PW_K | ^\xi P_d W) = \sum_a \xi_{a|d} P(P_a W_K |P_{ad} W) \nonumber \\
= \left( \sum_a \xi_{a|d} \frac{\bar{m}_{aK}}{\bar{m}_a} \right) P(W_K) 
=\frac{\tilde{\bar{m}}_{K}}{\tilde{\bar{m}}}  P(W_K), 
\end{eqnarray}

\noindent
which we write as a Typical Be on set $\tilde{P}$ defined by Eqs. (\ref{tilde-n_a-intof-kappa_a}, \ref{def_kappas}).  This is the same expression as for a Be in Eq. (\ref{Prisoner_P(SgivenD)}) with the elements from set $\tilde{P}$.  Note that if we condition on a subset $a$, the selection within that subset is typical (all atypicality comes from nontrivial weighting of the different subsets $P_a$), thus  
$P(^\xi P_a W_K | ^\xi P_{ad} W) = P(P_a W_K | P_{ad} W)$.

\subsection{Atypical example}

Let's see how this atypical notation works in an example using prisoners of two types.  Suppose half the cellblocks are filled with humans ($a=h$) and half filled with zombies ($a=z$).  Humans are distributed as in the Prisoner Problem, ${\bar n_{hL}}= \rho  {\bar n_{hS}}$ and ${\bar m_{hL}}= {\bar m_{hS}}$.  Zombies have the same distribution in cells, ${\bar n_{zL}}= \rho  {\bar n_{zS}}$, but let us assume that all zombies who can think well enough to formulate a question, think they meet datum $d$, \emph{i.e.}, ${\bar m_{zL}}= \rho  {\bar m_{zS}}$.  If you think it is equally likely that you are a human or a zombie (because half the prisoners are humans, half zombies), and for simplicity you assume $P(W_S)=P(W_L)=1/2$, then you calculate the Typical Be posterior probabilities,

\begin{eqnarray}
\label{typical_example}
P(PW_S | P_d W) = \frac{\bar{m}_{S}}{\bar{m}}  P(W_S)
= \frac{2}{3+\rho}, \\
P(PW_L | P_d W) = \frac{\bar{m}_{L}}{\bar{m}}  P(W_L)
= \frac{1+\rho}{3+\rho}.
\end{eqnarray}

\noindent
Thus, unlike the Prisoner Problem, there \emph{is} an observer selection effect $R_{P/W}=(1+\rho)/2$, favoring that you are in $W_L$, because there are more zombies matching $d$ in $W_L$.

But suppose you think it is quite unlikely that you are a zombie, say because zombies don't usually use Bayesian reasoning.  For simplicity, you take $\kappa_{h|d}=1$ and set $\kappa_{z|d}$ to be some very small number $\kappa$---one zombie out of every $\kappa$ thinks well enough to calculate the probabilities we have been discussing (the ratio of chances you are a zombie to you are a human is $\kappa$, not $1$). Then you calculate the Atypical Be,

\begin{eqnarray}
\label{atypical_example}
P(^\xi PW_S | ^\xi P_d W) &=& \frac{\tilde{\bar{m}}_{S}}{\tilde{\bar{m}}}  P(W_S)
= \frac{1+\kappa}{2+\kappa(1+\rho)}, \\
P(^\xi PW_L | ^\xi P_d W) &=& \frac{\tilde{\bar{m}}_{L}}{\tilde{\bar{m}}}  P(W_L)
= \frac{1+\kappa\rho}{2+\kappa(1+\rho)}.
\end{eqnarray}

\noindent
There is still an observer selection effect $R_{P/W}=(1+\kappa \rho)/(1+\kappa)$, favoring $W_L$, but note that when $\kappa \to 0$, $R_{P/W} \to 1$, because there is no OSE due to the human prisoners.  If you \emph{assume} you are not a zombie, then you take $\kappa=0$ and all probabilities spring from $P_h$---in fact if you are going to do that, you might as well drop the label $h$ and ignore the zombies

\subsection{Redefine the conditional}

Another way of addressing an atypical selection which is due to different subsets $a$ meeting the conditional with different relative frequencies, is to redefine the conditional so the weights are the same.  For example, in the case above, we deweighted zombies by a factor $\kappa$ because only that fraction of zombies could formulate the question.  So why not limit the sets $P$ and $P_d$ to the subset $P_Q$ of $P$ of people who have formulated the Bayesian question in the first place?  As we discuss in Appendix A, adding such a conditional is not just another label, but actually redefining the set $P$ as set $[P_Q]$.  Then all we need to do is define set $\tilde{P} \equiv [P_Q]$, and typical selection on $\tilde{P}$ gives the probabilities for those atypical people who ask the question.

\subsection{Boltzmann brains}

Normal observers are necessarily far from equilibrium and experience an arrow of time of increasing entropy \cite{Aaronson_consciousness}. Fortunately, the observable Universe is in a relatively low entropy state \cite{Penrose1979,WALD2006394}.  How did it get that way? Ludwig Boltzmann argued that a low entropy `world' could arise as a stupendously rare fluctuation within a higher entropy world \cite{Boltzmann1897,Carroll:2017BB}.  The prevailing theory of cosmology is more subtle: that our Universe began within a patch of smooth spacetime, which inflated for a time at an exponential rate \cite{Inflation_Guth,*STAROBINSKY1980,*LINDE1982,*AlbrechtSteinhardt1982} (for a review, see \cite{TASI_inflation}).  Though inflation has ended here, it has likely not stopped everywhere in the larger Universe. Further, our observable Universe has seemingly entered another era of exponential expansion, and seems slated to approach de Sitter space (a spacetime with a positive cosmological constant $\Lambda$ and vanishing matter density) asymptotically.  

If so, the empty places greatly outnumber the places where normal observers can live.  Further, de Sitter space is a thermal state (with a temperature which depends only on the cosmological constant:  
$T=\sqrt{\Lambda/12 \pi^2}$) \cite{PhysRevD.15.2738}, and thus seems subject to worlds fluctuating into existence via stupendously rare fluctuations.  And one may not need such a large fluctuation, the size of a galaxy or a planet, to create observers, one may need only `Boltzmann brains'  \cite{schulman_1997,Dyson_Kleban_Susskind,AlbrechtSorbo}, which are spontaneously formed configurations of matter that, for a brief period, are self-aware, including ones that think they are having the thoughts you are having now.
Such events are still extremely improbable, occurring at a rate 
$\sim e^{-\Delta S}$, where $\Delta S$ is the reduction in entropy that the fluctuation represents.  For a brain-sized object, the timescale to form them, $\tau_{BB}$, will be enormous, of order $e^{10^{70}}$.   (Note that the units don't actually matter with numbers this large---switching from Planck times to Hubble times changes the googol-sized exponents by only about 140.)   But this is small compared to the timescale for a Hubble volume to fluctuate into existence, $\tau_{HV}$ of order $e^{10^{122}}$.
This is time enough to form googolplexes of Boltzmann brains, far more than the number of normal observers \cite{Carroll:2017BB}.   

One might ask why this is a problem.  We do not seem to be Boltzmann brains.  In fact, we need to assume that we are normal observers in order to do science.  And if one conditions on the assumption that we are normal observers, the probability of us being a freak observer is zero, no matter how common they are ($P(\rm{freak} | \rm{normal}) =0$).  The problem is that if freak observers outnumber us by a large enough factor, say a googolplex, there are many, many of them that think that they are experiencing any given moment that any normal observer does, and it is \emph{not} safe to assume that you are a normal observer.  So the problem is one of consistency: you need to assume that your observations reflect reality to do science, and thus it is a problem if the resulting science says that this assumption is very likely to be false. The problem is especially acute if there is an \emph{infinite} volume of spacetime which could spawn Boltzmann brains, and only a finite volume containing normal observers. This possibility led Don Page to argue that the Universe must decay rapidly, via bubbles of vacuum decay \cite{Coleman_false_vacuum}, so as to avoid any infinite patches of spacetime, leading him to predict a lifetime of our Universe shorter than about 20 billion years \cite{Page_20_billion_yrs, *Page_astronomical_rate}. Many papers have been written with less drastic proposed solutions, such as having the physical `constants' vary over time  \cite{Carlip_transient}.

We would like to know if our analysis of typicality has any impact on the Boltzmann brain problem.  Since freak observers may be fooled into thinking that they are normal only for a small fraction of their `life,' we use \emph{observer moments} instead of observers.
Let us assume that there are two types of observer moments per comoving Hubble volume, normal ($n$), and freak ($f$), with $\bar{n}_f=\rho \bar{n}_n$ for some constant $\rho$ which now 
can be any nonnegative real number,
and $\bar{n}=\bar{n}_n+ \bar{n}_f$ is the total number of observer moments per comoving Hubble volume. The probability to Be a normal observer moment is just the fraction of observer moments per comoving Hubble volume which are normal:

\begin{equation}
\label{P_normal}
P(P_n)= \frac{\bar{n}_n}{\bar{n}} = \frac{1}{1+\rho},
\end{equation}

\noindent
which is not close to $1$ unless $\rho \to 0$.  But what we really want is the fraction of observer moments in which the observer is self-aware and could ask a question like ``Am I normal?" in the first place.  The typical freak observer moment which superficially seems like a normal observer moment might not pass that test.   Let us assume that freak observer moments are $\kappa$ times likely as normal moments to do so.  Then we are interested in the atypical selection $P(^\xi P_n)$, which is a typical selection on set $\tilde P$, scaled from $P$ by $\kappa$ on the freak observer moments,

\begin{equation}
\label{atypical_P_normal}
P(^\xi P_n)=P(\tilde P_n)= \frac{\tilde{\bar{n}}_n}{\tilde{\bar{n}}} = \frac{1}{1+\kappa \rho}.
\end{equation}

\noindent
This probability can go to 1 even if $\rho$ is large, if $\kappa$ is sufficiently small.  But if $\rho$ is huge, as the recurrence time of de Sitter space argues, the probability of being in a normal observer moment is near one only if there is an argument that $\kappa$ is zero.

Boddy \emph{et al.} \cite{Boddy2016} make such a case.  They argue that if the theory is unitary (``many worlds"), de Sitter space is in a stationary state.  Fluctuations do occur, including ones which correspond to Boltzmann brains, but they do not actually correspond to self-aware freak observer moments because nothing happens in a stationary state---there is no decoherence corresponding to the splitting of worlds.  If true, this is akin to setting $\kappa=0$, since being a self-aware freak observer moment is not only atypical, it does not happen.
Obviously if $\kappa=0$, $P(^\xi P_n)=1$ independent of how big $\rho$ is. 

How might this argument be affected by the fact that our Universe contains matter? Well, rarely, stable matter could play the role of an `environment' by interacting with a Boltzmann brain, causing decoherence. Such atypical Boltzmann brains might thus actually be self-aware. How rare is rare? An upper bound to the fraction $\kappa$ of such atypical matter-interacting fluctuations is the fraction of Hubble volumes which contain even a single matter particle. Let's define the entropy of a Hubble-volume-sized fluctuation entropy change, 

\begin{equation}
\label{def_cal_S}
\mathcal{S}\equiv 10^{122},
\end{equation}

\noindent
so that the fluctuation time $\tau_{HV}$ for Hubble volumes is $\sim e^\mathcal{S}$ and the fluctuation time for Boltzmann brains $\tau_{BB}$ is `about' $e^{\sqrt{\mathcal{S}}}$ (more accurately, $\sim e^{\mathcal{S}^{0.57}}$). Then the number of freak observers is huge: $\bar{n}_f \sim \tau_{HV}/\tau_{BB} \sim e^\mathcal{S}$. The number of normal observers per comoving Hubble volume is proportional to the volume of spacetime in which they can occur. A healthy upper bound on $\bar{n}_n$ is $\mathcal{S}$ (\emph{e.g.}, $10^{20}moments/lyr^3 s\times 10^{31} lyr^3 \times 10^{64} yrs \times 10^7 s/yr$), so that 

\begin{equation}
\label{rho_freak}
\rho \equiv \frac{\bar{n}_f } {\bar{n}_n} \sim e^\mathcal{S},
\end{equation}

\noindent
\emph{i.e.} the number of freak observer moments is so vast that the number of normal observer moments is irrelevant. Then the probability of being normal vanishes: $P(P_n) \simeq 0$ to a \emph{very} good approximation, yielding a seemingly serious consistency problem. \emph{But} only fraction $\kappa$ of freak observers actually can be self-aware by the argument above, where $\kappa$ must be smaller than the fraction of Hubble volumes with any matter in them. de Sitter space expands exponentially fast, so soon there is fewer than one particle per Hubble volume. By the time of the first Boltzmann brains, the fraction of Hubble volumes with a single matter particle is 

\begin{equation}
\label{kappa_freak}
\kappa < e^{-\tau_{BB}} < e^{-e^{\sqrt{\mathcal{S}}}}.
\end{equation}

\noindent
This is exponentially smaller than $\rho$ is big, and $\kappa \rho$ \emph{does} go to zero so that the relevant probability that we are normal observers,  $P(^\xi P_n)$, goes to 1. In summary, by this argument Boltzmann brains are overwhelmingly plentiful, but those which are atypically self-aware are very rare, and thus not a problem.
That matter effects are negligible is unlikely to come as a surprise to those already convinced by the arguments of Ref. \cite{Boddy2016}.  We do think it is interesting that there is a typicality factor so strong that it overwhelms even an exponentially large factor like the ratio of freak to normal observers ($\kappa \rho \ll 1$).
%

\subsection{Scarce observers}
Thus far we have assumed that observers in models are not rare. In fact, we have assumed that there is one observer per `cell'.  What if we relax this assumption and assume cells are filled only with probability $p_{\mathcal{F}}$? Hartle, Hertog, and Srednicki show that there is a different kind of OSE called `first-person probabilities' \cite{Hartle_Hertog_1p,*Srednicki_Hartle_1p}. Consider a set of models $\Theta_K$. If $p_{\mathcal{F}}$ is small enough, it is possible for there to be no observers in some or all of them (we do not necessarily think that assuming `scarce observers' is a reasonable hypothesis, we are merely considering the consequences of that assumption).  First-person probabilities weight models by the probability, $p^{\geq1}$, that there is at least one observer in the model—one cannot be an observer in a model if there are no observers in it.  
If there are $n_J$ observer locations (\emph{e.g.}, cells in a prison block or Hubble volumes in a Universe) which contain observers with probability $p_{\mathcal{F}}$, then the probability that there are no observers in the model is $(1-p_{\mathcal{F}})^{n_K}$, and the probability that there is at least one observer in the model is \cite{Hartle_Hertog_1p},
 
\begin{equation}
\label{p_gteq1}
p_K^{\geq1} = 1 - (1-p_{\mathcal{F}})^{n_K}.
\end{equation}

Now, the inclusive probability $P(P\Theta_K | P\Theta)$ (\emph{i.e.} multiple theories are realized---a theoryverse) is not affected by $p_K^{\geq1}$ because we are conditioning on there being one observer (the `$P\Theta$'), and the weighting by the number of observers in each model, $p_{\mathcal{F}} n_K$, already takes that into account. So we have,

\begin{equation}
\label{scarce_Be}
P(P \Theta_K)_{p_{\mathcal{F}}} = \frac{p_{\mathcal{F}} n_K P(\Theta_K)} {\sum_J p_{\mathcal{F}} n_J P(\Theta_J)}
=\frac{n_K}{\langle n \rangle} P(\Theta_K),
\end{equation}

\noindent
where $\langle n \rangle= \sum_J n_J P(\Theta_J)$ is the average number of observer cells per model.  Models with more observer cells are favored because it is more likely for an observer to be in such a model, as expected from our previous results. In a cosmological model this corresponds to \emph{volume weighting} \cite{Page_volume} where models with greater volume for observers are favored.

What about the exclusive probability $P(P \Pick \Theta_K | P \Pick \Theta)$, which is how one generally selects between competing models?  Condition `$P \Pick \Theta$' ensures that there is at least one observer in one of the models, but to ensure that a given model meets that criterion, we need to weight the models by $p_K^{\geq1}$\cite{Hartle_Hertog_1p,*Srednicki_Hartle_1p}:

\begin{eqnarray}
\label{scarce_Pick}
P(P \Pick \Theta_K)_{p_{\mathcal{F}}} = 
\frac{(1 - (1-p_{\mathcal{F}})^{n_K}) P(\Theta_K)} {\sum_J (1 - (1-p_{\mathcal{F}})^{n_J}) P(\Theta_J)}.
\end{eqnarray}

\noindent
There are two interesting limits: where observers are common or rare.
First, if $p_{\mathcal{F}} n_K$ is large for some models and tiny in others, then $p_K^{\geq1}$ are close to 1 for the former models, and they have observers. Define these models that certainly have observers by subset $\Theta_{obs}$ and normalization factor $\mathcal{N}\equiv \sum_{J \in \Theta_{obs}} P(\Theta_J)$. Then the probability becomes,

\begin{eqnarray}
\label{scarce_Pick_large}
P(P \Pick \Theta_K)_{common} \simeq \frac{1}{\mathcal{N}} P(\Theta_K).
\end{eqnarray}

\noindent
Note that models either `pass' (are in $\Theta_{obs}$) or `fail' (are not in $\Theta_{obs}$). If all models we consider pass ($\Theta_{obs}=\Theta$) then $\mathcal{N}=1$, and we obtain the usual expression for a Pick probability.

If, on the other hand, all the $p_{\mathcal{F}} n_K$ are small, so there are no models that certainly have observers ($\Theta_{obs}=\emptyset$), then $p_K^{\geq1} \simeq p_{\mathcal{F}} n_K$ (because $(1-p)^n=1-np+\mathcal{O}((np)^2)$) and the Pick probability becomes,

\begin{eqnarray}
\label{scarce_Pick_small}
P(P \Pick \Theta_K)_{rare} \simeq  \frac{p_{\mathcal{F}} n_K P(\Theta_K)} {\sum_J p_{\mathcal{F}} n_J P(\Theta_J)}
=\frac{n_K}{\langle n \rangle} P(\Theta_K).
\end{eqnarray}

\noindent
This is the same as the inclusive probability! 
Even though we are Picking between mutually exclusive models $K$, there is nonetheless a volume weighting factor, not just a pass/fail selection, due to it being less likely that scarce observers are in a model with few places for them to be.  So this `first-person' effect of Hartle, Hertog, and Srednicki is somewhat orthogonal to the observer effect we have been discussing: ours assumes observers in every `cell', $p_{\mathcal{F}}=1$, and comes from the difference between inclusive and exclusive selection, while theirs assumes the limit where observers are scarce, $p_{\mathcal{F}} \ll 1$, and is the \emph{same} for inclusive and exclusive selection in that limit.


This `first-person' analysis can be used in the context of freak observers.  Suppose we consider two models, $S$ and $L$, which differ only in the volume of spacetime in which freak observers occur. We could assign probability $p_n$ for `you' to arise normally per unit volume of spacetime and $p_f$ for a `freak' observer that thinks they are you (\emph{i.e.} after any typicality effects have been folded in). Let the volume of spacetime where normal observers can arise be $m_K$, and the volume where freaks could arise be $n_K$, which is usually much larger.  We want the case where you exist within the model, $(1-(1-p_n)^{m_K})$, \emph{and} that no freak versions of you exist, $(1-p_f)^{n_K}$ (as we argued before, you want to rule out cases where you might be a freak observer for self-consistency). Let's refer to this as `$1n,0f$'.  Then the ratio of exclusive probabilities is,

\begin{eqnarray}
\label{R_P_Theta}
R^f_{P \Pick \Theta} &\equiv& \frac{P(P_{1n,0f} \Pick \Theta_L)} {P(P_{1n,0f} \Pick \Theta_S)} \nonumber \\
&=& \frac{(1-(1-p_n)^{m_L})} {(1-(1-p_n)^{m_S})} \frac {(1-p_f)^{n_L}} {(1-p_f)^{n_S}} \frac{P(\Theta_L)} {P(\Theta_S)} \\
&=&  (1-p_f)^{n_L- n_S} R_\Theta \nonumber \\
&\to&  e^{-p_f (n_L - n_S) }R_\Theta, \nonumber 
\end{eqnarray}

\noindent
where $R_\Theta \equiv P(\Theta_L)/P(\Theta_S)$ and we have assumed $m_S=m_L$ (\emph{i.e.} that the models do not differ in the volume of spacetime available to normal observers). The last line follows for large $n$. 

We can neglect $n_S$ for $n_S \ll n_L$.  Then there are two interesting limits. If $p_f n_L$ is small, freak observers are scarce, and the `first-person' ratio $R_{P\Theta}$ is only slightly smaller than the `third-person' one:

\begin{equation}
\label{R_P_Theta_small}
R^f_{P \Pick \Theta}  \big\rvert_{p_f n_L \to 0} \simeq (1 - p_f n_L) R_\Theta.
\end{equation}

\noindent
This is a slight preference for $S$ models over $L$ ones, but for $p_f n_L \ll 1$ the preference is negligible. The other limit of interest is when both models have problems with freak observers because $p_f n_K$ is large. Then each theory is deweighted by the factor $(1-p_f)^{n_K}$ which goes to 0, but the factor for $L$ falls much faster and we have,

\begin{equation}
\label{R_P_Theta_large}
R^f_{P \Pick \Theta}  \big\rvert_{p_f n_L \to 1} \simeq e^{-p_f n_L} R_\Theta,
\end{equation}

\noindent
strongly favoring $S$ over $L$. So under the criterion of `no freaks like me', if there are \emph{no} models without significant probability for freak observers, the ones which minimize the volume for them to spawn are strongly preferred. Of course any model which has no freak observers would, by that criterion, be preferred over those.
 
\Section{Gott analysis}  \label{Section_Gott}

J. Richard Gott III wrote about what seems to be an entirely different kind of observer selection effect \cite{Gott1993}.  He argued that simply by knowing how long some finite-lifetime entity has been observed, one can bound the probability of it lasting a long time.  For example, if you live at time $t$ after the start of a civilization, his argument says that simply assuming you are a random observer implies that the probability of the civilization lasting $40t$ is only $1/40$, or $2.5\%$.

There are a number of problems with this argument, as we shall see.  The first is that Gott's analysis did not make use of a prior \cite{Buch1994}, which Gott then addressed \cite{Gott1994}.  This point was echoed by Carleton Caves \cite{Caves_crit_Gott},  who found that the prior probability for a world having lifetime $T$ needed to obtain Gott's result is the Jeffereys prior, which goes as $1/T$.  However, as we shall see, this corresponds to a Pick-selection. The prior needed to obtain the probability Gott finds to Be in a civilization lasting time $T$ is \emph{not} the Jeffereys prior, 
but a prior that goes as $1/T^2$ \cite{CavesNote,Caves2008}.  Caves argued that the analysis was also flawed because it assumed that the observer had to live only during the timespan of the `world', and that once one relaxes that assumption, the effect goes away.  
(This is really about what set of observer moments it is reasonable for one to consider that the moment at hand is randomly drawn from. For Gott's example of the Berlin wall, one could assert that his observation of the wall was drawn randomly from possible moments during the existence of the wall when he could ponder the question of the duration of its existence, rather than a random moment from his lifetime that predates and postdates the wall. It is then a question of whether that assumption is reasonable. It is certainly problematic in many cases. For example, it is hard to argue that the observer moment in which you ponder the lifetime of an architectural construction is randomly drawn from all the moments during its existence if you were born before it was built---for a long-lived construction you are necessarily seeing only its earliest moments.) But it should not be a problem in the narrow case of interest to us: where we assign probabilities for the lifetime of the world in which we were born---we are necessarily alive only during the world in which we are born, and so random observer moments in our lifetime are necessarily within the time window of the world's existence.

We will first explain the Gott argument in his notation and then ours.  Then we will show how to incorporate a prior, derive  results for different priors, and determine which one gives Gott's results.  Then we show that Gott's results do not actually represent an OSE, and we trace the source of the effect.  Finally, we consider the exclusive case, where one lifetime is picked.

\subsection{Gott's argument}

Suppose we are a random intelligent observer of some `world' of lifetime $T$ which has existed so far for time $t$.  We do not know $T$ and we want to know if knowing $t$ tells us anything about $T$, other than $T \ge t$. Gott gives a few examples \cite{Gott1993}, but they are of two types: things on which our existence does not depend, such as the timespan for which the Berlin wall existed, and things on which it does depend, such as the civilization in which we were born.  We will not consider the former further, except to note that 
the second critique of Caves may apply to those situations.
Thus, since we assume we live during the world, we can without loss of generality define Gott's quantities as

\begin{eqnarray}
\label{Gottmapping}
&t_{begin} &\rightarrow 0 \nonumber \\
&t_{end} &\rightarrow T \nonumber \\
&t_{now} &\rightarrow t    \\
&t_{future} &\rightarrow T_{fut} \equiv T-t, \nonumber \\
\end{eqnarray}

 \noindent 
where we take as a precondition that $t$ is in the range $[0,T]$.  This world could refer to our planet (in which case $t \sim 10^9$ years), the era of \emph{homo sapiens} ($t \sim 10^5$ years), our civilization ($t \sim 10^4$ years), or civilization since Bayesian questions like this have been asked ($t \sim 40$ years). One could even try to argue that it refers to the metastable electroweak vacuum ($t\sim 10^{10}$). Now, going back to the original assumption, it is not at all clear that we qualify as a random observer in \emph{any} of these `worlds', but nevertheless let us assume that we do.  

First, Gott argues each value of $t$ in the range $[0,T]$ is equally likely.  This is true if there is an equal number of observers at each time $t$ in $[0,T]$ (unreasonable in most cases---really $t$ and $T$ are better thought of as the current and final tally of observers in the world) and one selects them at random.  This can be loosely written,

\begin{equation}
\label{Gott1}
``P(t) = \rm{constant}/T."
\end{equation}

\noindent
Further, this means that $t/T$ is a random number between 0 and 1, so,

\begin{equation}
\label{Gott2}
``P(t/T) = \rm{constant}."
\end{equation}

\noindent
Finally, if we sum up the probabilities for our expectation for the remaining time left for the world, $T_{fut} \equiv T-t$, we obtain that it is overwhelmingly likely to be of roughly of order $t$ (neither much greater nor smaller than $t$),

\begin{equation}
\label{Gott3}
``P(\frac{1}{39} t < T_{fut} < 39 t) = 0.95,"
\end{equation}

\noindent
or focusing on the upper end and using $T_{fut} \equiv T-t$ to write this more generally,

\begin{equation}
\label{Gott3K}
``P(T >  Kt) = 1/K," 
\end{equation}

\noindent
where $K>1$.  Note that for $K=40$, we get the probability of $T_{fut} = T-t$ being greater than $39t$ is 1/40, or 2.5\%, in agreement with \Eq{Gott3} (the upper and lower tails are equally probable).  Further, note that these are scale-invariant probabilities: they depend on the ratio $t/T$ independent of whether the scale is decades or millennia.

Gott seemingly found a way to argue that our datum $t$ not only tells us something about our world's eventual lifetime $T$, but argued that $T$ is unlikely to be more than a few times $t$, no matter the scale.

Is this right?

\subsection{Our argument}

As usual, we have a set of observers $P$ and a set of worlds $W$. As Gott does, we will for simplicity assume that 
the number of observers at each time is the same.  We will use the compact notation outlined at the end of Appendix A, \emph{i.e.}

\begin{equation}
\label{PWcompact}
P(P_\alpha W_\beta)  \equiv  P(\alpha \beta),\ P(P_\alpha \Pick W_\beta)  \equiv   P(\alpha \Pick \beta)
\end{equation}

\noindent
where $\alpha$ and $\beta$ can be `null,' \textit{e.g.}, $P(  T | t ) \equiv P( P  W_T  | P_t  W)$ and $P( \Pick T | t \Pick) \equiv P( P \Pick W_T | P_t \Pick W)$.  Let us then define the probability density to Be  in a world at time $t$ (for a moment lasting $dt$):

\begin{equation}
p(t) \equiv P(P_{\left[t, t+ dt\right]}W)/dt.
\end{equation}

\noindent
The probability density to Be in a world of lifetime $T$ (one again needs a finite range $\left[T, T+dT\right]$), and to Pick a world of lifetime $T$ are

\begin{eqnarray}
p(T) \equiv P(PW_{\left[T, T+dT\right]})/dT, \\
p(\Pick T) \equiv P(P \Pick W_{\left[T, T+dT\right]})/dT.
\end{eqnarray}

\noindent
Note that the probability density to Be in a world is weighted as before by the total number of observers who will ever live in the world, which by assumption is proportional to $T$, so,

\begin{equation}
\label{ourGottBeTtimesPick}
p(T)  \sim T p(\Pick T).
\end{equation}

What we are going to do is start with a prior probability density for our world having lifetime $T$, $p(\Pick T)$, the likelihood density of being in our world at time $t$ given that it will exist for time $T$, $p(t|T)$, and we will use Bayes' theorem to calculate the posterior probability density of our world living time $T$ given our datum $t$, $p(T|t)$.  


The likelihood density is, as Gott said, a constant, independent of $t$,

\begin{equation}
\label{ourGottlikelihood}
p(t|T)\equiv P(P_{\left[t, t+dt\right]}W | PW_{\left[T, T+dT\right]})/dt= \frac{1}{T}.
\end{equation}

\noindent
Note that if we integrate this probability density over all values of $t$ in $[0,T]$, $P((0\leq t \leq T) | T)=\int_0^T p(t|T)dt$ we get $1$.  This is essentially the same expression as \Eq{Gott1} which we used to express Gott's words, except that here we are explicitly writing a likelihood  density conditioned on lifetime $T$.


The key problem with Gott's analysis is that he jumps right to a probability for $t/T$ without a prior.  Let us examine three possible priors, and see which gives the results Gott found.  We need the prior probability density for Picking a world of lifetime $T$, $p(\Pick T)$,  because it should contain all factors \emph{other} than our existence.  This is parallel to what we did in the Prisoner scenario, though there we needed only  probabilities $P(W_S)$ and $P(W_L)$, whereas here we need a function of $T$ over its range.  This brings up an important point: we need to define minimum and maximum plausible values of lifetime $T$ for the world we are in, $T_-$ and $T_+$ respectively. They allow us to properly normalize our expressions, but $T_\pm$  play a more subtle role too, as we shall see.  It must end up being the case that $T_+$ is greater than both $t$ and $T$, and that $T_-$ be smaller than $T$, so if we really tried to define $T_\pm$ without any idea of the timescales involved, we might fail in that.  And our expectations for the timescale might change with $t$.  For example, today we might see $T_+=10^6$ years as reasonable, but if civilization somehow survives for a million years, that $T_+$ will be too low.  This is less of an issue for $T_+$ because we will be able take it to infinity in our final expressions.  But $T_-$ is trickier.

Three reasonable choices for our prior $p(\Pick T)$, are constant, $\sim  1/T$  (Jeffereys), and $\sim 1/T^2$.  The normalized priors to Pick a world of lifetime $T \in [T_-,T+]$ are:

\begin{eqnarray}
\label{OurGottPriors}
&p(\Pick T) \big\vert_{const}&= \frac{1}{T_+ - T_-}, \nonumber \\
&p(\Pick T) \big\vert_{1/T}&= \frac{1}{T \ln{(T_+/T_-)}}, \\
&p(\Pick T) \big\vert_{1/T^2}&= \frac{1}{T^2 (1/T_- - 1/T_+)}, \nonumber
\end{eqnarray}

\noindent
which lead to corresponding probability densities to Be in such a world (again assuming the number of observers at each time is constant and \Eq{ourGottBeTtimesPick}):

\begin{eqnarray}
\label{OurGottPriorsBe}
&p(T) \big\vert_{const}&= \frac{T}{\frac{1}{2}(T_+^2 - T_-^2)}, \nonumber \\
&p(T) \big\vert_{1/T}&= \frac{1}{T_+ - T_-}, \\
&p(T) \big\vert_{1/T^2}&= \frac{1}{T \ln{(T_+/T_-)}}. \nonumber
\end{eqnarray}


Next we plug the likelihood density $p(t|T)$ in \Eq{ourGottlikelihood} and our Be priors $p(T)$ in \Eq{OurGottPriorsBe} into Bayes' theorem,

\begin{equation}
\label{OurGottBayesThm}
p(T|t) = \frac{p(t|T) p(T)}{p(t)}.
\end{equation}

\noindent
We can calculate $p(t)$ by integrating $p(t|T) p(T) dT$ over $T$.  We need to be a little careful about the limits of integration because we have defined $T \ge t$ and $T \ge T_-$, but at the moment it is ambiguous whether $t$ is greater than $T_-$ or not.  So let us define the lower limit on $T$ to be the maximum of the two: $t_m \equiv \max{(t, T_-)}$. 
For the three different priors, we obtain three posterior probability densities for $T \in [t_m, T_+]$:

\begin{eqnarray}
\label{OurGottPosteriors}
&p(T|t) \big\vert_{const}&=  \frac{1}{T_+ - t_m} \to \sim {\rm constant},
\nonumber \\
&p(T|t) \big\vert_{1/T}&= \frac{1}{T \ln{(T_+/t_m)}} \to \sim \frac{1}{T}, \\
&p(T|t) \big\vert_{1/T^2}&= \frac{1}{T^2 (1/t_m - 1/T_+)} \to \frac{t_m}{T^2}, \nonumber
\end{eqnarray}

\noindent
where the righthand side is the limit where $T_+ \to \infty$. 
Note that these are the same expressions as the priors in \Eq{OurGottPriors} with $T_-$ replaced by $t_m$.  In other words, the only effect of the datum here is the trivial replacement of the lower bound on $T$ because it is necessarily at least equal to $t$.  So if we quantify the OSE by taking the ratio of the posterior to the prior, 

\begin{equation}
\label{GottRsubTdef}
R_T \equiv \frac{p(T|t)}{p(\Pick T)},
\end{equation}

\noindent
we obtain for the three priors,

\begin{eqnarray}
\label{GottRsubT}
&R_T \big\rvert_{const}&=  \frac{T_+ - T_-}{T_+ - t_m} \to 1, \nonumber \\
&R_T \big\rvert_{1/T}&= \frac{\ln{(T_+/T_-)}}{\ln{(T_+/t_m)}} \to 1, \\
&R_T \big\rvert_{1/T^2}&=\frac{1/T_- - 1/T_+}{1/t_m - 1/T_+} \to \frac{t_m}{T_-}, \nonumber
\end{eqnarray}

\noindent
where again the righthand side is for $T_+ \to \infty$.  In that limit, the first two priors yield $R_T=1$ even if we include the replacement effect of $T_- \to t_m$. To evaluate the third prior, we need to discuss the value of $t_m$. There are three possible values:

\begin{itemize}
\item $t<T_-$, so $t_m = T_-$, and our lower bound on $T$ does \emph{not} increase.
\item $t=T_-$,  so $t_m = t=T_-$, and our lower bound on $T$ does \emph{not} increase.
\item $t>T_-$, so $t_m =t$, and our lower bound on $T$ \emph{does} increase.
\end{itemize}

\noindent
The first case means that prior to our using our datum $t$ we assumed that the minimum value of $T$ was larger, asserting that there is zero probability for our world to end between now, $t$, and $T_-$. The third case means that prior to taking note of $t$, we thought that the lower bound on $T$ was $T_-$, and so datum updates our knowledge, raising that lower bound---yet somehow we are still confident in our prior assumed probability density despite being wrong about its endpoint. The second case strikes us as the most reasonable, because we should already know that $T_- \geq t$ and cannot know that $T_- > t$, so we should assume $T_-=t$. Nevertheless, let us consider all three cases.

For the first two cases, $t \leq T_-$, all three priors lead to $R_T = 1$. For $t> T_-$ and the $1/T^2$ prior, $R_T=t/T_-$, which is $>1$. This is an upward shift due to the fact that the posterior probability density is nonzero over a smaller range, 
$[t,T_+]$, than the prior probability density $[T_-,T_+]$. We will call this a `boundary condition OSE'. It is \emph{not} due to the number of elements in the set of observers, $P$, as in OSEs we considered previously. Rather, it is simply due  
to raising the lower bound on $T$ from $T_-$ to $t$.


So, given that there is only at best a boundary condition OSE here, can we reproduce Gott's result?  We can.   To compare to Gott's result, we have to integrate these functions of $T$ from $Kt$ to $T_+$ for fixed $t$ (and assume $Kt \in [T_-,T_+]$).  This yields probabilities for $T$ in the range of $Kt$ to $T_+$:

\begin{eqnarray}
\label{OurGott_gtK}
&P(T>Kt|t) \big\vert_{const}&= \frac{T_+ - Kt}{T_+ - t_m} \to 1,  \nonumber \\
&P(T>Kt|t) \big\vert_{1/T}&=  \frac{\ln{(T_+ /Kt)}}{\ln{(T_+ /t_m)}} \to 1, \\
&P(T>Kt|t) \big\vert_{1/T^2}&=   \frac{1}{K} \frac{t_m}{t} \frac{T_+-Kt}{T_+ - t_m} \to  \frac{1}{K}  \frac{t_m}{t}, \nonumber
\end{eqnarray}

\noindent
where we again take the limit that $T_+ \to \infty$.  We see that for the constant and Jeffereys priors, the probability of $T>Kt$ goes to 1.  This is not surprising;  if we assume the maximum on $T$ is much greater than $Kt$, the probability that $T>Kt$ approaches 1, unless our prior falls very fast.  For the prior $p(\Pick T) \sim 1/T^2$ it \emph{does} fall fast enough.  If  $t \geq T_-$ then $t_m =t$ so that 

\begin{equation}
\label{OutGottoneoverK}
P(T>Kt|t) \vert_{1/T^2, \ t \geq T_-} \to \frac{1}{K}, 
\end{equation}

\noindent
and we have obtained Gott's expression in \Eq{Gott3K}. (For $t < T_-$, this integrated probability is \emph{larger}. We shall see what that means shortly.)

So even though there is only a boundary-condition OSE, we have reproduced the result of Gott, seemingly disfavoring long-term worlds. How is that possible?


\subsection{Why does Gott seem to find an OSE?}

To answer this, consider the situation \emph{before} we know datum $t$  and where we Pick a world at random.  We know by assumption that with probability 1, $T \in [T_-,T_+]$ (integrate $p(\Pick T)$ from $T_-$ to $T_+$ and we get 1).  Suppose we ask what the probability is for this world to last $K$ times its minimum, \emph{i.e.,} for $T>K T_-$.  We simply integrate $p(\Pick T)$ from $KT_-$ to $T_+$.  This gives

\begin{equation}
\label{GottKTminus}
P(\Pick (T > K T_-)) \big\vert_{1/T^2} = \frac{1}{K} \frac{T_+ - K T_-}{T_+ - T_-} \to \frac{1}{K} .
\end{equation}

\noindent
For fixed $K T_-$ and $T_+ \to \infty$ this gives $1/K$!  In other words, the effect that Gott found has nothing to do with the datum $t$, but just the rapidly falling prior to which his result corresponds.

Still, it is useful to define a metric which manifestly shows that there is no OSE.  For that, let us define the ratio of probability densities integrated over $T$.  Dividing Eqs. (\ref{OurGott_gtK}) by (\ref{GottKTminus}) we see that for the $1/T^2$ prior,

\begin{equation}
\label{RintegrateT}
R_{\int T} \Big\vert_{1/T^2} \equiv \frac{P(T>Kt|t)} {P(\Pick (T > K T_-))} \Bigg\vert_{1/T^2} = \frac{t_m}{t}.
\end{equation}

\noindent
For $t\geq T_-$, the cases where we obtained Gott's result, we see that this equals 1---that the posterior probability is the same as we obtained using the prior lower bound, and there is no OSE of any kind. For the case $t<T_-$ this ratio is \emph{larger} than 1 (note that the righthand side cannot exceed $K$ because if $Kt<T_-$ then $P(T>Kt|t)=1$). What that means is that from our prior, we assumed that large $T$ worlds were disfavored, but upon learning that $t<T_-$, our expectation is \emph{less} negative due to not having reached the lower bound in the world's lifetime, $T_-$.

So in the inclusive case, there is no $1/T$ OSE.  For a fast falling prior we can obtain Gott's $1/K$ result, but it is not an OSE either, just a manifestation of the fast-falling prior we assumed. The only OSE that remains in any of these cases is if we assumed a fast-falling $1/T^2$ prior, thinking that worlds with $T > KT_-$ were very unlikely, but then finding out that $t<T_-$, making our posterior probability \emph{less} dire than our prior.


\subsection{Picking hypothesis $T$} \label{Section_Gott_Picking}

Suppose that instead of Being in a set of worlds of various lifetimes $T$, we assert that there is precisely one world, with one future, one lifetime $T_{*}$, and we have a set of hypotheses $\Theta_T$ for what $T_*$ is.  This is an exclusive case, and we are interested in the posterior probability density,

\begin{eqnarray}
p(\Pick T | t \Pick) &\equiv& P(P \Pick \Theta_{[T,T+dT]} | P_{[t,t+dt]} \Pick \Theta)/dT  \nonumber \\
&=&\frac{p(t \PickNeutered| \PickNeutered T) p(\Pick T)}{p(t \Pick)}.
\end{eqnarray}
 
 \noindent
 The key difference from our analysis above is that the prior that goes into Bayes' theorem is the Pick probability density $p(\Pick T)$ instead of the Be probability density $p(T)$ (and the corresponding denominator $p(t \Pick))$.  The likelihood is not affected, as in the Warden case, because the Pick is neutered.  The upshot is that the posterior probabilities go as $\sim 1/T$ times those in the Be case in \Eq{OurGottPosteriors},

 \begin{eqnarray}
&p(\Pick T| t \Pick) \big\vert_{const}&= \frac{1}{T \ln{(T_+/t_m)}} \to \sim \frac{1}{T}, \nonumber \\
&p(\Pick T|t \Pick) \big\vert_{1/T}&= \frac{t_m}{T^2} \frac{T_+}{T_+ - t_m} \to  \frac{t_m}{T^2},
 \end{eqnarray}
 
 \noindent
 which means there \emph{is} an OSE for this pick-a-hypothesis-$T_*$:
 
\begin{equation}
\label{R_PickT}
R_{\Pick T} \equiv \frac{p(\Pick T|t \Pick)}{p(\Pick T)} \to \sim \frac{1}{T}.
\end{equation}

\noindent
Specifically,

 \begin{eqnarray}
\label{GottRsubTPick}
&R_{\Pick T} \big\rvert_{const}&= \frac{1}{T}   \frac{T_+ - T_-}{\ln{(T_+/t_m)}} \to \sim \frac{1}{T}, \nonumber \\
&R_{\Pick T} \big\rvert_{1/T}&= \frac{1}{T} \frac{\ln{(T_+/T_-)}}{1/t_m - 1/T_+} \to \sim \frac{t_m}{T}.
\end{eqnarray}

But as with $R_T$, $R_{\Pick T}$ is not an ideal metric of OSE, so we should consider the probabilities resulting from integrating over $T$:
 
\begin{eqnarray}
\label{OurGott_gtKPick}
&P(\Pick(T>Kt)|t \Pick) \Big\vert_{const}&= \frac{\ln{(T_+ /Kt)}}{\ln{(T_+ /t_m)}} \to 1,   \\
&P(\Pick(T>Kt)|t \Pick) \Big\vert_{1/T}&= \frac{1}{K} \frac{t_m}{t} \frac{T_+-Kt}{T_+ - t_m} \to  \frac{1}{K}  \frac{t_m}{t}, \nonumber
\end{eqnarray}
 
\noindent 
and we obtain the same $1/K$ expression as Gott, now for the $1/T$ prior and $T_-=t$ (the expression is the same as the Gott case, but his description of the problem seems like a Be, and thus corresponds to \Eq{OutGottoneoverK}).
 
 As we did in the Be case, we define an OSE metric as the ratio of integrated probability densities,
 
 \begin{equation}
 R_{\int \Pick T}  \equiv \frac{P(\Pick (T>Kt)|t \Pick)} {P(\Pick (T > K T_-))},
 \end{equation}

\noindent
which yields for the two priors we consider here, 
 
\begin{eqnarray}
\label{RintegrateTPick}
R_{\int \Pick T} \Big\vert_{const}   &\to& 1, \nonumber \\
R_{\int \Pick T} \Big\vert_{1/T}   &\to& \frac{1}{K} \frac{t_m}{t}.
\end{eqnarray}
  
What this means is that there is a true OSE in the $t\geq T_-$ Pick case for the $1/T$ prior which manifests itself as a factor of $1/K$ in that ratio of the integrated posterior to prior probability densities.  In other words, the posterior probability density falls with $T$ faster than the prior probability density due to an OSE, which manifests itself in $R_{\int \Pick T}$ being smaller than one. If $t<T_-$ this is mitigated by the $T_-/t$ factor, and is completely erased if $Kt <T_-$, yielding $R_{\int \Pick T}=1$.

For the constant prior case, there is an OSE in the ratio of probability densities ($R_{\Pick T} \sim 1/T$) but it is washed out when one integrates over $T$ (the posterior probability density falls faster with $T$ than the prior probability density, but both fall slowly enough that their integrated probabilities go to 1, hence their ratio, $R_{\int \Pick T}$, is also 1).

So in the exclusive case there is a real OSE, but only if the prior falls fast enough and $t$ is not much less than $T_-$.
  
\Section{Doomsday Argument}   \label{Section_Doomsday}

We are now finally ready to discuss the Doomsday argument.  The question is, 

\begin{quote}
Do observer selection effects increase the probability that our world will be short-lived?
\end{quote}

First, this is a very strange thing to ask.  This would entail laying out all the factors which we might use to assign a probability for the world ending soon, and separate out the datum of what year it is.  But all of the factors are intertwined.  For the purpose of the argument below, we need to make the somewhat unreasonable assumption that we can put all factors ($e.g.$, our estimate for the probability of nuclear war) other than that datum  into some prior---which is somewhat unreasonable because such a calculation usually depends on temporal information ($e.g.$, the survival probability per year was surely lower in the early days of nuclear weapons than at other times).  In any case, we make this assumption for the arguments below.

As it is usually stated, the question is whether the probability that we live in a short-lived world (world type $S$) or a long-lived one (world type $L$) is changed given the information about the date (datum $d$).  Clearly this is a Be selection---we are born in this world without the need for that world to be picked. So the zeroth order analysis is that the case is like our very first example, the Prisoner Problem, where there was no OSE, and thus no Doomsday effect.  The posterior probability of being in a short-lived world is just given by \Eq{Prisoner_P(SgivenD)}, and equals the prior probability of picking such a world, so that the ratio of posterior probabilities to their priors, $R_{P/W}$, is one:

\begin{eqnarray}
P(PW_S|P_dW)&&=P(W_S), \\
R_{P/W}&&= 1.
\end{eqnarray}

But we need to be careful just what our assumptions are regarding any larger sets $PW$ are embedded in.  For example, if we treat the world types as mutually exclusive hypotheses for short-lived and long-lived worlds, $\Theta_S$ and $\Theta_L$, then there is a Pick at that level and there is an OSE akin to that in \Eq{Prisoner_P(SgivenD)_Pick}, 

\begin{eqnarray}
P(PW \Pick \Theta_S | P_d W \Pick\Theta)&&=
\frac{P(\Theta_S)}
{P(\Theta_S) + \frac{1}{\rho}  P(\Theta_L)}, \\
R_{P \Pick /\Theta} &&=\frac{1}{\rho}.
\end{eqnarray}

Note that here we are saying that either hypothesis $\Theta_S$ or $\Theta_L$ is realized, but not both.  This is reasonable only if one assumes that there is only one relevant planet (the Earth) because there are no relevant exoplanets (we are not asking about the inhabitants of inhabitable worlds, just of the Earth), nor copies of the Earth nor multiple futures of this one Earth (in a partial or complete multiverse of some sort, such as in unitary quantum mechanics).  Again, there is an OSE given these assumptions because we are saying that there are multiple hypotheses ($\Theta_S$ and $\Theta_L$), but only one of them can be realized.

This also assumes that we are typical observers.  This, too, can depend on assumptions or on how the problem is stated.  For example, by saying that you are equally likely to be any human throughout history fails to take into account the fact that only a tiny fraction of humans throughout history might have asked the Doomsday question, at least as stated.  For example, humans before 1763 could not have phrased a question in terms of Bayes' Theorem \cite{BayesTheorem}, and the question ``will our civilization last until the year 2500?'' will become moot in 500 years.  Similarly, the question ``will our civilization last another 100 years?" changes character as the centuries we survive accrue, since a century becomes a smaller and smaller fraction of the civilization's lifetime.  We need to phrase the question in such a way that it would be just as reasonable for a current and future inhabitant of the civilization to ask it.

We argue that the question framed by Gott is actually best, because ``will our world last $K$ times its present age?" is somewhat timescale invariant.  There are still issues with assigning a starting point for the world, and a prior probability density for a world of lifetime $T$, $p(\Pick T)$ ($e.g.$, neglecting the problem of lumping all other factors into the prior in a time-independent way), but at least it is reasonable for future observers to ask that same question.

So, to be specific, we should ask whether the current age of our world, $t$, should affect our estimate for the lifetime of the world, $T$.   As we discussed in Section \ref{Section_Gott}
the selection in $PW$ is a Be, and there is just a boundary condition OSE: the effect of replacing the lower bound on $T$, $T_-$, with $t$, for $t >T_-$.  We further argued that it is not reasonable to have chosen $T_-$ either greater or smaller than $t$, and that for $T_-=t$, the prior and posterior probability densities are equal, so there is no OSE at all:
  
\begin{equation}
p(T | t) \big\rvert_{T_-=t} = p(\Pick T) \implies R_T \big\rvert_{T_-=t}=1.
 \end{equation}

We then integrate these probability densities over $T$ to obtain the probability of Being in a world with $T>Kt$ given $t$.  As we said in Section \ref{Section_Gott} this goes to 1 unless the prior falls quickly, see \Eq{OurGott_gtK}.  Even in the case of such a fast falling prior, the $1/K$ effect is \emph{not} an OSE, but just an artifact of that prior.  We quantified that by taking the ratio of integrated probabilities in \Eq{RintegrateT},

\begin{equation}
R_{\int T} \big\rvert_{T_-=t}=1,
 \end{equation}

\noindent
which shows that there is no OSE at all in the Be case.

Is there any somewhat reasonable set of assumptions which leads to a Doomsday effect?  Yes.  If we assert, as we did in Section \ref{Section_Gott_Picking}, that there is a unique lifetime for the world, $T_*$, and we have hypotheses $T$ for what that $T_*$ is, then there is a Pick on the nested set, $P \Pick \Theta$, and there \emph{is} an OSE given by \Eq{R_PickT}:

\begin{equation}
p(\Pick T | t\Pick) \sim \frac{1}{T} p(\Pick T) \implies R_{\Pick T} \sim \frac{1}{T}.
 \end{equation}

\noindent
But even then, if we choose a constant prior probability density $p(\Pick T)$, the posterior probability that the world will last $K$ times longer than it has so far, goes to one as in \Eq{OurGott_gtKPick}.  However, if we start with a $1/T$ prior, the OSE is \emph{not} washed out in \Eq{OurGott_gtKPick}, and the OSE survives in the ratio of integrated probabilities, \Eq{RintegrateTPick}:

\begin{equation}
R_{\int \Pick T} \Big\vert_{1/T,\ T_-=t}   \to \frac{1}{K}.
\end{equation}

\noindent
This \emph{is} a Doomsday effect.  It says that given the assumptions above, even if we include our timescale in setting the minimum lifetime ($T_-=t$), and integrate our probability densities over $T$, and normalize to that integrated probability for the prior, there \emph{is} an OSE in the Pick case for a falling prior---that our datum $t$, by itself, should cause us to reduce our posterior probability that our world will last substantially longer than it has.

So, in summary, there \emph{can} be a Doomsday effect, but to have one requires a set of assumptions like this:

\begin{itemize}
\item All factors other than the current age of the world, $t$, can be separated out into a prior, which is a simple function of the world's lifetime $T$.
\item You are typical of observers throughout the lifetime of the world, including in what question is being asked.
\item There is exactly one true value of the lifetime, $T_*$, because you consider only one world with one fixed future---so you view the values of $T$ to be mutually exclusive hypotheses for the value of $T_*$, resulting in a Pick.  It is not enough to assume an exclusiverse, it has to a be universe with only one manifestation of the world so that there is only one true lifetime $T_*$.
\item The prior probability density falls as a function of $T$ so that the integration over $T$ does not wash out the OSE.
\end{itemize}

\noindent
Absent a set of assumptions like these, there is no Doomsday effect.  All of these strike us as somewhat unreasonable, except the last. Thus, one can probably not argue that our `world', be it the era of Bayesian reasoning or of the stable electroweak vacuum, is doomed to end soon on the basis of datum $t$.

\Section{Universal Doomsday Argument}  \label{Section_Universal}

In addition to the Doomsday argument, which concerns our world, some authors have discussed a `Universal Doomsday' argument \cite{Knobe-Olum-Vilenkin,Gerig-Olum-Vilenkin}, which says that not only does our datum imply that our world is doomed to die sooner than our priors for its lifetime, due to some OSE, but that all worlds are also doomed to die out sooner due to our datum.  Some authors argue that `Universal Doomsday' can occur even when the Doomsday effect is not present.  This cannot be.  If there is a Doomsday effect due to a temporal datum, that lowered posterior probability \emph{can} affect our posterior probability for the lifetimes of other worlds, but it should be clear that if there is no Doomsday effect, if we gain no information from our datum about our own world, our posteriors for other worlds must be unchanged as well.

What we are interested in is how the datum affects an ensemble of worlds, $E$, as we consider in the inclusive and exclusive cases of Sections \ref{Section_Inclusive} and \ref{Section_Exclusive}.  In particular, here are the posterior probability densities for ensembles of type $y$ given datum $d$, in the inclusive case where there is no Doomsday effect, and in the exclusive case where there  can be one:

\begin{eqnarray}
p(y|d)\equiv P(PWE_{[y,y+dy]} | P_dWE)/dy, \\
p(\Pick y|d \Pick)\equiv P(PW \Pick E_{[y,y+dy]} | P_dW \Pick E)/dy.
\end{eqnarray}

\noindent
We ask whether these differ from the prior probability density for $y$,

 \begin{equation}
 \label{p_Pick_y}
 p(\Pick y) \equiv P(PW \Pick E_{[y,y+dy]})/dy.
 \end{equation}

\noindent
Universal Doomsday is the claim that it does.  If the probability distribution function for $y$ changes, so does 
our estimate for the average fraction $y$ of worlds of type $S$.  Our prior estimate is the average of $y$ weighted by the prior $p(\Pick y)$,

\begin{equation}
\label{avg_y_0} 
\left<y\right> \equiv \int_0^1 y \, p(\Pick y) dy.
 \end{equation}
 
\noindent
After taking our datum $d$ into account, our posterior estimates for that average in the inclusive and exclusive cases are weighted by the posterior probability distribution functions $p(y|d)$ and $p(\Pick y|d \Pick)$, respectively,

\begin{eqnarray}
\label{avg_y_d} 
\left<y\right>_d &\equiv& \int_0^1 y \, p(y|d) dy, \\
\label{avg_y_dPick} 
\left<y\right>_{d\Pick} &\equiv& \int_0^1 y \, p(\Pick y|d \Pick) dy.
\end{eqnarray}

\noindent
For reasons that will become clear in a moment, let's define metrics for Universal Doomsday,

\begin{eqnarray}
\label{RPW_UD}
R_W^{UD} &\equiv& \frac{1- \left<y\right>} {\left<y\right>}, \nonumber \\
R_P^{(\Pick)UD} &\equiv& \frac{1- \left<y\right>_{d(\Pick)}} {\left<y\right>_{d(\Pick)}},  \\
R_{P/W}^{(\Pick)UD} &\equiv& \frac{R_P^{(\Pick)UD}} {R_W^{UD}}, \nonumber
\end{eqnarray}

\noindent
where the ``$\Pick$'' is there in the exclusive case but not the inclusive case. 

It turns out we have already come across these averages.  The prior average fraction $\left<y\right>$ in \Eq{avg_y_0} is equal to the prior probability of worlds of type $S$:

\begin{eqnarray}
\label{P_S_eq_avg_y_0} 
P(W_S \Pick E) &=& \int_0^1 P(W_S \PickNeutered E | W \PickNeutered E_y) p(W \Pick E_y) dy \nonumber \\
&=& \int_0^1 y \, p(\Pick y) dy =\left<y\right>.
 \end{eqnarray}

\noindent
Note that if we assume that $\bar{N}_y/\bar{N}=1$, \emph{i.e.} that the ensembles differ only by fraction of worlds type S, $y$, not their number, then $p(W_S \Pick E)=p(W_S E)$, so that this is the prior probability of worlds of type $S$ in both the exclusive and inclusive cases.  What about $\left<y\right>_d$ and $\left<y\right>_{d\Pick}$?  They turn out to be simply equal to the posterior probabilities for being in an $S$ world, given datum $d$, in the inclusive and exclusive cases, respectively:

\begin{eqnarray}
\label{P_S_d_eq_avg_y_d} 
P(&P&W_S E | P_d WE) \nonumber \\
&=& \int_0^1 P(PW_SE | P_d WE_y) p(P_d W E_y | P_d WE) dy \nonumber \\
&=& \int_0^1 y \, p(y | d) dy =\left<y\right>_d, \\
\label{P_S_d_eq_avg_y_dPick} 
P(&P&W_S \Pick E | P_d W \Pick E)  \nonumber \\
&=& \int_0^1 P(PW_S \PickNeutered E | P_d W \PickNeutered E_y) p(P_d W \Pick E_y | P_d W \Pick E) dy \nonumber \\
&=& \int_0^1 y \, p(\Pick y | d \Pick) dy =\left<y\right>_{d\Pick}.
 \end{eqnarray}

\noindent
These are just the expressions for the posterior probabilities for worlds of type $S$.  In fact we see that,

\begin{eqnarray}
\label{p_y_d}
p(y | d) &=& p(\Pick y), \nonumber \\
p(\Pick y | d \Pick) &=& \frac{p(\Pick y)} {y +(1-y) \rho}
\left< \frac{1} {y +(1-y) \rho} \right>^{-1}.
 \end{eqnarray}

\noindent
Thus, we see that the metrics for Universal Doomsday are exactly the same as for Doomsday,

\begin{eqnarray}
\label{RPW_UD_vs_Doom}
R_W^{UD}= R_W^{(\Pick) E}, \nonumber \\
R_P^{(\Pick)UD} =  R_P^{(\Pick)E}, \\
R_{P/W}^{(\Pick)UD} = R_{P/W}^{(\Pick)E}. \nonumber
\end{eqnarray}

\noindent
In the inclusive case, $\left<y\right>_d =  \left<y\right>$, $R^E_P=R^E_W$, and $R^E_{P/W}=1$ for both Doomsday and Universal Doomsday.  So one cannot have one without the other.  For the exclusive case, $\left<y\right>_{d \Pick} \neq \left<y\right>$, and $R^{\Pick E}_P \neq R^{\Pick E}_W$, but the values for these metrics and  $R^{\Pick E}_{P/W}$ for Universal Doomsday and Doomsday are the same.  There is a fundamental reason for this: any Doomsday effect, from our data on being in a world selected from ensemble $E$, can be written as a Universal Doomsday change in our weighting of the ensemble, $i.e.$, taking $p(\Pick y) \rightarrow p((\Pick) y | d (\Pick))$.  So Universal Doomsday and Doomsday are two different ways of expressing the same effect, or lack thereof.

\Section{Sleeping Beauty Problem}  \label{Section_Sleeping}
Let us apply what we have learned to an observer thought experiment called the `Sleeping Beauty Problem' \cite{Elga_SBP}, which has generated disagreement to the point that  philosophers have separated into two camps called `Halfers' \cite{Lewis_SBP,Arntzenius,Pust2008} and `Thirders' \cite{Elga_SBP,Papineau01012009,Rosenthal,Horgan2008}:

\textit{Suppose Sleeping Beauty
is put to sleep on Sunday.  She is woken on Monday, questioned, then put back to sleep, and all her memories of that day are deleted.  A fair coin is flipped.  If it lands tails, she is also woken on Tuesday, and again questioned, put back to sleep and her memory deleted.  If it lands heads, she is not woken on Tuesday.  In either case she awakes on Wednesday after the experiment concludes.  Beauty is aware of all of the above.  She is asked each time she is woken for the probability that the coin flip results in ``heads."}

So-called Halfers argue that she should answer ``1/2" (each time) because it is a fair coin and she learns nothing from being awakened, and the question is the same as ``what is the probability you are in a heads world?" (\emph{i.e.} a world where the coin landed heads). 
So-called Thirders argue that she should say ``1/3" because there is one observer moment associated with a head flip, which we will call Mon-$H$, and there are two associated with tails, Mon-$T$ and Tue-$T$, and the question is effectively the same as ``what is the probability you are in a heads observer moment?"
There are a number of other papers advocating one side or the other, but none of them specify whether the situation corresponds to inclusive or exclusive selection, which we will see is key.  A number of authors assume the SIA, which as we have pointed out is an unfortunate kludge that leads to the Presumptuous Philosopher problem.  All authors seem to argue that if Beauty learns that it is Monday, her estimate for ``heads" should go up.  As we will see, that is not always true. There are also arguments about what wagers she should be willing to accept and whether that reasoning should affect her probability estimate, which we address at the end of the section.

For our formalism, we need two sets. We need a set of worlds, $W= \{ W_H,W_T\}$, in which the coin came up $H$ or $T$.  For a fair coin, the probability of picking each world is the same: $P(W_H)=P(W_T)=1/2$.  Nested inside $W$ is the set, $P$, of Sleeping Beauty observer moments, $P=P_{\text{Mon},H} \cup P_{\text{Mon},T} \cup P_{\text{Tue},T} = \{\text{Mon-}H,\ \text{Mon-}T,\  \text{Tue-}T\}$, where the first element belongs to $P_H$ (which is nested in $W_H$), and the other two to $P_T$ (nested in $W_T$).  If Beauty does not know the day, all three of these observer moments are indistinguishable to her.

First, let's look at Beauty's viewpoint within the inclusive case.  The probability that she should assign for the coin coming up heads within the world associated with her observer moment is given by the Be probability for a heads observer moment,

\begin{equation}
\label{SBP1_Be}
P(P W_H | PW) = P(P_H) = \frac{n_{,H}}{n}= \frac{1}{3}.
\end{equation}

\noindent
That is, in the inclusive case ``she" is in all three observer moments, only one of which is a heads observer moment.  

If she learns the day is Monday, the set of observer moments is $[P_{\text{Mon}}]$ instead of $P$, and her probability for ``heads" increases because ``she" could be in only two Monday observer moments: 

\begin{equation}
\label{SBPMon1_Be}
P([P_{\text{Mon}}] W_H \, | \, [P_{\text{Mon}}] W) 
= \frac{[n_{\text{Mon}}]_{,H}}{[n_{\text{Mon}}]}= \frac{1}{2}.
\end{equation}

\noindent
Thus, in the inclusive case, learning that it is Monday \textit{does} increase her probability estimate that the coin came up heads, and both of these probabilities correspond to those of the Thirder camp.

Next, let's look at Beauty's viewpoint with exclusive selection.  If she does not know the day, her probability estimate is the same as that of an outside observer, such as the coin flipper, where a single world (coin flip) result is Picked first:

\begin{equation}
\label{SBP1_Pick}
P(P \Pick W_H | P \Pick W) = P(W_H)= \frac{1}{2}.
\end{equation}

\noindent
In other words, if she assumes there is one world, it has a 1/2 chance of being an $H$-world, and her being awake in an observer moment and not knowing the day brings her no new information.  This is the Halfer point of view.  

Now, suppose she learns it is a Monday.  One might think that this information should increase her credence in ``heads".  And in fact, if you were to Pick a single recording of a random day in the experiment (Mon-$H$ in an $H$-world, Mon-$T$ or Tue-$T$ in a $T$-world), and the recording turned out to be from a Monday, you should increase your credence that the coin came up heads, as the Halfer camp claims,

\begin{eqnarray}
\label{SBPMon1dayPicked}
P&&(\text{``Record Picked= Mon-\textit{H}"} | \text{``Mon"})= \nonumber \\
&&P(P \Pick W_H \, | \, P_{\text{Mon}} \Pick W) =\nonumber \\
&&\frac{P(P_{\text{Mon}} \PickNeutered W | P \PickNeutered W_H)P(W_H)}{P(P_{\text{Mon}} \Pick W | P \Pick W)} = \nonumber\\
&&\frac{P(W_H)}{\sum_{F=\{H,T\}}P(P_{\text{Mon},F}|P_{,F})P(W_F)}=\nonumber \\
&&\frac{1/2}{1/2+1/4}= \frac{2}{3},
\end{eqnarray}

\noindent
\textit{but that's not what Beauty does.}  Instead, if the coin comes up tails, she experiences \textit{both} Mon-$T$ and Tue-$T$, so the fact that one of them is on a Monday adds no new information.  In our formalism, the way to see this is that the set of observer moments is $[P_{\text{Mon}}]$ instead of $P$, and her estimate for the probability of heads is just,

\begin{equation}
\label{SBPMon1_Pick}
P([P_{\text{Mon}}] \Pick W_H \, | \, [P_{\text{Mon}}] \Pick W) 
= P(W_H) = \frac{1}{2}.
\end{equation}

\noindent
So if Beauty assumes exclusive selection, learning that it is Monday does not increase her credence that she is in an $H$-world because she is sure to experience a Monday whatever the coin flip.  
(The reader might note that if Beauty learns that it is a Tuesday, she should assign zero probability to $H$, but that fact does not affect her probability for $H$ in the case where she learns it is Monday because in a tails world `she' experiences both days.)
This is good, because if she knows it is a Monday, then the amnesia drug is irrelevant, it is the same situation if you ask anyone what the odds a fair coin will come up heads, and there had better be no difference between inclusive and exclusive selection: they both conclude that the probability of heads is 1/2, as they do in Eqs. (\ref{SBPMon1_Be}) and (\ref{SBPMon1_Pick}).

Now, it is interesting to consider what happens if we run the experiment multiple times, once a week for $w$ weeks. We will assume she does not know the day, so the amnesia drug does matter.  If Beauty knows the week, she can treat each of the $w$ experiments like a copy of the original experiment, and she should come to the Thirder (Halfer) probability in the inclusive (exclusive) case.  If she does \textit{not} know the week, the inclusive probability is unchanged, but something interesting happens in the exclusive case: we get the result Nick Bostrom calls a `hybrid model' \cite{Bostrom_SBP}.

In this exclusive situation, there is one fixed set of coin flips $F=\{F_1,F_2,... F_w\}$ which actually occurs.  The set of worlds can be broken into $2w$ subsets specifying exactly one flip, such as $W_{F_1}$, where the coin in week 1 came up heads for $F_1=H_1$ and tails for $F_1=T_1$, and we do not specify what happened in the other weeks.   We can also break $W$ down into subsets with the flips in multiple weeks specified, including the $2^w$ subsets where they are all specified: $W_{F_1 F_2 ... F_w}$.  There is a third way to partition the set $W$, by the total number of heads, $h$, in set $F$, $W_h$.
If $w=1$, we have $P(P \Pick W_{H_1})=1/2$ because she is in either $W_{H_1}$ or $W_{T_1}$ with equal probability. But, if $w>1$, although she reasons she can experience exactly one sequence of coin flips, \textit{e.g.,} $\{H_1,T_2\}$, she also reasons that in that world she should lump observer moment Mon$_1$-$H_1$ with Mon$_2$-$T_2$ and Tue$_2$-$T_2$, since she has no way to tell them apart.  So for sequences with half the flips heads, $h = w/2$, she will come up with a probability of 1/3 for the coin having been heads in a given observer moment.  For a sequence with a total of $h$ heads out of $w$ flips, the probability of her being in a heads observer moment is $h/(h+2(w-h))$.  Thus she just needs to weight this probability by the probability that the sequence that occurs has $h$ heads, $P(W_h)$, which is $\frac{1}{2^w} {w \choose h}$:

\begin{eqnarray}
\label{SBP_multiple}
P&&((P_1 \Pick W_{H_1}) \lor ... (P_w \Pick W_{H_w}) | P \Pick W) \nonumber \\
=&&\sum_{h=0}^w P((P_1 \Pick W_{H_1}) \lor ... (P_w \Pick W_{H_w}) | P \Pick W_h) 
P(W_h)  \nonumber \\
=&&\sum_{h=0}^w \frac{h}{h + 2 (w-h)} 
\frac{1}{2^w} {w \choose h}.
\end{eqnarray}

\noindent
For $w=1$, this is $1/2$, for $w=2$ it is $5/12$, which is midway between $1/2$ and $1/3$, and for $w=10$, the probability of heads drops to about $0.35$.  For larger and larger $w$,$P(W_h)$ is approximately a narrower and narrower Gaussian centered on $h=w/2$, and the probability for Beauty's heads observer moments gets closer and closer to $1/3$.  In other words, exclusive selection with a large number of indistinguishable trials becomes indistinguishable from inclusive selection. 

Let us consider what happens if we ask Beauty to wager on whether the coin will come up heads or tails.  Can she distinguish whether she is in a reality that corresponds to the inclusive or exclusive case?  The answer is  ``no", because they lead to the same result, though for seemingly different reasons.  Suppose she is offered $x$:1 odds that the coin landed heads.  We will consider the cases where she bets at every awakening, or only on Mondays.  First, consider how Beauty would see the situation on Wednesday, after the experiment is over.  Whether she is in  the inclusive or exclusive case, she calculates that she has a 1/2 chance of being in a world where the coin came up heads and she won $x$ on Monday, and a 1/2 chance of being in a world where the coin came up tails and she lost 1 on both Monday and Tuesday, so she calculates her average winnings to be,

\begin{equation}
\label{Delta}
\Delta= \frac{1}{2}(x-2).
\end{equation}

\noindent
Thus, she will break even ($\Delta=0$) if she is given 2:1 odds.  If the betting occurs only on Mondays, then, whether she is in the inclusive or exclusive case, she calculates that she has a 1/2 chance of being in a world where the coin came up heads and she won $x$ on Monday, and a 1/2 chance of being in a world where the coin came up tails and she lost 1 on Monday.  Thus Beauty after the experiment calculates her average winnings on Mondays to be,

\begin{equation}
\label{DeltaMon}
\Delta_{\text{Mon}}= \frac{1}{2}(x-1),
\end{equation}

\noindent
and she will break even ($\Delta_{\text{Mon}}=0$) on Monday bets if she is given even money, 1:1 odds.  

How can her winnings be the same for the inclusive or exclusive case when her credence for heads differs for them (if she does not know the day)? If she assumes she is in the exclusive case, then her reasoning during the experiment is exactly the same as afterwards.  She has a 1/2 chance of being in a world where the coin comes up heads and she wins $x$ on Monday, and a 1/2 chance of being in a world where the coin comes up tails and she loses 1 on both Monday and Tuesday.  Thus she calculates her winnings for betting each day (on Mondays) to be Eq. (\ref{Delta})  (Eq. (\ref{DeltaMon})).  The exclusive case and Wednesday results are the same because they both refer to head and tail worlds.

If she assumes she is in an inclusive case, then ``she" is in all three of the observer moments, $\{\text{Mon-}H,\ \text{Mon-}T,\  \text{Tue-}T\}$, and so if she bets in each, her winnings per observer moment are, 

\begin{equation}
\label{Deltamoment}
\Delta^{\text{moment}}= \frac{1}{3}(x-2).
\end{equation}

\noindent
and if she bets only on the two Monday moments, then her winnings per observer moment are,

\begin{equation}
\label{DeltaMonmoment}
\Delta^{\text{moment}}_{\text{Mon}}= \frac{1}{2}(x-1).
\end{equation}

\noindent 
But to compare apples to apples, we need to know what she thinks the winnings per \emph{world} will be, which just changes the normalization factor for \Eq{Deltamoment} by the number of observer moments per world, which is $3/2$: 
$\Delta=\frac{3}{2} \Delta^{moment}=\frac{1}{2}(x-2)$.  For the Monday case, the number of observer moments and worlds is the same, so $\Delta_{\text{Mon}}=\Delta^{\text{moment}}_{\text{Mon}}=\frac{1}{2}(x-1)$, and we again get Eq. (\ref{Delta}-\ref{DeltaMon}).  

So an inclusive Beauty calculates the same winnings per world as an exclusive Beauty.  Inclusive Beauty needs 2:1 odds to break even because she wins in only one observer moment out of three.  Exclusive Beauty needs 2:1 odds to break even because although she has a $1/2$ probability of a heads world picked out by the coin flip, whenever she is in a tails world she loses twice.  What this means is that there is no practical difference between the inclusive and exclusive case in this thought experiment, and no way to tell them apart.  

The question, ``what credence do you assign to heads?"  has answer ``1/3" if Beauty sees herself as being in all three observer moments, and ``1/2" if she sees herself as living in an $H$-world or a $T$-world.  So, in the end, the only difference between inclusive Beauty (Thirder position) and exclusive Beauty (Halfer position) is that the former sees `herself'  in all three observer moments with equal probability, and the latter sees `herself' in one of two worlds with equal probability.  For the Halfer, the person in Mon-T and Tue-T is the same, a temporal continuation of one being, but not the same person as Mon-H because they are mutually exclusive timelines. For the Thirder, all three observer moments correspond to the same person, an inclusive viewpoint.  Neither of these is inherently right or wrong, it is a matter of how 
we define `self'--- we do not give an answer about which camp is ‘right’ because they are each right given a reasonable set of assumptions. 
We can analyze the problem with either definition, but there is no physical difference between them, as shown by the identical betting odds for the Halfer and Thirder viewpoints.

Note that one can rephrase the single-run Sleeping Beauty Problem as several equivalent problems, such as the Sailor's Child Problem \cite{Sailors_Child}, but the answer is the same: for the inclusive case the probability is $1/3$, and for the exclusive case it is $1/2$, and there is no way to tell them apart with betting.

Finally, it is possible to construct a similar Gedankenexperiment where betting \emph{can} distinguish between inclusive and exclusive cases. Motivated by Nick Bostrom's Incubator problem \cite{Bostrom_book}, Scott Aaronson suggests the following scenario \cite{Scott_private1}: If a fair coin comes up heads, Beauty H-One is cloned into existence; if tails Beauties T-One and T-Two are cloned into existence. If you find yourself to be one of these people, what odds would you need to bet that the coin comes up heads? One needs to be extra careful when observers are created like this. In the exclusive case, if $H$, then you are H-One and you win $x$; if $T$ you are either T-One or T-Two, and you lose 1, so $x=1$, you are willing to take 1:1 odds. For the inclusive case, you need to specify your assumptions about personhood. H-One wins $x$, and T-One and T-Two each lose 1, but which of them are `you'? Here are three possibilities:

\begin{enumerate}
\item You are exactly one of the three. You have $1/3$ chance of winning $x$ and $2/3$ chance of losing 1, so $x=2$, you need 2:1 odds.
\item You are one person each world. If heads you are H-One, if tails you are \emph{one} of T-One or T-Two. You have $1/2$ chance of winning $x$ and $1/2$ chance of losing 1, so $x=1$, you need 1:1 odds.
\item You are all three. You have $1/3$ chance of winning $x$ and $2/3$ chance of losing 1, so $x=2$, you need 2:1 odds.
\end{enumerate}

\noindent
So with the first and the third assumptions, the inclusive case \emph{differs} from the exclusive one, whereas it does not for  the second assumption. As we have stressed throughout this work, carefully specifying assumptions is crucial.

\Section{Heuristic Summary and Future Directions}  \label{Section_Summary}

We fully recognize that some readers interested in the topic of observer selection effects (OSE) are not used to as much math as we used.  To that end, we provide a heuristic summary of our main results.  We end by pointing to some directions in which this line of research may proceed.

Our central goal was to study the claim that there is a Doomsday effect---that by taking into account one's temporal location in a world, that datum leads one to conclude that the world will end sooner than one otherwise would have thought.  Along the way, we built the tools needed to investigate that claim, laid out arguments about when the Doomsday effect holds, and discussed related issues, such as the problems in cosmology due to Boltzmann brains.

Throughout the paper, we discussed probabilities of selecting `people' from some set $P$.  Usually the people were the observers in the problem.  The key distinguishing element about whether there is an OSE or not is if the selection is a Pick or a Be---whether one first \emph{picks} a `world' that the person belongs to, or whether no such picking is needed because the person just \emph{is} in the world.  

In Section \ref{Section_Prisoner}, we explored the latter via the Prisoner Problem.  If you are a prisoner in a cell, no one has to select that cell, or cellblock, or prison, for you to experience an observer moment there.  You just \emph{are} there. As a result, you are more likely to Be in a cellblock type $L$, which has more prisoners than a cellblock of type $S$, and that effect exactly cancels the effect of learning rank information $d$, which would otherwise favor you in being in a cellblock type $S$ (see the left half of Fig. 1).

Contrast that to Section \ref{Section_Warden}, where we considered the Warden Problem, where a warden has to \emph{pick} a cellblock before selecting a prisoner.  This is the way things usually work when not selecting observers:
when the entity being selected is in an enclosing set, such as a prisoner in a cell within a cellblock, to select them one has to pick the outer set, such as the cellblock, first.  The effect of this Pick is to nullify the counteracting effect, seen in the Be case, due to the number of prisoners.  The result is that the rank information $d$ \emph{does} tell you that if you are picked by the warden, you are more likely to be in a cellblock type $S$ (see the right half of Fig. 1).

Actually, to be more precise, the issue is whether there is \emph{any} selection beyond the one needed on the innermost (leftmost, in our notation) set, and not whether that selection is a Be or Pick.  If the selection on the leftmost set is the only one, we call it \emph{inclusive} selection.  If there is a selection on one or more of the enclosing sets, then we call it \emph{exclusive} selection.  In most of the inclusive cases we considered the selection of the innermost set was a Be. This is unsurprising, because in order to physically select elements of a set within some set of `worlds', one usually must pick the `world' (urn, cellblock, civilization...) first.  (We did give a counter-example, the Warden Cafeteria Problem, where the warden directly picks a prisoner in the cafeteria, circumventing the enclosing set $W$ (the prisoners are still labeled by the `world' that they belong to, just not constrained to be selected via that world).  And it is also possible to have a Be-selection on a set other than the leftmost set by making $P$ an enclosing set for some other set which the observer picks from, and then the situation will necessarily be exclusive.)

We then explored the concepts of inclusive and exclusive selection by extending our analysis of the Prisoner Problem to the largest physical enclosing set in the problem, which we call $E$.  For our problem, this corresponds to a set of prisons containing various fractions of $S$ and $L$ cellblocks. In the inclusive case (Section \ref{Section_Inclusive}), the only selection is on the leftmost set (a Be-selection of set $P$). 
We then considered exclusive selection (Section \ref{Section_Exclusive}), where there is selection on $E$ in addition to the Be-selection on $P$.
As in the Prisoner Problem, we found that there is no OSE in the inclusive case.  In the exclusive case, there is an OSE, but its magnitude depends on our prior assumptions.  One can find effects which range from nearly no OSE to an OSE as large as in the Warden case (see Fig. 2).  The larger the differential between the choices one picks from, the larger the OSE.  We can generalize $E$ to comprise `everything', a set of all possible universes.  Inclusive selection then corresponds to \emph{the inclusiverse}, that we also later called \emph{the complete multiverse}, 
which simply means that we assume all possibilities are realized.  Exclusive selection corresponds to \emph{an exclusiverse}, where only some possibilities are realized.

Next, in Section \ref{Section_Theories}, we added an enclosing set of theories, $\Theta$.  
We tend to view theories and hypotheses as mutually exclusive: one must \emph{pick} one and then analyze the resulting scenario.  But that Pick introduces an OSE because now the selection is exclusive, so one should be careful not to promote coexisting possibilities to hypotheses, such as ``I am in an $S$ cellblock".   Instead, one should say that there are multiple physical cellblocks, and we are in one of them with some probability for being in an $S$ cellblock.  If we really want to have coexisting hypotheses,  we would need to have inclusive selection on $\Theta$, a ``theoryverse" if you will.  That is not as unreasonable as it seems.  For example, the string landscape predicts multiple coexisting theories.  Another avenue we took in this set-of-theories analysis was to ask if we can probe whether we live in the inclusiverse or an exclusiverse. It is not generally possible, because it is usually impossible to disentangle other effects.
We also briefly discussed the Presumptuous Philosopher problem.  It is not a problem for us because we do \emph{not} make use of something called the Self-Indication Assumption, 
and argue against its use.
(We noted in several places that if we use the SIA---where a weighting factor for observers is put in by hand instead of it arising naturally out of typicality and keeping track of how observers are selected---then we get the wrong answer when there is exclusive selection. The Presumptuous Philosopher problem is an example of this.)

Thus far, we had assumed that whatever selection was done, was ``typical'', that is, corresponding to what one would get by random selection of a given subset of entities from a set.  We relaxed that assumption, and found that any atypical selection can be made typical by a simple redefinition of the relevant sets.  This allowed us to address the question of Boltzmann brains, which are hypothetical freak observer moments which arise from very rare fluctuations.  They are a problem in a stupendously large universe where it is possible for them to dominate normal observers, which are confined to a small subset of the spacetime.  This is a consistency problem because we must assume that we are not freak observers for us to argue that we have a correct understanding of the world, so that understanding is inconsistent if it predicts that we are freak observers. We examined an argument by Boddy \emph{et al.} \cite{Boddy2016} that there are no self-aware freak observers because at late times the Universe will be an empty exponentially expanding de Sitter space with no decoherence to split into ``many worlds.'' We argued that there could be decoherence effects from diluted matter, but an upper bound on the typicality of that is so small that it counters the huge number of future freak observers such that, by this argument, there are essentially no self-aware freak observers. We also used the analysis of Hartle, Hertog, and Srednicki to demonstrate a `first-person probability' effect which is somewhat orthogonal to ours---that when models with observers are scarce, models with more places for them to be are favored, even with exclusive selection. Conversely, if all viable models allow potentially many freak observers, those with fewer places for those freak observers to fluctuate into existence are favored.

We then considered the analysis of J. Richard Gott III in Section \ref{Section_Gott}, which seems to constitute a different kind of OSE.  He argued that one can bound the probability of a world lasting time $T$ using an observer's time $t$ since the start of the world; this is strange because the selection seems to be inclusive: just the Be-selection of the observer. One problem is that his original treatment did not include a prior, which is essential. We showed that one needs a fast falling ($\sim 1/T^2$) prior to reproduce his results. Then there is an effect, but it is \emph{not} an OSE, rather just an artifact of the fast-falling prior.  However, if we consider a scenario with a Pick-selection of a unique lifetime for the world, and the prior falls with $T$, then there \emph{is} an OSE.

All of this prepared us to address, in Section \ref{Section_Doomsday}, the Doomsday question, ``Do observer selection effects increase the probability that our world will be short- lived?”  The answer is, ``Probably not."  One must first write the question in a scale-invariant way, by which we mean that it makes just as much sense to ask at any timescale during the world.  A question that could work is, ``Will our world last $K$ times its present age?'', which naturally leads to using the formalism we developed in Section \ref{Section_Gott} for the Gott analysis.  There are scenarios where it is reasonable for the selection there to be exclusive, and it \emph{is} possible to conclude that there is a Doomsday effect, but only under a set of assumptions akin to those listed at the end of Section \ref{Section_Doomsday}.

Several papers have argued for a Universal Doomsday effect, which says that our data imply that worlds on average are probably more short-lived than we would have estimated without our data.  We showed that Universal Doomsday and Doomsday are inextricably linked because if our expectation for the fraction of short-lived worlds changes as a result of our data, so does our expectation for the lifetime of our world, and vice versa.  So the assumptions needed for a Universal Doomsday effect are the same as those needed for a Doomsday effect. 

We then applied our formalism to a somewhat different scenario called the Sleeping Beauty Problem.  Beauty is woken once or twice during an experiment, depending on a coin flip, and her memory of each awakening is deleted.  What probability should she assign to the coin having come up `heads'?  This would seem to be trivial, but has led to philosophers dividing into two camps, `Halfers', who would assign probability 1/2, and `Thirders', who would assign probability 1/3.  It turns out that they are both right.  The problem is that the question is insufficiently clearly posed and each answer is right, given a particular question.  If Beauty views `herself' as occupying the three equally likely observer-moments, the inclusive case, then she agrees with the Thirders.  If, on the other hand, she views `herself' as being in one of two possible timelines: in the one waking session of the `heads world', \emph{or} the two waking sessions of the `tails world', she will agree with the Halfers.  These are both reasonable ways of interpreting who `she' is.
They might also be interpreted as implying whether the world is a multiverse (in the inclusive case) or not (in the exclusive case), though this is an extrapolation---all she is really doing is assuming one or the other definition of self.  Anyway, the two cases are physically indistinguishable.  For example, we showed that both cases lead to precisely the same betting outcomes, though Beauty arrives at the same correct odds of winning in each case for different reasons.  We also discussed multiple trials, and the creation of observers, which may help extend the formalism of the paper to more general problems.

So, we have explored multiple ways in which it matters how observers are selected.  The key factor is whether the selection is inclusive or exclusive. There can be an OSE in the latter case but not the former, at least for the problems we considered.  Inclusive selection means that all events considered actually occur, though you may not experience them, such as prisoners being in an $S$ and an $L$ cellblock.  Exclusive selection means assigning nonzero probabilities to some events which do not occur, such as picking an $S$ or $L$ cellblock.  So,

\begin{quote}
Observer Selection Effects arise from assuming that there are some possibilities which are not realized.
\end{quote}

\noindent
Among other things, to have a Doomsday effect requires such an exclusive selection, which we wrote as, ``There is exactly one true value of the lifetime, $T_*$, because you consider only one world with one fixed future.''  It is thus crucial that one carefully lays out all of one's assumptions, because whether there is an OSE or not  depends upon them.

Finally, we lay out some possible future directions for this work.

A simple direction to go in is to relax some of the assumptions we made, such as $\rho$ being constant across the ensemble of possibilities, or that the subsets are nonoverlapping (see Appendix A) to generalize our results.

Almost all of our analysis was classical.  It would be interesting to explore further the quantum context.  One consequence is clear: if quantum theory corresponds to something like the Many Worlds Interpretation, then we are in a multiverse with inclusive selection of events.  If there is ``wave-function collapse," so that there is only one reality, then there is an exclusive selection.  But a comprehensive evaluation of our discussion in the quantum context may turn up interesting results.  For example, what of ‘quantum observers’ which comprise superpositions of observer states?

Another avenue of inquiry is how to analyze a theoryverse, such as the string landscape.  Is it reasonable to assume the inclusive case? In other words, should we sum probabilities of ``observers like us” from different parts of the string landscape which contain observers similar to us despite operating with different physical laws?  If so, then it is not the probability of a given vacuum in the landscape that matters, but that probability times its effective number of observer moments.

Finally, while we discussed atypical observers, and the problem of Boltzmann brains, there is perhaps more to learn from studying what one might call ‘freak observers’—any observer who happens to experience freakish conditions.  There are many metrics for ‘number of observers’ in addressing the problem of Boltzmann brains, and it would be useful to see if our results shed any light on them.  Also,  in a multiverse there are otherwise normal observers who happen to experience statistical fluctuations of many standard deviations who draw erroneous conclusions.  How do we treat such observers, especially with the recognition that it is not impossible in a multiverse that we are one of them? 

%
\begin{acknowledgments}
We thank Scott Aaronson, Jim Hartle, Carl Caves, Steve Carlip, and the anonymous referees for useful suggestions.
\end{acknowledgments}
%
\bibliography{OSE}{}

\begin{thebibliography}{54}%
\makeatletter
\providecommand \@ifxundefined [1]{%
 \@ifx{#1\undefined}
}%
\providecommand \@ifnum [1]{%
 \ifnum #1\expandafter \@firstoftwo
 \else \expandafter \@secondoftwo
 \fi
}%
\providecommand \@ifx [1]{%
 \ifx #1\expandafter \@firstoftwo
 \else \expandafter \@secondoftwo
 \fi
}%
\providecommand \natexlab [1]{#1}%
\providecommand \enquote  [1]{``#1''}%
\providecommand \bibnamefont  [1]{#1}%
\providecommand \bibfnamefont [1]{#1}%
\providecommand \citenamefont [1]{#1}%
\providecommand \href@noop [0]{\@secondoftwo}%
\providecommand \href [0]{\begingroup \@sanitize@url \@href}%
\providecommand \@href[1]{\@@startlink{#1}\@@href}%
\providecommand \@@href[1]{\endgroup#1\@@endlink}%
\providecommand \@sanitize@url [0]{\catcode `\\12\catcode `\$12\catcode
  `\&12\catcode `\#12\catcode `\^12\catcode `\_12\catcode `\%12\relax}%
\providecommand \@@startlink[1]{}%
\providecommand \@@endlink[0]{}%
\providecommand \url  [0]{\begingroup\@sanitize@url \@url }%
\providecommand \@url [1]{\endgroup\@href {#1}{\urlprefix }}%
\providecommand \urlprefix  [0]{URL }%
\providecommand \Eprint [0]{\href }%
\providecommand \doibase [0]{http://dx.doi.org/}%
\providecommand \selectlanguage [0]{\@gobble}%
\providecommand \bibinfo  [0]{\@secondoftwo}%
\providecommand \bibfield  [0]{\@secondoftwo}%
\providecommand \translation [1]{[#1]}%
\providecommand \BibitemOpen [0]{}%
\providecommand \bibitemStop [0]{}%
\providecommand \bibitemNoStop [0]{.\EOS\space}%
\providecommand \EOS [0]{\spacefactor3000\relax}%
\providecommand \BibitemShut  [1]{\csname bibitem#1\endcsname}%
\let\auto@bib@innerbib\@empty
\bibitem [{\citenamefont {Noether}(1918)}]{noether1918}%
  \BibitemOpen
  \bibfield  {author} {\bibinfo {author} {\bibfnamefont {E.}~\bibnamefont
  {Noether}},\ }\bibfield  {title} {\enquote {\bibinfo {title} {Invariante
  variationsprobleme},}\ }\href@noop {} {\bibfield  {journal} {\bibinfo
  {journal} {Nachrichten von der Gesellschaft der Wissenschaften zu
  G{\"o}ttingen, Mathematisch-Physikalische Klasse}\ }\textbf {\bibinfo
  {volume} {1918}},\ \bibinfo {pages} {235--257} (\bibinfo {year} {1918})},\
  \bibinfo {note} {{E}nglish reprint:
  http://dx.doi.org/10.1080/00411457108231446.}\BibitemShut {Stop}%
\bibitem [{\citenamefont {Carter}(1974)}]{carter_1974}%
  \BibitemOpen
  \bibfield  {author} {\bibinfo {author} {\bibfnamefont {Brandon}\ \bibnamefont
  {Carter}},\ }\bibfield  {title} {\enquote {\bibinfo {title} {Large number
  coincidences and the anthropic principle in cosmology},}\ }\href {\doibase
  10.1017/S0074180900017368} {\bibfield  {journal} {\bibinfo  {journal}
  {Symposium - International Astronomical Union}\ }\textbf {\bibinfo {volume}
  {63}},\ \bibinfo {pages} {291--298} (\bibinfo {year} {1974})}\BibitemShut
  {NoStop}%
\bibitem [{\citenamefont {Weinberg}(1989)}]{Weinberg:CC}%
  \BibitemOpen
  \bibfield  {author} {\bibinfo {author} {\bibfnamefont {Steven}\ \bibnamefont
  {Weinberg}},\ }\bibfield  {title} {\enquote {\bibinfo {title} {The
  cosmological constant problem},}\ }\href {\doibase 10.1103/RevModPhys.61.1}
  {\bibfield  {journal} {\bibinfo  {journal} {Reviews of Modern Physics}\
  }\textbf {\bibinfo {volume} {61}},\ \bibinfo {pages} {1--23} (\bibinfo {year}
  {1989})}\BibitemShut {NoStop}%
\bibitem [{\citenamefont {Carter}\ and\ \citenamefont
  {McCrea}(1983)}]{Carter1983}%
  \BibitemOpen
  \bibfield  {author} {\bibinfo {author} {\bibfnamefont {B.}~\bibnamefont
  {Carter}}\ and\ \bibinfo {author} {\bibfnamefont {W.~H.}\ \bibnamefont
  {McCrea}},\ }\bibfield  {title} {\enquote {\bibinfo {title} {The anthropic
  principle and its implications for biological evolution [and discussion]},}\
  }\href {\doibase 10.1098/rsta.1983.0096} {\bibfield  {journal} {\bibinfo
  {journal} {Philosophical Transactions of the Royal Society A: Mathematical,
  Physical and Engineering Sciences}\ }\textbf {\bibinfo {volume} {310}},\
  \bibinfo {pages} {347--363} (\bibinfo {year} {1983})}\BibitemShut {NoStop}%
\bibitem [{\citenamefont {Dieks}(1992)}]{DieksSIA}%
  \BibitemOpen
  \bibfield  {author} {\bibinfo {author} {\bibfnamefont {Dennis}\ \bibnamefont
  {Dieks}},\ }\bibfield  {title} {\enquote {\bibinfo {title} {Doomsday--{Or}:
  The dangers of statistics},}\ }\href {\doibase 10.2307/2220450} {\bibfield
  {journal} {\bibinfo  {journal} {The Philosophical Quarterly}\ }\textbf
  {\bibinfo {volume} {42}},\ \bibinfo {pages} {78} (\bibinfo {year}
  {1992})}\BibitemShut {NoStop}%
\bibitem [{\citenamefont {Bostrom}(1996)}]{Bostrom96}%
  \BibitemOpen
  \bibfield  {author} {\bibinfo {author} {\bibfnamefont {Nick}\ \bibnamefont
  {Bostrom}},\ }\href
  {http://www.anthropic-principle.com/preprints/inv/investigations.html}
  {\enquote {\bibinfo {title} {Investigations into the doomsday argument},}\
  }\bibinfo {howpublished} {(unpublished)} (\bibinfo {year} {1996})\BibitemShut
  {NoStop}%
\bibitem [{\citenamefont {Bostrom}(2002)}]{Bostrom_book}%
  \BibitemOpen
  \bibfield  {author} {\bibinfo {author} {\bibfnamefont {Nick}\ \bibnamefont
  {Bostrom}},\ }\href@noop {} {\emph {\bibinfo {title} {Anthropic bias:
  Observation selection effects in science and philosophy}}}\ (\bibinfo
  {publisher} {Routledge},\ \bibinfo {year} {2002})\BibitemShut {NoStop}%
\bibitem [{\citenamefont {Olum}(2002)}]{OlumSIA}%
  \BibitemOpen
  \bibfield  {author} {\bibinfo {author} {\bibfnamefont {Ken~D.}\ \bibnamefont
  {Olum}},\ }\bibfield  {title} {\enquote {\bibinfo {title} {The doomsday
  argument and the number of possible observers},}\ }\href {\doibase
  10.1111/1467-9213.00260} {\bibfield  {journal} {\bibinfo  {journal} {The
  Philosophical Quarterly}\ }\textbf {\bibinfo {volume} {52}},\ \bibinfo
  {pages} {164--184} (\bibinfo {year} {2002})}\BibitemShut {NoStop}%
\bibitem [{\citenamefont {Bostrom}\ and\ \citenamefont
  {Cirkovic}(2003)}]{BostromSIA}%
  \BibitemOpen
  \bibfield  {author} {\bibinfo {author} {\bibfnamefont {Nick}\ \bibnamefont
  {Bostrom}}\ and\ \bibinfo {author} {\bibfnamefont {Milan~M.}\ \bibnamefont
  {Cirkovic}},\ }\bibfield  {title} {\enquote {\bibinfo {title} {The doomsday
  argument and the self-indication assumption: Reply to {Olum}},}\ }\href
  {\doibase 10.1111/1467-9213.00298} {\bibfield  {journal} {\bibinfo  {journal}
  {The Philosophical Quarterly}\ }\textbf {\bibinfo {volume} {53}},\ \bibinfo
  {pages} {83--91} (\bibinfo {year} {2003})}\BibitemShut {NoStop}%
\bibitem [{\citenamefont {Knobe}\ \emph {et~al.}(2006)\citenamefont {Knobe},
  \citenamefont {Olum},\ and\ \citenamefont {Vilenkin}}]{Knobe-Olum-Vilenkin}%
  \BibitemOpen
  \bibfield  {author} {\bibinfo {author} {\bibfnamefont {Joshua}\ \bibnamefont
  {Knobe}}, \bibinfo {author} {\bibfnamefont {Ken~D.}\ \bibnamefont {Olum}}, \
  and\ \bibinfo {author} {\bibfnamefont {Alexander}\ \bibnamefont {Vilenkin}},\
  }\bibfield  {title} {\enquote {\bibinfo {title} {Philosophical implications
  of inflationary cosmology},}\ }\href {\doibase 10.1093/bjps/axi155}
  {\bibfield  {journal} {\bibinfo  {journal} {The British Journal for the
  Philosophy of Science}\ }\textbf {\bibinfo {volume} {57}},\ \bibinfo {pages}
  {47--67} (\bibinfo {year} {2006})}\BibitemShut {NoStop}%
\bibitem [{\citenamefont {Gerig}\ \emph {et~al.}(2013)\citenamefont {Gerig},
  \citenamefont {Olum},\ and\ \citenamefont {Vilenkin}}]{Gerig-Olum-Vilenkin}%
  \BibitemOpen
  \bibfield  {author} {\bibinfo {author} {\bibfnamefont {Austin}\ \bibnamefont
  {Gerig}}, \bibinfo {author} {\bibfnamefont {Ken~D.}\ \bibnamefont {Olum}}, \
  and\ \bibinfo {author} {\bibfnamefont {Alexander}\ \bibnamefont {Vilenkin}},\
  }\bibfield  {title} {\enquote {\bibinfo {title} {Universal doomsday:
  analyzing our prospects for survival},}\ }\href
  {http://stacks.iop.org/1475-7516/2013/i=05/a=013} {\bibfield  {journal}
  {\bibinfo  {journal} {Journal of Cosmology and Astroparticle Physics}\
  }\textbf {\bibinfo {volume} {2013}},\ \bibinfo {pages} {013} (\bibinfo {year}
  {2013})}\BibitemShut {NoStop}%
\bibitem [{\citenamefont {Garriga}\ and\ \citenamefont
  {Vilenkin}(2008)}]{GarrigaVilenkin2008}%
  \BibitemOpen
  \bibfield  {author} {\bibinfo {author} {\bibfnamefont {J.}~\bibnamefont
  {Garriga}}\ and\ \bibinfo {author} {\bibfnamefont {A.}~\bibnamefont
  {Vilenkin}},\ }\bibfield  {title} {\enquote {\bibinfo {title} {Prediction and
  explanation in the multiverse},}\ }\href {\doibase
  10.1103/PhysRevD.77.043526} {\bibfield  {journal} {\bibinfo  {journal} {Phys.
  Rev. D}\ }\textbf {\bibinfo {volume} {77}},\ \bibinfo {pages} {043526}
  (\bibinfo {year} {2008})}\BibitemShut {NoStop}%
\bibitem [{\citenamefont {Carroll}(2017)}]{Carroll:2017BB}%
  \BibitemOpen
  \bibfield  {author} {\bibinfo {author} {\bibfnamefont {Sean~M.}\ \bibnamefont
  {Carroll}},\ }\bibfield  {title} {\enquote {\bibinfo {title} {Why {B}oltzmann
  brains are bad},}\ }\href {https://arxiv.org/abs/1702.00850} {\  (\bibinfo
  {year} {2017})},\ \Eprint {http://arxiv.org/abs/1702.00850} {1702.00850}
  \BibitemShut {NoStop}%
\bibitem [{\citenamefont {III}(1993)}]{Gott1993}%
  \BibitemOpen
  \bibfield  {author} {\bibinfo {author} {\bibfnamefont {J.~Richard~Gott}\
  \bibnamefont {III}},\ }\bibfield  {title} {\enquote {\bibinfo {title}
  {Implications of the {C}opernican principle for our future prospects},}\
  }\href@noop {} {\bibfield  {journal} {\bibinfo  {journal} {Nature}\ }\textbf
  {\bibinfo {volume} {363}},\ \bibinfo {pages} {315} (\bibinfo {year}
  {1993})}\BibitemShut {NoStop}%
\bibitem [{end()}]{endnotenotprior}%
  \BibitemOpen
  \href@noop {} {}\bibinfo {note} {{T}his is not a prior probability for a
  world of type $S$, but rather the probability of being in a world of type
  $S$---so making use of the information of the fraction of observers in such
  worlds but prior to making use of rank information data.}\BibitemShut {Stop}%
\bibitem [{Sco()}]{Scott_private1}%
  \BibitemOpen
  \href@noop {} {\ }\bibinfo {note} {Scott Aaronson, private
  communication.}\BibitemShut {Stop}%
\bibitem [{\citenamefont {Hartle}\ and\ \citenamefont
  {Srednicki}(2007)}]{HartleSrednicki_Typical}%
  \BibitemOpen
  \bibfield  {author} {\bibinfo {author} {\bibfnamefont {James~B.}\
  \bibnamefont {Hartle}}\ and\ \bibinfo {author} {\bibfnamefont {Mark}\
  \bibnamefont {Srednicki}},\ }\bibfield  {title} {\enquote {\bibinfo {title}
  {Are we typical?}}\ }\href {\doibase 10.1103/PhysRevD.75.123523} {\bibfield
  {journal} {\bibinfo  {journal} {Phys. Rev. D}\ }\textbf {\bibinfo {volume}
  {75}},\ \bibinfo {pages} {123523} (\bibinfo {year} {2007})}\BibitemShut
  {NoStop}%
\bibitem [{\citenamefont {Srednicki}\ and\ \citenamefont
  {Hartle}(2010)}]{SrednickiHartle_LargeUniverse}%
  \BibitemOpen
  \bibfield  {author} {\bibinfo {author} {\bibfnamefont {Mark}\ \bibnamefont
  {Srednicki}}\ and\ \bibinfo {author} {\bibfnamefont {James}\ \bibnamefont
  {Hartle}},\ }\bibfield  {title} {\enquote {\bibinfo {title} {Science in a
  very large universe},}\ }\href {\doibase 10.1103/PhysRevD.81.123524}
  {\bibfield  {journal} {\bibinfo  {journal} {Phys. Rev. D}\ }\textbf {\bibinfo
  {volume} {81}},\ \bibinfo {pages} {123524} (\bibinfo {year}
  {2010})}\BibitemShut {NoStop}%
\bibitem [{Aar()}]{Aaronson_consciousness}%
  \BibitemOpen
  \href@noop {} {}\bibinfo {note} {This is true almost by definition---we
  define an observer as a self-aware entity that can process external
  information (such as datum $d$). But it is an interesting question whether
  participation in the arrow of time is a requirement for consciousness---see
  Scott Aaronson ``Could a Quantum Computer Have Subjective Experience?"
  https://www.scottaaronson.com/blog/?p=1951.}\BibitemShut {Stop}%
\bibitem [{\citenamefont {Penrose}(1979)}]{Penrose1979}%
  \BibitemOpen
  \bibfield  {author} {\bibinfo {author} {\bibfnamefont {Roger}\ \bibnamefont
  {Penrose}},\ }\bibfield  {title} {\enquote {\bibinfo {title} {Singularities
  and time-asymmetry},}\ }in\ \href@noop {} {\emph {\bibinfo {booktitle}
  {General Relativity, an Einstein Centennary Survey}}},\ \bibinfo {editor}
  {edited by\ \bibinfo {editor} {\bibfnamefont {S.W.}\ \bibnamefont {Hawking}}\
  and\ \bibinfo {editor} {\bibfnamefont {W.}~\bibnamefont {Israel}}}\ (\bibinfo
   {publisher} {Cambridge University Press},\ \bibinfo {year}
  {1979})\BibitemShut {NoStop}%
\bibitem [{\citenamefont {Wald}(2006)}]{WALD2006394}%
  \BibitemOpen
  \bibfield  {author} {\bibinfo {author} {\bibfnamefont {Robert~M.}\
  \bibnamefont {Wald}},\ }\bibfield  {title} {\enquote {\bibinfo {title} {The
  arrow of time and the initial conditions of the universe},}\ }\href {\doibase
  https://doi.org/10.1016/j.shpsb.2006.03.005} {\bibfield  {journal} {\bibinfo
  {journal} {Studies in History and Philosophy of Science Part B: Studies in
  History and Philosophy of Modern Physics}\ }\textbf {\bibinfo {volume}
  {37}},\ \bibinfo {pages} {394 -- 398} (\bibinfo {year} {2006})}\BibitemShut
  {NoStop}%
\bibitem [{\citenamefont {Boltzmann}(1897)}]{Boltzmann1897}%
  \BibitemOpen
  \bibfield  {author} {\bibinfo {author} {\bibfnamefont {L.}~\bibnamefont
  {Boltzmann}},\ }\bibfield  {title} {\enquote {\bibinfo {title} {Zu {H}rn.
  {Z}ermelo's {A}bhandlung: {U}eber die mechanische {E}rkl{\"a}rung
  irreversibler {V}org{\"a}nge},}\ }\href@noop {} {\bibfield  {journal}
  {\bibinfo  {journal} {Annalen der Physik}\ }\textbf {\bibinfo {volume}
  {296}},\ \bibinfo {pages} {392} (\bibinfo {year} {1897})},\ \bibinfo {note}
  {trans. in \emph{Kinetic Theory}, ed. S. G. Brush (Oxford, 1966), p.
  412.}\BibitemShut {Stop}%
\bibitem [{\citenamefont {Guth}(1981)}]{Inflation_Guth}%
  \BibitemOpen
  \bibfield  {author} {\bibinfo {author} {\bibfnamefont {Alan~H.}\ \bibnamefont
  {Guth}},\ }\bibfield  {title} {\enquote {\bibinfo {title} {Inflationary
  universe: A possible solution to the horizon and flatness problems},}\ }\href
  {\doibase 10.1103/PhysRevD.23.347} {\bibfield  {journal} {\bibinfo  {journal}
  {Phys. Rev. D}\ }\textbf {\bibinfo {volume} {23}},\ \bibinfo {pages}
  {347--356} (\bibinfo {year} {1981})}\BibitemShut {NoStop}%
\bibitem [{\citenamefont {Starobinsky}(1980)}]{STAROBINSKY1980}%
  \BibitemOpen
  \bibfield  {author} {\bibinfo {author} {\bibfnamefont {A.A.}\ \bibnamefont
  {Starobinsky}},\ }\bibfield  {title} {\enquote {\bibinfo {title} {A new type
  of isotropic cosmological models without singularity},}\ }\href {\doibase
  https://doi.org/10.1016/0370-2693(80)90670-X} {\bibfield  {journal} {\bibinfo
   {journal} {Physics Letters B}\ }\textbf {\bibinfo {volume} {91}},\ \bibinfo
  {pages} {99 -- 102} (\bibinfo {year} {1980})}\BibitemShut {NoStop}%
\bibitem [{\citenamefont {Linde}(1982)}]{LINDE1982}%
  \BibitemOpen
  \bibfield  {author} {\bibinfo {author} {\bibfnamefont {A.D.}\ \bibnamefont
  {Linde}},\ }\bibfield  {title} {\enquote {\bibinfo {title} {A new
  inflationary universe scenario: A possible solution of the horizon, flatness,
  homogeneity, isotropy and primordial monopole problems},}\ }\href {\doibase
  https://doi.org/10.1016/0370-2693(82)91219-9} {\bibfield  {journal} {\bibinfo
   {journal} {Physics Letters B}\ }\textbf {\bibinfo {volume} {108}},\ \bibinfo
  {pages} {389 -- 393} (\bibinfo {year} {1982})}\BibitemShut {NoStop}%
\bibitem [{\citenamefont {Albrecht}\ and\ \citenamefont
  {Steinhardt}(1982)}]{AlbrechtSteinhardt1982}%
  \BibitemOpen
  \bibfield  {author} {\bibinfo {author} {\bibfnamefont {Andreas}\ \bibnamefont
  {Albrecht}}\ and\ \bibinfo {author} {\bibfnamefont {Paul~J.}\ \bibnamefont
  {Steinhardt}},\ }\bibfield  {title} {\enquote {\bibinfo {title} {Cosmology
  for grand unified theories with radiatively induced symmetry breaking},}\
  }\href {\doibase 10.1103/PhysRevLett.48.1220} {\bibfield  {journal} {\bibinfo
   {journal} {Phys. Rev. Lett.}\ }\textbf {\bibinfo {volume} {48}},\ \bibinfo
  {pages} {1220--1223} (\bibinfo {year} {1982})}\BibitemShut {NoStop}%
\bibitem [{\citenamefont {Baumann}(2011)}]{TASI_inflation}%
  \BibitemOpen
  \bibfield  {author} {\bibinfo {author} {\bibfnamefont {Daniel}\ \bibnamefont
  {Baumann}},\ }\bibfield  {title} {\enquote {\bibinfo {title} {{Inflation}},}\
  }in\ \href {\doibase 10.1142/9789814327183_0010} {\emph {\bibinfo {booktitle}
  {{Physics of the large and the small, TASI 09, proceedings of the Theoretical
  Advanced Study Institute in Elementary Particle Physics, Boulder, Colorado,
  USA, 1-26 June 2009}}}}\ (\bibinfo  {publisher} {World Scientific Publishing
  Company},\ \bibinfo {address} {Hackensack, NJ},\ \bibinfo {year} {2011})\
  pp.\ \bibinfo {pages} {523--686},\ \Eprint {http://arxiv.org/abs/0907.5424}
  {arXiv:0907.5424 [hep-th]} \BibitemShut {NoStop}%
\bibitem [{\citenamefont {Gibbons}\ and\ \citenamefont
  {Hawking}(1977)}]{PhysRevD.15.2738}%
  \BibitemOpen
  \bibfield  {author} {\bibinfo {author} {\bibfnamefont {G.~W.}\ \bibnamefont
  {Gibbons}}\ and\ \bibinfo {author} {\bibfnamefont {S.~W.}\ \bibnamefont
  {Hawking}},\ }\bibfield  {title} {\enquote {\bibinfo {title} {Cosmological
  event horizons, thermodynamics, and particle creation},}\ }\href {\doibase
  10.1103/PhysRevD.15.2738} {\bibfield  {journal} {\bibinfo  {journal} {Phys.
  Rev. D}\ }\textbf {\bibinfo {volume} {15}},\ \bibinfo {pages} {2738--2751}
  (\bibinfo {year} {1977})}\BibitemShut {NoStop}%
\bibitem [{\citenamefont {Schulman}(1997)}]{schulman_1997}%
  \BibitemOpen
  \bibfield  {author} {\bibinfo {author} {\bibfnamefont {Lawrence~S.}\
  \bibnamefont {Schulman}},\ }\href {\doibase 10.1017/CBO9780511622878} {\emph
  {\bibinfo {title} {Time's Arrows and Quantum Measurement}}}\ (\bibinfo
  {publisher} {Cambridge University Press},\ \bibinfo {year}
  {1997})\BibitemShut {NoStop}%
\bibitem [{\citenamefont {Dyson}\ \emph {et~al.}(2002)\citenamefont {Dyson},
  \citenamefont {Kleban},\ and\ \citenamefont
  {Susskind}}]{Dyson_Kleban_Susskind}%
  \BibitemOpen
  \bibfield  {author} {\bibinfo {author} {\bibfnamefont {Lisa}\ \bibnamefont
  {Dyson}}, \bibinfo {author} {\bibfnamefont {Matthew}\ \bibnamefont {Kleban}},
  \ and\ \bibinfo {author} {\bibfnamefont {Leonard}\ \bibnamefont {Susskind}},\
  }\bibfield  {title} {\enquote {\bibinfo {title} {Disturbing implications of a
  cosmological constant},}\ }\href
  {http://stacks.iop.org/1126-6708/2002/i=10/a=011} {\bibfield  {journal}
  {\bibinfo  {journal} {Journal of High Energy Physics}\ }\textbf {\bibinfo
  {volume} {2002}},\ \bibinfo {pages} {011} (\bibinfo {year}
  {2002})}\BibitemShut {NoStop}%
\bibitem [{\citenamefont {Albrecht}\ and\ \citenamefont
  {Sorbo}(2004)}]{AlbrechtSorbo}%
  \BibitemOpen
  \bibfield  {author} {\bibinfo {author} {\bibfnamefont {Andreas}\ \bibnamefont
  {Albrecht}}\ and\ \bibinfo {author} {\bibfnamefont {Lorenzo}\ \bibnamefont
  {Sorbo}},\ }\bibfield  {title} {\enquote {\bibinfo {title} {Can the universe
  afford inflation?}}\ }\href {\doibase 10.1103/PhysRevD.70.063528} {\bibfield
  {journal} {\bibinfo  {journal} {Physical Review D}\ }\textbf {\bibinfo
  {volume} {70}},\ \bibinfo {pages} {063528} (\bibinfo {year}
  {2004})}\BibitemShut {NoStop}%
\bibitem [{\citenamefont {Coleman}(1977)}]{Coleman_false_vacuum}%
  \BibitemOpen
  \bibfield  {author} {\bibinfo {author} {\bibfnamefont {Sidney}\ \bibnamefont
  {Coleman}},\ }\bibfield  {title} {\enquote {\bibinfo {title} {Fate of the
  false vacuum: Semiclassical theory},}\ }\href {\doibase
  10.1103/PhysRevD.15.2929} {\bibfield  {journal} {\bibinfo  {journal} {Phys.
  Rev. D}\ }\textbf {\bibinfo {volume} {15}},\ \bibinfo {pages} {2929--2936}
  (\bibinfo {year} {1977})}\BibitemShut {NoStop}%
\bibitem [{\citenamefont {Page}(2008{\natexlab{a}})}]{Page_20_billion_yrs}%
  \BibitemOpen
  \bibfield  {author} {\bibinfo {author} {\bibfnamefont {Don~N.}\ \bibnamefont
  {Page}},\ }\bibfield  {title} {\enquote {\bibinfo {title} {Is our universe
  likely to decay within 20 billion years?}}\ }\href {\doibase
  10.1103/PhysRevD.78.063535} {\bibfield  {journal} {\bibinfo  {journal} {Phys.
  Rev. D}\ }\textbf {\bibinfo {volume} {78}},\ \bibinfo {pages} {063535}
  (\bibinfo {year} {2008}{\natexlab{a}})}\BibitemShut {NoStop}%
\bibitem [{\citenamefont {Page}(2008{\natexlab{b}})}]{Page_astronomical_rate}%
  \BibitemOpen
  \bibfield  {author} {\bibinfo {author} {\bibfnamefont {Don~N.}\ \bibnamefont
  {Page}},\ }\bibfield  {title} {\enquote {\bibinfo {title} {Is our universe
  decaying at an astronomical rate?}}\ }\href {\doibase
  http://dx.doi.org/10.1016/j.physletb.2008.08.039} {\bibfield  {journal}
  {\bibinfo  {journal} {Physics Letters B}\ }\textbf {\bibinfo {volume}
  {669}},\ \bibinfo {pages} {197 -- 200} (\bibinfo {year}
  {2008}{\natexlab{b}})}\BibitemShut {NoStop}%
\bibitem [{\citenamefont {Carlip}(2007)}]{Carlip_transient}%
  \BibitemOpen
  \bibfield  {author} {\bibinfo {author} {\bibfnamefont {S.}~\bibnamefont
  {Carlip}},\ }\bibfield  {title} {\enquote {\bibinfo {title} {Transient
  observers and variable constants or repelling the invasion of the
  {B}oltzmann's brains},}\ }\href
  {http://stacks.iop.org/1475-7516/2007/i=06/a=001} {\bibfield  {journal}
  {\bibinfo  {journal} {Journal of Cosmology and Astroparticle Physics}\
  }\textbf {\bibinfo {volume} {2007}},\ \bibinfo {pages} {001} (\bibinfo {year}
  {2007})}\BibitemShut {NoStop}%
\bibitem [{\citenamefont {Boddy}\ \emph {et~al.}(2016)\citenamefont {Boddy},
  \citenamefont {Carroll},\ and\ \citenamefont {Pollack}}]{Boddy2016}%
  \BibitemOpen
  \bibfield  {author} {\bibinfo {author} {\bibfnamefont {Kimberly~K.}\
  \bibnamefont {Boddy}}, \bibinfo {author} {\bibfnamefont {Sean~M.}\
  \bibnamefont {Carroll}}, \ and\ \bibinfo {author} {\bibfnamefont {Jason}\
  \bibnamefont {Pollack}},\ }\bibfield  {title} {\enquote {\bibinfo {title} {De
  {S}itter space without dynamical quantum fluctuations},}\ }\href {\doibase
  10.1007/s10701-016-9996-8} {\bibfield  {journal} {\bibinfo  {journal}
  {Foundations of Physics}\ }\textbf {\bibinfo {volume} {46}},\ \bibinfo
  {pages} {702--735} (\bibinfo {year} {2016})}\BibitemShut {NoStop}%
\bibitem [{\citenamefont {Hartle}\ and\ \citenamefont
  {Hertog}(2017)}]{Hartle_Hertog_1p}%
  \BibitemOpen
  \bibfield  {author} {\bibinfo {author} {\bibfnamefont {James}\ \bibnamefont
  {Hartle}}\ and\ \bibinfo {author} {\bibfnamefont {Thomas}\ \bibnamefont
  {Hertog}},\ }\bibfield  {title} {\enquote {\bibinfo {title} {{The Observer
  Strikes Back}},}\ }in\ \href {\doibase 10.1017/9781316535783.010} {\emph
  {\bibinfo {booktitle} {{The Philosophy of Cosmology}}}}\ (\bibinfo
  {publisher} {Cambridge University Press},\ \bibinfo {address} {New York,
  NY},\ \bibinfo {year} {2017})\ pp.\ \bibinfo {pages} {181--205},\ \Eprint
  {http://arxiv.org/abs/1503.07205} {arXiv:1503.07205 [gr-qc]} \BibitemShut
  {NoStop}%
\bibitem [{\citenamefont {Srednicki}\ and\ \citenamefont
  {Hartle}(2013)}]{Srednicki_Hartle_1p}%
  \BibitemOpen
  \bibfield  {author} {\bibinfo {author} {\bibfnamefont {Mark}\ \bibnamefont
  {Srednicki}}\ and\ \bibinfo {author} {\bibfnamefont {James}\ \bibnamefont
  {Hartle}},\ }\bibfield  {title} {\enquote {\bibinfo {title} {{The Xerographic
  Distribution: Scientific Reasoning in a Large Universe}},}\ }\bibfield
  {booktitle} {\emph {\bibinfo {booktitle} {{Proceedings, 6th International
  Symposium on Quantum Theory and Symmetries (QTS6): Lexington, Kentucky, USA,
  July 20-25, 2009}}},\ }\href {\doibase 10.1088/1742-6596/462/1/012050}
  {\bibfield  {journal} {\bibinfo  {journal} {J. Phys. Conf. Ser.}\ }\textbf
  {\bibinfo {volume} {462}},\ \bibinfo {pages} {012050} (\bibinfo {year}
  {2013})},\ \Eprint {http://arxiv.org/abs/1004.3816} {arXiv:1004.3816
  [hep-th]} \BibitemShut {NoStop}%
\bibitem [{\citenamefont {Page}(1997)}]{Page_volume}%
  \BibitemOpen
  \bibfield  {author} {\bibinfo {author} {\bibfnamefont {Don~N.}\ \bibnamefont
  {Page}},\ }\bibfield  {title} {\enquote {\bibinfo {title} {Space for both
  no-boundary and tunneling quantum states of the universe},}\ }\href {\doibase
  10.1103/PhysRevD.56.2065} {\bibfield  {journal} {\bibinfo  {journal} {Phys.
  Rev. D}\ }\textbf {\bibinfo {volume} {56}},\ \bibinfo {pages} {2065--2072}
  (\bibinfo {year} {1997})}\BibitemShut {NoStop}%
\bibitem [{\citenamefont {Buch}(1994)}]{Buch1994}%
  \BibitemOpen
  \bibfield  {author} {\bibinfo {author} {\bibfnamefont {P.}~\bibnamefont
  {Buch}},\ }\bibfield  {title} {\enquote {\bibinfo {title} {Future prospects
  discussed},}\ }\href {http://dx.doi.org/10.1038/368107b0} {\bibfield
  {journal} {\bibinfo  {journal} {Nature}\ }\textbf {\bibinfo {volume} {368}},\
  \bibinfo {pages} {107} (\bibinfo {year} {1994})}\BibitemShut {NoStop}%
\bibitem [{\citenamefont {Gott}(1994)}]{Gott1994}%
  \BibitemOpen
  \bibfield  {author} {\bibinfo {author} {\bibfnamefont {J.~Richard}\
  \bibnamefont {Gott}},\ }\bibfield  {title} {\enquote {\bibinfo {title}
  {Future prospects discussed---{G}ott {R}eplies},}\ }\href@noop {} {\bibfield
  {journal} {\bibinfo  {journal} {Nature}\ }\textbf {\bibinfo {volume} {368}},\
  \bibinfo {pages} {108} (\bibinfo {year} {1994})}\BibitemShut {NoStop}%
\bibitem [{\citenamefont {Caves}(2000)}]{Caves_crit_Gott}%
  \BibitemOpen
  \bibfield  {author} {\bibinfo {author} {\bibfnamefont {Carlton~M.}\
  \bibnamefont {Caves}},\ }\bibfield  {title} {\enquote {\bibinfo {title}
  {Predicting future duration from present age: A critical assessment},}\
  }\href {\doibase 10.1080/001075100181105} {\bibfield  {journal} {\bibinfo
  {journal} {Contemporary Physics}\ }\textbf {\bibinfo {volume} {41}},\
  \bibinfo {pages} {143--153} (\bibinfo {year} {2000})},\ \Eprint
  {http://arxiv.org/abs/https://doi.org/10.1080/001075100181105}
  {https://doi.org/10.1080/001075100181105} \BibitemShut {NoStop}%
\bibitem [{Cav()}]{CavesNote}%
  \BibitemOpen
  \href@noop {} {}\bibinfo {note} {{C}aves later \cite{Caves2008} gave a
  geometric argument that the prior must go $\sim 1/T^2$ to obtain {G}ott's
  result, though he did not differentiate between a Pick and Be
  prior.}\BibitemShut {Stop}%
\bibitem [{\citenamefont {Caves}(2008)}]{Caves2008}%
  \BibitemOpen
  \bibfield  {author} {\bibinfo {author} {\bibfnamefont {Carlton~M.}\
  \bibnamefont {Caves}},\ }\href@noop {} {\enquote {\bibinfo {title}
  {Predicting future duration from present age: {R}evisiting a critical
  assessment of {G}ott's rule},}\ } (\bibinfo {year} {2008}),\ \Eprint
  {http://arxiv.org/abs/arXiv:0806.3538} {arXiv:0806.3538} \BibitemShut
  {NoStop}%
\bibitem [{\citenamefont {Bayes}(1763)}]{BayesTheorem}%
  \BibitemOpen
  \bibfield  {author} {\bibinfo {author} {\bibfnamefont {Thomas}\ \bibnamefont
  {Bayes}},\ }\bibfield  {title} {\enquote {\bibinfo {title} {Lii. an essay
  towards solving a problem in the doctrine of chances. {B}y the late {R}ev.
  {M}r. {B}ayes, {F. R. S.} communicated by {M}r. {P}rice, in a letter to
  {J}ohn {C}anton, {A. M. F. R. S.}}}\ }\href {\doibase 10.1098/rstl.1763.0053}
  {\bibfield  {journal} {\bibinfo  {journal} {Philosophical Transactions}\
  }\textbf {\bibinfo {volume} {53}},\ \bibinfo {pages} {370--418} (\bibinfo
  {year} {1763})},\ \Eprint
  {http://arxiv.org/abs/http://rstl.royalsocietypublishing.org/content/53/370}
  {http://rstl.royalsocietypublishing.org/content/53/370} \BibitemShut
  {NoStop}%
\bibitem [{\citenamefont {Elga}(2000)}]{Elga_SBP}%
  \BibitemOpen
  \bibfield  {author} {\bibinfo {author} {\bibfnamefont {Adam}\ \bibnamefont
  {Elga}},\ }\bibfield  {title} {\enquote {\bibinfo {title} {Self-locating
  belief and the {S}leeping {B}eauty problem},}\ }\href {\doibase
  10.1111/1467-8284.00215} {\bibfield  {journal} {\bibinfo  {journal}
  {Analysis}\ }\textbf {\bibinfo {volume} {60}},\ \bibinfo {pages} {143--147}
  (\bibinfo {year} {2000})}\BibitemShut {NoStop}%
\bibitem [{\citenamefont {Lewis}(2001)}]{Lewis_SBP}%
  \BibitemOpen
  \bibfield  {author} {\bibinfo {author} {\bibfnamefont {David}\ \bibnamefont
  {Lewis}},\ }\bibfield  {title} {\enquote {\bibinfo {title} {Sleeping
  {B}eauty: reply to {E}lga},}\ }\href {\doibase 10.1111/1467-8284.00291}
  {\bibfield  {journal} {\bibinfo  {journal} {Analysis}\ }\textbf {\bibinfo
  {volume} {61}},\ \bibinfo {pages} {171--76} (\bibinfo {year}
  {2001})}\BibitemShut {NoStop}%
\bibitem [{\citenamefont {Arntzenius}(2002)}]{Arntzenius}%
  \BibitemOpen
  \bibfield  {author} {\bibinfo {author} {\bibfnamefont {Frank}\ \bibnamefont
  {Arntzenius}},\ }\bibfield  {title} {\enquote {\bibinfo {title} {Reflections
  on {S}leeping {B}eauty},}\ }\href {\doibase 10.1111/1467-8284.00330}
  {\bibfield  {journal} {\bibinfo  {journal} {Analysis}\ }\textbf {\bibinfo
  {volume} {62}},\ \bibinfo {pages} {53--62} (\bibinfo {year} {2002})},\
  \Eprint
  {http://arxiv.org/abs/https://onlinelibrary.wiley.com/doi/pdf/10.1111/1467-8284.00330}
  {https://onlinelibrary.wiley.com/doi/pdf/10.1111/1467-8284.00330}
  \BibitemShut {NoStop}%
\bibitem [{\citenamefont {Pust}(2008)}]{Pust2008}%
  \BibitemOpen
  \bibfield  {author} {\bibinfo {author} {\bibfnamefont {Joel}\ \bibnamefont
  {Pust}},\ }\bibfield  {title} {\enquote {\bibinfo {title} {Horgan on
  {S}leeping {B}eauty},}\ }\href@noop {} {\bibfield  {journal} {\bibinfo
  {journal} {Synthese}\ }\textbf {\bibinfo {volume} {160}},\ \bibinfo {pages}
  {97--101} (\bibinfo {year} {2008})}\BibitemShut {NoStop}%
\bibitem [{\citenamefont {Papineau}\ and\ \citenamefont
  {Dur{\`a}-Vil{\`a}}(2009)}]{Papineau01012009}%
  \BibitemOpen
  \bibfield  {author} {\bibinfo {author} {\bibfnamefont {David}\ \bibnamefont
  {Papineau}}\ and\ \bibinfo {author} {\bibfnamefont {V{\'\i}ctor}\
  \bibnamefont {Dur{\`a}-Vil{\`a}}},\ }\bibfield  {title} {\enquote {\bibinfo
  {title} {A thirder and an {E}verettian: a reply to {L}ewis's `{Q}uantum
  {S}leeping {B}eauty'},}\ }\href {\doibase 10.1093/analys/ann012} {\bibfield
  {journal} {\bibinfo  {journal} {Analysis}\ }\textbf {\bibinfo {volume}
  {69}},\ \bibinfo {pages} {78--86} (\bibinfo {year} {2009})}\BibitemShut
  {NoStop}%
\bibitem [{\citenamefont {Rosenthal}(2009)}]{Rosenthal}%
  \BibitemOpen
  \bibfield  {author} {\bibinfo {author} {\bibfnamefont {Jeffrey~S.}\
  \bibnamefont {Rosenthal}},\ }\bibfield  {title} {\enquote {\bibinfo {title}
  {A mathematical analysis of the {S}leeping {B}eauty problem},}\ }\href@noop
  {} {\bibfield  {journal} {\bibinfo  {journal} {The Mathematical
  Intelligencer}\ }\textbf {\bibinfo {volume} {31}},\ \bibinfo {pages} {32--37}
  (\bibinfo {year} {2009})}\BibitemShut {NoStop}%
\bibitem [{\citenamefont {Horgan}(2008)}]{Horgan2008}%
  \BibitemOpen
  \bibfield  {author} {\bibinfo {author} {\bibfnamefont {Terry}\ \bibnamefont
  {Horgan}},\ }\bibfield  {title} {\enquote {\bibinfo {title} {Synchronic
  {B}ayesian updating and the {S}leeping {B}eauty problem: reply to {P}ust},}\
  }\href {\doibase 10.1007/s11229-006-9121-1} {\bibfield  {journal} {\bibinfo
  {journal} {Synthese}\ }\textbf {\bibinfo {volume} {160}},\ \bibinfo {pages}
  {155--159} (\bibinfo {year} {2008})}\BibitemShut {NoStop}%
\bibitem [{\citenamefont {Bostrom}(2007)}]{Bostrom_SBP}%
  \BibitemOpen
  \bibfield  {author} {\bibinfo {author} {\bibfnamefont {Nick}\ \bibnamefont
  {Bostrom}},\ }\bibfield  {title} {\enquote {\bibinfo {title} {Sleeping
  {B}eauty and self-location: {A} hybrid model},}\ }\href@noop {} {\bibfield
  {journal} {\bibinfo  {journal} {Synthese}\ }\textbf {\bibinfo {volume}
  {157}},\ \bibinfo {pages} {59--78} (\bibinfo {year} {2007})}\BibitemShut
  {NoStop}%
\bibitem [{\citenamefont {Neal}(2006)}]{Sailors_Child}%
  \BibitemOpen
  \bibfield  {author} {\bibinfo {author} {\bibfnamefont {Radford~M.}\
  \bibnamefont {Neal}},\ }\bibfield  {title} {\enquote {\bibinfo {title}
  {Puzzles of anthropic reasoning resolved using full non-indexical
  conditioning},}\ }\href@noop {} {\bibfield  {journal} {\bibinfo  {journal}
  {arXiv:math/0608592}\ } (\bibinfo {year} {2006})}\BibitemShut {NoStop}%
\end{thebibliography}%

\appendix
\Section{Notation}  \label{Section_Notation}
Consider two sets, $A$ and $B$.
We will write $AB$ to mean the compound set that consists of set $B$, and of set $A$ that is \emph{nested} in $B$, by which we mean that
every element of $A$ is associated with exactly one element of $B$.
If, for example, $A$ is a set of nuts and $B$ is a set of jars, then $AB$ is a set of jars with nuts in them.
Formally, every element $A_i \in A$ has a secondary label $j$ which corresponds to a specific element $B_j \in B$.  So $A_{i,j}$ is an element of $A$ which is associated with (or, usually, ``in'') an element $B_j$ of $B$; and $A_{,j}$ denotes all elements in $A$ which correspond to a given $B_j$.
But we do not usually refer to labels for individual elements.  Instead we focus on subsets.  Let us  
define subsets $A_a$ and $B_b$ of sets $A$ and $B$ by properties $a$ and $b$, such as the subset of all nuts which are peanuts or cashews, or the subset of large or small jars.  We will assume that these subsets are nonoverlapping and form a complete basis, \textit{i.e.}, 

\begin{equation}
\label{complete_basis}
A = \bigcup_a A_a,\ A_a \neq \emptyset,
\ A_a \cap A_{a' \neq a} =\emptyset, 
\end{equation}

\noindent 
and the same for $B$ and $B_b$ (in our example above, all the nuts are peanuts or cashews, all the jars large or small).  Further, we can define $A_{a,b}$ to be the subset of $A$ whose elements all belong to $A_a$ and correspond to some element in subset $B_b$, \textit{e.g.} all cashews in small jars, $A_{c,S}$, are in the set of cashews $A_c$ and are ``in'' a small jar (they correspond to an element in $B_S$).  Note that the set $A$ is the union of all its nonoverlapping subsets: $A = \bigcup_{a,b} A_{a,b}$.  Further, the subset $A_{,b}$ is the union of all subsets corresponding to label $b$, independent of $a$, \emph{i.e.}, $A_{,b} = \bigcup_{a} A_{a,b}$.  For example, $A_{,S}$ is the set of all nuts in small jars, which is the union of peanuts in small jars ($A_{p,S}$) and cashews in small jars ($A_{c,S}$).

Let us define the number of elements of $A$, $A_a$, $A_{,b}$, and $A_{a,b}$, to be $n$, $n_a$, $n_{,b}$ and $n_{a,b}$, and the number of elements of $B$ and $B_b$ to be $N$ and $N_b$.  Note that since $A$ is the union of nonoverlapping subsets $A_{a,b}$, we have $n=\sum_a n_a=\sum_{b} n_{,b}=\sum_{a,b} n_{a,b}$, and since $B$ is the union of the nonoverlapping subsets $B_b$, we have $N=\sum_b N_b$.  We also define the number of elements in a subset \emph{normalized} by the number of elements in its next enclosing set with an overbar:

\begin{eqnarray}
\label{normalized_number}
\label{define_bar_n}
\bar{n} &\equiv& \frac{n}{N} =\sum_{b} \bar{n}_{,b} \frac{N_b}{N}, \\ 
\label{define_bar_n_a}
\bar{n}_{a}  &\equiv& \frac{n_{a}}{N}=\sum_{b} \bar{n}_{a,b} \frac{N_b}{N}, \\
\label{define_bar_n_b}
\bar{n}_{,b}  &\equiv& \frac{n_{,b}}{N_b}, \\ 
\label{define_bar_n_ab}
\bar{n}_{a,b} &\equiv& \frac{n_{a,b}}{N_b}.
\end{eqnarray}
 
 \noindent
(Note that all $N_b\neq0$ by definition, see \Eq{complete_basis}.) For example, $\bar{n}_{c,S}=n_{c,S}/N_S$ is the average number of cashews per small jar, which is the number of cashews in small jars divided by the number of small jars; and $\bar{n}_{c}$ is the average number of cashews per jar, which is the sum of the average number of cashews in each type of jar weighted by the fraction of jars that are of that type: $\bar{n}_{c}=\sum_{J}^{L,S} \bar{n}_{c,J} (N_J/N)$ ($J$ is summed over $S$ and $L$).

In most of the problems we consider, the leftmost set will be $P$, a set of people, and the set it is nested in, $W$, is a set of worlds of some kind.  The main subset of the leftmost set we will be interested in is `$d$', those people matching datum $d$.  Since we will often contrast the number of people, $n$, with the number of people matching datum $d$, $n_d$, and that is the only subset we need to worry about, we drop the comma before nesting subset label $b$, and define $m$:

\begin{eqnarray}
\label{definem}
n_{b} \equiv n_{,b},\ m \equiv n_d, \ m_b \equiv n_{d,b}.
\end{eqnarray}

We are interested in the probability of selecting an element of some set that belongs to a subset of that set.  We will assume that the selection is random and the same for each element, so that the probability of selection is equal to the fraction of elements in the subset (if this is not the case, we can always make it so by weighting the number of elements of the subsets by some scaling factors---see Section \ref{Section_Typicality} on typicality).  Let us define $P(A_a)$, to mean ``the probability that a randomly selected element of set $A$ belongs to subset $A_a$."  Note that $P(A)=1$, since an element selected from $A$ belongs to $A$ by definition.  So $P(A_a)=P(A_a | A)$ because the conditional $A$ just means that ``an element was randomly selected from $A$", which is already part of the definition of $A_a$.  With these assumptions,

\begin{equation}
\label{P(Aa)}
P(A_a)=P(A_a|A)=\frac{n_a}{n}=\frac{\bar{n}_a}{\bar{n}}, \ P(B_b)=\frac{N_b}{N}.
\end{equation}

\noindent
Note that we can thus replace $(N_b/N)$ in Eqs. (\ref{define_bar_n}, \ref{define_bar_n_a}) with $P(B_b)$.   For example, if $A$ is the set of cards in a deck, $P(A_{clubs})=1/4$, 
and $P(A_{aces})=1/13$.

So long as we are selecting from one set only, there is no ambiguity.  But if we are selecting from compound set $AB$ with set $A$ nested in set $B$, there are two possibilities: either we first select an element of $B_j$ of $B$, and then an element $A_{i,j}$ which corresponds to (is ``in'') element $B_j$, which we call to \textit{Pick}; \textit{or} we directly select the element $A_i$, despite being nested in set $B$, which we define as to \textit{Be}.  One has to \textit{pick} a nut from a jar: select a jar $B_j$ and then select a nut from within the jar.  But if the elements of $A$ are themselves observers, say prisoners in specific cellblocks, there is another way to select: You can \textit{be} a prisoner in a cellblock without having to perform a cellblock selection---you are just there.  (It is possible to Pick directly from set $A$ even if it is nested in $B$, if the correspondence between $A_{i,j}$ and $B_j$ is not really to be ``in'' it.  For example, set $B$ could correspond to a label, $S$ or $L$, we place on each nut, and we toss them all together and randomly select one.  No jar selection is needed to do that, yet the nesting is preserved by the labeling.  We mention this briefly in Section \ref{Section_Warden} with the Warden Cafeteria Problem.)

\textit{Be} probabilities are simple, just the fraction of elements in the inner set meeting the criteria:

\begin{eqnarray}
\label{P(AB)} 
P(AB) & = &P(A)=\frac{n}{n}=1, \nonumber \\
\label{P(ABb)} 
P(A B_b) &=& P(A_{,b})
=\frac{n_{,b}}{n}
=\frac{\bar{n}_{,b}}{\bar{n}} P(B_b), \nonumber \\
\label{P(AaB)} P(A_a B) &=& P(A_a)
=\frac{n_a}{n}
=\frac{\bar{n}_a}{\bar{n}},  \\
\label{P(AaBb)} 
P(A_a B_b) &=&P(A_{a,b})
=\frac{n_{a,b}}{n}
=\frac{\bar{n}_{a,b}}{\bar{n}} P(B_b). \nonumber
\end{eqnarray}

\textit{Pick} probabilities are weighted by the selection that first must be made on set $B$.  We use a superscripted vertical bar $\Pick$ to indicate a Pick from the set immediately to its right.   It is akin to a conditional within the statement, \emph{e.g.}, ``$A_a \Pick B_b$'' means ``we pick an element of type $b$ from set $B$ and then from the elements of $A$ corresponding to that element of $B$ we select an element of $A$ that is in subset $A_a$.''  This is the same as saying ``we picked an element in $A_a$ from $A$ given that we picked an element of $B_b$ from $B$.''  If there are no subset labels indicated to the left of a Pick, then the situation is as if we are ignoring that set.  So $P(A \Pick B_b)=P(B_b)$ because after we pick an element type $b$ from $B$ with probability $P(B_b)$, it is certain that the element we pick from $A$ is from subset $A$ (which is just the whole set $A$).  (We assume that there is some such element of $A$, \emph{i.e.}, $A_{a,b} \neq \emptyset$.)  If there are subsets specified to the left of the Pick, such as in $P(A_a \Pick B_b)$, we can write it as a product of conditional probabilities defined below, $P(A_a \Pick B_b | A \Pick B)=P(A_a \PickNeutered B_b | A \PickNeutered B_b) P( A \Pick B_b)$.  Note that we have put a slash through the Picks in the first term of the righthand side.  We will call such Picks \emph{neutered} because we are conditioning on the fact that an element was chosen from subset $B_b$, and thus no action is needed before selecting the element from $A$.  Thus, the probability with a neutered Pick is the same as for a Be, \emph{e.g.},

\begin{equation}
P(A_a \PickNeutered B_b | A \PickNeutered B_b) = P(A_a  B_b | A  B_b) = \frac{\bar{n}_{a,b}}{\bar{n}_{,b}}.
\end{equation}

\noindent
For example, the probability of picking a small jar and then picking a cashew \emph{given} that one picked a small jar, is the same as picking a cashew \emph{given} that one picked a small jar.  So the Pick probabilities are,

\begin{eqnarray}
P(A \Pick B) &=& 1 \nonumber \\
\label{P(APickBb)}
P(A \Pick B_b)&=&P(B_b), \nonumber \\
\label{P(AaPickB)} 
P(A_a \Pick B) &=& \sum_{b} P(A_a \Pick B_{b})= \sum_{b} \frac{\bar{n}_{a,b}}{\bar{n}_{,b}} P(B_{b}),  \\
\label{P(AaPickBb)}
P(A_a \Pick B_b)&=&P(A_a \PickNeutered B_b | A \PickNeutered B_b) P(A \Pick B_b)
= \frac{\bar{n}_{a,b}}{\bar{n}_{,b}} P(B_b). \nonumber
\end{eqnarray}

\noindent
The astute reader may wonder why the selection on the leftmost set differs from the selection of the sets to its right.  Actually, it does not, and we could put a ``$\Pick$" to the left of every leftmost set.  But our notation \emph{assumes} that there is a selection on the leftmost set.  So really ``$\Pick$" means a selection done on a set other than the leftmost set.  
(Note that one can have a set to the left of an observer, and then one needs to insert a selection ``$\Pick$" to the left of the observers set, \emph{e.g.}, $C \Pick P$, where $C$ are cards and $P$ are observers, and although that observer is Be-selected (\emph{i.e.}, just \emph{is}), this is exclusive selection since there is a selection other than on the innermost set.)

Let us explore conditional probabilities, such as the ones we employed above, where there is one set of selections \emph{given} another.  Here are the nontrivial possibilities (keeping in mind that $P(A_a B | A B_b)=P(A_a B_b | A B_b)$ etc):

\begin{enumerate}
\item 
$P(A_a B | A B_b)$: the probability that we select an element of type $a$ from $A$ nested in $B$ \emph{given} that we select an element of $A$ that corresponds to an element of $B$ of type $b$ .

\item
$P(A B_b | A_a B)$: the probability that we select an element of $A$ that corresponds to an element of $B$ of type $b$ \emph{given} that we select an element of type $a$ from $A$ nested in $B$.

\item
$P(A_a \Pick B | A \Pick B_b)$: the probability that we select an element of B and then select an element type $a$ from $A$ which is associated with that element of $B$ \emph{given} that we select an element of $B$ of type $b$ and then select an element of $A$ associated with that element of $B$.

\item
$P(A \Pick B_b | A_a \Pick B)$:  the probability  that we select an element of $B$ of type $b$ and then select an element of $A$ associated with that element of $B$ \emph{given}  that we select an element of B and then select an element type $a$ from $A$ which is associated with that element of $B$.
\end{enumerate}

\noindent
For example $P(A \Pick B_S | A_c \Pick B)$ is the probability to pick a small jar and then pick a nut from that jar \emph{given} that we pick some jar and then pick a cashew from it.  There are actually only three nontrivial possibilities because the first and the third are equal since the selection in the third is neutered:

\begin{eqnarray}
P(A_a B | A B_b) &=& P(A_a \PickNeutered B | A \PickNeutered B_b)= 
\frac{P(A_{a,b})}{P(A_{,b})}=\frac{\bar{n}_{a,b}}{\bar{n}_{,b}}, \nonumber \\
\label{P(conditional)}
P(A B_b | A_a B) &=& \frac{P(A_{a,b})}{P(A_{a})}=\frac{\bar{n}_{a,b}}{\bar{n}_{a}} P(B_b), \\
P(A \Pick B_b | A_a \Pick B) &=& \frac{P(A_{a} \Pick B_b)}{P(A_{a} \Pick B)}=
\frac{\frac{\bar{n}_{a,b}}{\bar{n}_{,b}} P(B_b)}{\sum_{b'} \frac{\bar{n}_{a,b'}}{\bar{n}_{,b'}} P(B_{b'})}.
\nonumber
\end{eqnarray}

In Eq. (\ref{P(AaPickB)}) we showed that $P(A_{a} \Pick B_b)$ is not in general equal to $P(B_b)$, because the selection of an element of type $a$ adds a nontrivial weighting factor.  That is because there is an implied conditional $A \Pick B$: we take it as a given that we pick some element of $B$ and then some element associated with that element from the whole set $A$, \emph{i.e.}, $P(A_{a} \Pick B_b)$ means $P(A_{a} \Pick B_b | A \Pick B)$.
But sometimes we want to redefine the set $A$ we select from so that it is some subset of qualifying elements.  For example, if our jars contain peanuts, cashews, and pebbles, but our selection process ensures that only nuts are picked, then we are really concerned with the subset $A_{nut}$ of cashews and peanuts.  To help clarify such situations, we write redefined sets with square brackets $[A_{re}]$.  This new set then has subsets $[A_{re}]_{a,b}$, and we can write the number of elements in these as $[n_{re}]$ and $[n_{re}]_{a,b}$, etc.  Now set $[A_{re}]$ acts like $A$ did in Eq. (\ref{P(AaPickB)}),

\begin{eqnarray}
\label{AaPickBbGivenAaPickB}
P([A_{re}] \Pick B) &=& 1 , \\
P([A_{re}] \Pick B_b)&=& P([A_{re}] \Pick B_b | [A_{re}] \Pick B) =P(B_b), \nonumber \\
P([A_{re}]_a \Pick B)&=& P([A_{re}]_a \Pick B | [A_{re}] \Pick B) \nonumber \\
&=& \sum_{b} \frac{[\bar{n}_{re}]_{a,b}} {[\bar{n}_{re}]_{,b}} P(B_{b}),  \nonumber \\
P([A_{re}]_a \Pick B_b)&=& P([A_{re}]_a \Pick B_b | [A_{re}] \Pick B) \nonumber \\
&=& \frac{[\bar{n}_{re}]_{a,b}} {[\bar{n}_{re}]_{,b}} P(B_b), \nonumber 
\end{eqnarray}

\noindent 
since one selects some element of $[A_{re}]$ with certainty.

Now, one might object that there is a lot of redundant information in the above notation, namely the set labels $A$ and $B$.  We think it is important to retain those labels if there is any confusion about which sets are considered, which subset labels correspond to which set, and which sets have a Pick on them---an issue if there are more than two nested sets.  But if there are only two nested sets which are the same throughout some calculation, and the subscript labels are unique to a set, we can use a compact notation by omitting the set names while preserving the order of any subscript labels and selection bars:

\begin{equation}
\label{shorthandalpha}
P(\alpha\beta) \equiv P(A_\alpha B_\beta), \ P(\alpha \Pick \beta) \equiv P(A_\alpha \Pick B_\beta),
\end{equation}

\

\noindent
where $\alpha$ and $\beta$ can be `null,' \textit{e.g.}, $P(  b | a ) \equiv P( A  B_b | A_a  B)$ and $P( \Pick b | a \Pick) \equiv P( A \Pick B_b | A_a \Pick B)$.  For example, in compact notation, using Eq. (\ref{P(ABb)}-\ref{P(APickBb)}),

\begin{equation}
\label{shorthandBayesBe}
P(b) \equiv P(A B_b) =  \frac{\bar{n}_{,b}}{\bar{n}} P(B_b) = \frac{\bar{n}_{,b}}{\bar{n}} P(\Pick b),
\end{equation}

\noindent
and Bayes' Law with a Pick is,

\begin{equation}
\label{shorthandBayesPick}
P(\Pick b | a \Pick) = \frac{P(a \Pick | \Pick b) P(\Pick b)}{P(a \Pick)}.
\end{equation}

\noindent
We use the more verbose notation in most of the main text for clarity.  Here are the terse versions:  The posterior probability for a Be, \Eq{Prisoner_P(SgivenD)}, becomes

\begin{equation}
\label{Prisoner_P(SgivenD)_Be_short}
P(S|d)=\frac{P(d|S)P(S)}{P(d)}=P(\Pick S),
\end{equation}

\noindent
the posterior probability for a Pick \Eq{Prisoner_P(SgivenD)_Pick} becomes

\begin{eqnarray}
\label{Prisoner_P(SgivenD)_Pick_short}
P( \Pick S | d \Pick )&=&\frac{P(d \PickNeutered  |  \PickNeutered S) P( \Pick S)}{P(d \Pick)} \nonumber \\
&=& \frac{P(\Pick S)}
{P(\Pick S) + \frac{1}{\rho}  P(\Pick L)}.
\end{eqnarray}

We can use our compact formalism for three or more nested sets, but there is then an ambiguity about the location of the Pick.  Does $P(\Pick c)$ mean $P(A \Pick B C_c)$ or $P(A  B \Pick C_c)$?  To avoid this ambiguity, we use a double-Pick mark (and if need a triple-Pick mark) on inner sets, so $P(\DoublePick c) \equiv P(A \Pick B C_c)$ and $P(\Pick c) \equiv P(A B \Pick  C_c)$. For example, the probabilities in Section \ref{Section_Exclusive} using sets $P_d \DoublePick W_S \Pick E_y$ are,

\begin{eqnarray}
\label{Pickswithy}
P(d \Pick) &\equiv& P(P_d W \Pick E), \nonumber \\
P(S \Pick) &\equiv&  P(P W_S \Pick E), \\
P(\Pick y)  &\equiv&  P(P W \Pick E_y) = P(E_y), \nonumber \\
P( \DoublePick S ) &\equiv& P(P \Pick W_S E) = P(W_S E). \nonumber
\end{eqnarray}
We conclude with a table which summarizes our main results in compact notation:


\begingroup \squeezetable
\begin{table*}[htp]
\begin{center}
\caption{Summary of our major results using compact notation where the list of sets provides a key for the location of the Picks. Worlds $J=S$ or $L$. For three or more sets we use a double-Pick mark to avoid ambiguity. For `Probing a Multiverse' $h$ = $in$ or $ex$ (it's probably advisable not to use compact notation for four sets with controlled-Picks). The weighted averages are $\left<f(y)\right>\equiv  \int_0^1 f(y) \, p(\Pick y) dy$, $\left<f(y)\right>_{d^{(\vert)}} \equiv  \int_0^1 f(y) \, p(^{(\vert)}  y | d ^{(\vert)}) dy$. For the Gott case we take $t_m=t$.}
\begin{tabular}{|l|l|l|l|l|l|}
\hline
\bf{Section} & \bf{Description} & \bf{Sets} &  \bf{Input} & \bf{Output} & \bf{Result} \\
\hline

II \& IX & Be Selection  & $P_d \Pick W_{J}$ & $P(S) = \frac{\bar{n}_S}{\bar{n}} P(\Pick S)$ & $P(S|d)= P(\Pick S)$ & $R_{P/W} =1$ \\
III & Pick Selection  & $P_d \Pick W_J$ & $P(\Pick S)$ & $P(\Pick S|d \Pick)= \frac {P(\Pick S)} {P(\Pick S) + \frac{1}{\rho} P(\Pick L)} $ & $R_{P \Pick/W} =\frac{1}{\rho}$ \\

IV & Inclusive Selection & $P_d \DoublePick W_J \Pick E_y $ & $P(S) = \frac{\bar{n}_S}{\bar{n}} P(\DoublePick S)$ & $P(S|d)= P(\DoublePick S)=\left<y\right>$ & $R^E_{P/W} =1$ \\

V & Exclusive Selection &  $P_d \DoublePick W_J \Pick E_y $ & $P(S \Pick)=\left< \frac{y}{\rho - (\rho-1)y} \right>$
& $P(S \Pick | d \Pick )=\left< \frac{y}{\rho - (\rho-1)y} \right> \left< \frac{1}{\rho - (\rho-1)y} \right>^{-1}$ & $R^{\Pick E}_{P /W} 
\in \left[ \frac{1}{\rho},1 \right]$ \\
\hline

VI.A \& IX & Excl. Theory Selection  & $P_d W \Pick \Theta_J$ & $P(\Pick S)$ & $P(\Pick S|d \Pick)= \frac {P(\Pick S)} {P(\Pick S) + \frac{1}{\rho} P(\Pick L)} $ & $R_{P \Pick/\Theta} =\frac{1}{\rho}$
\\
\vspace{-10pt}
&&&&&\\

VI.B 
& Probing a Multiverse 
& $P_d  W \DoublePick \hspace{-2pt} \overleftarrow{\, E_y \Pick \Theta_h \hspace{-7pt}}\hspace{7pt}$ 
& 
$P_h=P( \DoublePick \hspace{-2pt} \overleftarrow{\  \Pick h \hspace{-2pt}} \hspace{2pt}),$
& $P_{h|d}=P( \DoublePick \hspace{-2pt} \overleftarrow{\  \Pick h \hspace{-2pt}} \hspace{2pt} |
\, d \DoublePick \hspace{-2pt} \overleftarrow{ \  \Pick } \,)$
& Prior-dependent\\
&&&$p=P( \DoublePick 1 \PickNeutered | \DoublePick \PickNeutered ex)$&&\\

\hline
VII.E 
&  Atypical Freak Observers
& $^{\xi} P_n$ & $P(n)=\frac{1}{1+ \rho}$ &$P(^{\xi} n)=\frac{1}{1+ \kappa \rho}$ & Need $\kappa\rho\ll1$ \\
VII.F 
& Rare Observers 
& $P \Pick \Theta_J$ 
& $p_J^{\geq1} = 1 - (1-p_{\mathcal{F}})^{n_J}$ 
& $P(\Pick J)_{rare}= P(J)_{p_{\mathcal{F}}} = \frac{n_J}{\langle n \rangle} P(\Pick J)$
& Rare-Pick = Be\\

VII.F
& Rare Freak Observers
& $P_{0 \hspace{-1pt} f} \hspace{-1pt} \Pick \Theta_J$ 
& $P(0 \hspace{-1pt} f \Pick | \Pick J)=(1-p_f)^{n_J}$
& $R^f_{P \Pick \Theta} =(1-p_f)^{n_L- n_S} R_\Theta$
& $R^f_{P \Pick / \Theta} \to e^{-p_f n_L} $\\

\hline
VIII \& IX 
& Incl. Gott \& 
& $P_t \Pick W_T$ 
&  $p(\Pick T)  \hspace{-1pt}  \sim \frac{1}{T^2}   \hspace{-1pt} \Rightarrow   \hspace{-1pt}   p(T)  \hspace{-1pt} \sim \frac{1}{T}$
& $p(T | t) \to  \frac{t}{T^2}$
& $R_T \to 1$ \\
&No Doomsday 
&&
& $P( (T> Kt) | t)\to \frac{1}{K}$ 
& $R_{\int T} \to1$ \\

\vspace{-8pt}
&&&&&\\

VIII \& IX 
& Excl. Gott \& 
& $P_t \Pick W_T$ 
&  $p(\Pick T)  \hspace{-1pt}  \sim \frac{1}{T} $
& $p( \Pick T | t \Pick ) \to  \frac{t}{T^2}$
& $R_{\Pick T} \to \frac{t}{T}$
 \\
&Doomsday 
&&
& $P( \Pick (T> Kt) | t \Pick) \to \frac{1}{K}$ 
& $R_{\int \Pick T} \to \frac{1}{K}$
 \\

\hline
X
& Incl. Universal Doomsday
& $P_d \DoublePick W_J \Pick E_y $
& $p(\Pick y) \Rightarrow \left< y \right>$
& $\left< y \right>_d = P(S | d)= \left< y \right>$ [same as Sec. IV]
& $R^{UD}_{P/W} = R^{E}_{P/W}$ \\

X
& Excl. Universal Doomsday
& $P_d \DoublePick W_J \Pick E_y $
& $p(\Pick y) \Rightarrow \left< y \right>$
& $\left< y \right>_{d \Pick} = P( S \Pick | d \Pick)=$ [same as Sec. V]
& $R^{\Pick UD}_{P /W} = R^{\Pick E}_{P/W}$ \\

\hline
XI & Beauty Thirder/Incl. &  $P_{\rm Day} \Pick W_{\rm flip}$ & 3 observer moments & 
$P(H)=\frac{1}{3}$ &Need 2:1 odds \\

XI &  Beauty Halfer/Excl. &  $P_{\rm Day} \Pick W_{\rm flip}$ & 2 observer timelines & 
$P(\Pick H)=\frac{1}{2}$ & Need 2:1 odds \\
XI & Mon. Beauty Thirder/Incl. &  $[P_{\rm Mon}] \Pick W_{\rm flip}$ & 2 observer moments & 
$P(H)=\frac{1}{2}$ &Need 1:1 odds \\

XI &  Mon. Beauty Halfer/Excl. &  $[P_{\rm Mon}] \Pick W_{\rm flip}$ & 2 observer timelines & 
$P(\Pick H)=\frac{1}{2}$ & Need 1:1 odds \\

\hline
\end{tabular}
\end{center}
\label{default}
\end{table*}
\endgroup

\end{document}